\documentclass[11pt]{article} 

\usepackage{amsmath,amsthm,latexsym,amssymb,amsfonts,epsfig,dsfont,graphicx,float,subcaption}
\usepackage[utf8]{inputenc}
\usepackage{color}
\usepackage{bbm}

\newcommand{\lr}{\left(}
\newcommand{\rr}{\right)}
\newcommand{\ls}{\left[}
\newcommand{\rs}{\right]}
\newcommand{\dummy}{\bigstar}

\newcommand{\be}{\begin{equation}}
\newcommand{\ee}{\end{equation}}
\newcommand{\ba}{\begin{eqnarray}}
\newcommand{\ea}{\end{eqnarray}}

\title{{\sf Reduced Loop Quantization with four Klein-Gordon Scalar Fields as Reference Matter}}
\author{
{\sf K. Giesel$^1$\thanks{{\sf 
kristina.giesel@gravity.fau.de}} , A. Vetter$^1$\thanks{{\sf 
almut.vetter@fau.de}} }
\\
{\sf $^1$ Institute for Quantum Gravity (IQG)} \\ {FAU Erlangen -- N\"urnberg,}\\
{\sf Staudtstr. 7, 91058 Erlangen, Germany}\\
}
\date{{\small\sf \today}}

\makeatletter
\@addtoreset{equation}{section}
\makeatother

\begin{document} 

\maketitle

{\sf
\begin{abstract}
We perform a reduced phase space quantization of gravity using four Klein-Gordon scalar fields as reference matter as an alternative to the Brown-Kuchar dust model in  \cite{Giesel:2007wn}, where dust scalar fields are used. We compare our results to an earlier model by Domagala et. al. \cite{Domagala:2010bm} where only one Klein-Gordon scalar field is considered as reference matter for the Hamiltonian constraint but the spatial diffeomorphism constraints are quantized using Dirac quantization. As a result we find that the choice of four conventional Klein-Gordon scalar fields as reference matter leads to a reduced dynamical model that cannot be quantized using loop quantum gravity techniques. However, we further discuss a slight generalization of the action for the four Klein-Gordon scalar fields and show that this leads to a model which can be quantized in the framework of loop quantum gravity. By comparison of the physical Hamiltonian operators obtained from the model by Domagala et. al. \cite{Domagala:2010bm} and the one introduced in this work we are able to make a first step towards comparing Dirac and reduced phase space quantization in the context of the spatial diffeomorphism constraints.
\end{abstract}
}

\section{Introduction}
\label{s1}
In the last years several different models describing the dynamics of loop quantum gravity (LQG) have been introduced
\cite{Giesel:2007wn,Domagala:2010bm,Husain:2011tk,Domagala:2012tq}. 
A common property of all these dynamical models is that they introduce additional matter fields which serve as reference matter
for either only the temporal coordinate or the temporal and spatial coordinates respectively. In the first case one needs one matter reference field, whereas in the latter one needs four of them, i.\,e. one matter field per constraint. In the framework of the relational formalism \cite{Rovelli:all,Dittrich04,Thiemann2004} these reference fields, also often called clocks, are used to construct Dirac observables with respect to either only the Hamiltonian or the Hamiltonian and spatial diffeomorphism constraints that are present in the ADM formulation of general relativity. In case only a partial reduction with respect to the Hamiltonian constraint is performed, the remaining spatial diffeomorphism constraints are quantized using Dirac quantization as done in  \cite{Domagala:2010bm}. The main difference between Dirac and reduced phase space quantization is that for a reduced phase space quantization only the reduced phase space involving only the physical degrees of freedom, but no gauge degrees of freedom, is quantized. Hence, after quantization one obtains directly the physical Hilbert space. On the other hand for Dirac quantization one quantizes the kinematical phase space, where the constraints under consideration have not yet been reduced and the corresponding gauge degrees of freedom are still present. After Dirac quantization one obtains the so-called kinematical Hilbert space. On this Hilbert space the vanishing of the classical constraints carries over to the requirement that physical states are annihilated by the associated constraint operators. Note that the notion kinematical might be misleading in case we consider a model where constraints are handled partially by Dirac and partially by reduced phase space quantization because the kinematical Hilbert space obtained is in general different from the one where all constraints are treated via Dirac quantization. Though in both cases we apply Dirac quantization and therefore we still end up with an intermediate kinematical Hilbert space which is not a physical Hilbert space yet. An example where a combination of Dirac and reduced quantization has been used and the corresponding Hilbert spaces have been analyzed in detail is the model in \cite{Domagala:2010bm}.

Following \cite{Giesel:2012rb} these models can be classified as type I and type II models. Models of type I are characterized by containing two pairs of four scalar fields and are usually a second class system. If one reduces the system with respect to the second class constraints one pair of the four scalar fields can be expressed in terms of the remaining degrees of freedom and one ends up with a first class system for which the remaining
four scalar fields can be used as reference matter. This first class system is then the starting point for the reduced phase space quantization. Thus, a full reduction with respect to the Hamiltonian and spatial diffeomorphism constraint is possible. An example for such a model is the Brown-Kuchar dust model, that has been introduced by Kucha\v{r} et al in their seminal papers  \cite{Kuchar:1995xn,Bicak:1997bx,Brown:1994py} and has been used in \cite{Giesel:2007wn} to perform a reduced phase space quantization of loop quantum gravity. On the other hand for models of type II only a partial reduction can be obtained for the reason that these models
include only one reference field usually used as reference matter associated with the Hamiltonian constraint. An example for a model of type II that has been applied in the context of loop quantum gravity is the model in  \cite{Domagala:2010bm}, where one Klein-Gordon scalar field has been considered as reference matter. The motivation for this model came from loop quantum cosmology where in the Ashtekar-Pawlowski-Singh (APS) model introduced in
\cite{Ashtekar:2006wn} also one Klein-Gordon scalar field is used as a clock. The model in  \cite{Domagala:2010bm} can be understood as a generalization of the APS-model to full loop quantum gravity. 

Now, when going over from the cosmological setting to the full theory, we have to decide how we deal with the spatial diffeomorphism constraints. In  \cite{Domagala:2010bm} these have been treated using Dirac quantization. In this work we want to extend the class of models of type I in such a way that we can consider a model for the full theory with four Klein-Gordon scalar fields as reference matter which allows in contrast to \cite{Domagala:2010bm} not only to apply for the Hamiltonian constraint but also for the spatial diffeomorphism constraints a reduced phase space quantization. In this sense we can understand the model presented here as the corresponding model of type I associated with the type II model presented in \cite{Domagala:2010bm}. Likewise to the model in \cite{Domagala:2010bm} the model presented in this work can also be seen as an equally justified extension of the APS-model to full loop quantum gravity. Particularly, analyzing the dynamical operators of the model presented here as well as the model in \cite{Domagala:2010bm} yields the possibility to get a first insight on possible differences in the quantum theory when either Dirac or reduced quantization is used for the spatial diffeomorphism constraints.

We will start with the most naive model of type I associated with the model of type II in \cite{Domagala:2010bm} that involves the Einstein-Hilbert action and four additional Klein-Gordon scalar fields. As we will show in section \ref{s2} this yields to a reduced model with a physical Hamiltonian which cannot be quantized using the loop quantum gravity representation. A Hamiltonian in the context of Dirac observables is called physical, since it generates the dynamics of the Dirac observables. The reason why our first physical Hamiltonian is not quantizable is the way how combinations of the observable corresponding to the geometric part of the spatial diffeomorphism constraints denoted by $C^{\rm geo}_j$, that involves the contribution from the gravitational degrees of freedom only,  enter into it. It turns out that this leads to a term, namely $\delta^{jk} C^{\rm geo}_j C^{\rm geo}_k$, which cannot be quantized because the infinitesimal generators of the spatial diffeomorphism constraints $C^{\rm geo}_j$ do not exist at the quantum level due to the lack of weak continuity for finite spatial diffeomorphisms in the loop quantum gravity representation, see also the end of section \ref{s24} for more details. As a consequence, comparing this naive model with the one in \cite{Domagala:2010bm} we get very different results. The model where spatial diffeomorphism constraints are treated via Dirac quantization works, whereas the corresponding reduced model cannot even be quantized in the context of loop quantum gravity. Hence, the conclusion of our work is that the naive model is not appropriate for performing a reduced phase space quantization using the usual loop quantum gravity representation.

With this result given we will consider in section \ref{s3} a slightly generalized model of the four Klein-Gordon scalar field case along the lines of other dust models of type I. The model in section \ref{s3} can be understood as model involving seven scalar fields. Likewise to other models of type I, this yields to a system that has second class constraints. We perform the reduction with respect to the second class constraints that reduces the three additional degrees of freedom and end up with a system that involves next to gravity one conventional and three (generalized) Klein-Gordon scalar fields and which has first class constraints only. The reason why we do not consider 8 scalar fields from the beginning as it has been done in in the dust models of type I in  \cite{Kuchar:1995xn,Bicak:1997bx,Brown:1994py} is that we want to be as close as possible to the model in \cite{Domagala:2010bm} and therefore want to choose for the reference field associated with the Hamiltonian constraint the same in both models. Thus, we only generalize the Klein-Gordon scalar fields that play the role of reference fields for the spatial diffeomorphism constraints. As we show in section \ref{s3} the corresponding reduced model leads to a physical Hamiltonian that can be quantized using loop quantum gravity techniques. There the physical Hamiltonian involves a contribution of the form $Q^{jk}C^{\rm geo}_jC^{\rm geo}_k$. It is exactly the appearance of the inverse spatial metric $Q^{jk}$ which makes it possible to quantize this  physical Hamiltonian in the loop quantum gravity representation.  In section \ref{s34} we present the details of the regularization and quantization of the physical Hamiltonian corresponding the generalized model in the usual loop quantum gravity framework as well as in the Algebraic Quantum Gravity (AQG) model from \cite{Giesel:2006uj}. Furthermore, we compare the physical Hamiltonian operator obtained in the model presented here with the one from \cite{Domagala:2010bm}. This provides the possibility to make first steps towards the comparison of Dirac and reduced phase space quantization in the context of the spatial diffeomorphism constraints at the level of the dynamical operators.

For the reason that we perform a reduced phase space quantization for both models the one in section \ref{s2} as well as the one in \ref{s3}, we briefly summarize the main three steps that need to performed in this approach below:
\begin{itemize}
\item{\underline{Step 1: Construction of Observables}}
\\
First, we need to perform a reduction with respect to the constraints of the system. 
Since loop quantum gravity is based on a formulation of general relativity in terms of Ashtekar variables
this includes the Hamiltonian, the spatial diffeomorphism as well as an additional SU(2) gauge constraint.
Note, that in all current available models the latter is solved by Dirac quantization and therefore not considered 
in the reduction of the classical theory.
We will follow the same line in our work here and derive the partially reduced phase space with respect to
the Hamiltonian and spatial diffeomorphism constraint and solve the Gauss constraint via Dirac quantization
at the quantum level. The classical reduction is obtained using the relational formalism that,
given a set of reference fields,  provides a formalism to construct observables.

\item{\underline{Step 2: Dynamics of the Observables on the Reduced Phase Space}}
\\
As a second step we have to derive the dynamics for the constructed observables.
Since by definition they Poisson commute with the Hamiltonian and spatial diffeomorphism constraints
their dynamics is no longer generated by the canonical ADM Hamiltonian.
This is also called the problem of time in the context of general relativity.
We will denote the generator of the dynamics of the observables physical Hamiltonian because,
as we will discuss below, it has similar properties than the Hamiltonian in unconstrained systems.

\item{\underline{Step 3: Reduced Phase Space Quantization}}
\\
Finally, given the reduced phase space, we want to obtain the corresponding quantum theory via canonical quantization.
For this purpose the algebra of observables needs to be computed and one has to find representations thereof.
In general the algebra of observables can be more complicated than the corresponding kinematical algebra.
However, for the existing models as well as for the model discussed here, the chosen reference matter has the feature
that the associated algebra of observables is isomorphic to the kinematical algebra. Hence, to find a representation
of this algebra, that corresponds to finding the physical Hilbert space, is not more difficult than quantizing
the kinematical theory. 
Furthermore, we are only interested in those representations for which the dynamics encoded in the physical Hamiltonian,
can be implemented as a well defined operator on the physical Hilbert space.
\end{itemize}
The paper is structured as follows:
\\
In section \ref{s2} we will discuss a model that includes four Klein-Gordon scalar fields and we perform the first two steps of the reduced quantization program in section \ref{s22} and \ref{s23} because as mentioned above the physical Hamiltonian obtained in the second step cannot be quantized using loop quantum gravity techniques. We discuss the latter result in section \ref{s24}.
In section \ref{s3} we will generalize the four Klein-Gordon scalar field model by adding in addition three more scalar fields. As we will show for this generalized model the reduced quantization program can be completed. After an analysis of the dynamics of the generalized model in section \ref{s31} the steps 1 and 2 of the reduced quantization can be found in section \ref{s32} and \ref{s33} respectively. In section \ref{s34} we present the technical details of step 3 involving the regularization and quantization of the physical Hamiltonian operator of the generalized model. Finally in section \ref{s4} we summarize our results and conclude. In addition we have moved longer calculations into the appendix. This involves a comparison between the reduced model and a corresponding gauge fixed model along the lines of the discussion in appendix H of \cite{Giesel:2007wi}, as well as details about the construction of the observables and some details on the stability analysis of the constraints in the generalized model.

\section{Four Klein-Gordon Scalar Fields as Reference Matter}
\label{s2}
The first model we want to discuss here is general relativity with four additional reference fields. 
This model can be understood as the natural type I  model associated with the one scalar field model in
\cite{Domagala:2010bm} which originally was considered because it is the full loop quantum gravity 
 generalization of the Ashtekar-Pawlowski-Singh (APS) model introduced in
\cite{Ashtekar:2006wn}\footnote{More recently a model by Dapor and Liegener \cite{Dapor:2017rwv} was introduced that derives the dynamics of the cosmological model using cosmological coherent states from full LQG.}. We assume that each of the reference fields is a Klein-Gordon scalar field.
Thus, the action of the \rm total system under consideration is given by
\begin{equation*}
S[g,\varphi^I]=\int\limits_M d^4X \sqrt{g}R^{(4)}
-\frac{1}{2}\int\limits_M d^4X \sqrt{g}\delta_{IJ}g^{\mu\nu}\varphi^I_{,\mu}\varphi^J_{,\nu}
=S^{\rm geo}+S^{\varphi},
\end{equation*}
where $g_{\mu\nu}$ is the space-time metric, $g := |\det(g_{\mu\nu})|$, $R^{(4)}$ denotes the four-dimensional Ricci scalar,
$\mu,\nu=0,\cdots,3$ are space-time indices and $I,J=0,\cdots, 3$ label the four Klein-Gordon scalar fields.
Note that the latter index is just an internal one labeling the reference matter fields and has no relation to the
space-time indices. We choose our signature convention for the space-time metric tensor $g_{\mu\nu}$ to be $(-,+,+,+)$.
We restrict our discussion to the ADM variables here. Since all the obtained results here can be straightforward 
carried over to case of Ashtekar variables. Applying the ADM formalism, where dot denotes the derivative with
respect to the time parameter $t$ in the ADM frame, we end up with the following canonical action
\begin{equation*}
S[q_{ab},p^{ab},n,p,n^a,p_a,\varphi^J,\pi_J]=\int\limits_{\mathbb{R}} dt\int\limits_{\chi} d^3x
\left(\dot{q}_{ab} p^{ab}+\dot{\varphi}^J \pi_J+\dot{n} p+\dot{n}^a p_a-\left[nc^{\rm tot}+n^ac_a^{\rm tot}+\nu{z}+\nu^a{z}_a\right]\right),
\end{equation*}
with primary Hamiltonian
\begin{eqnarray*}
 H_{\rm primary} &=& \int\limits_{\chi} \! d^3x \, h_{\rm primary}= \int\limits_{\chi} \! d^3x \,\left( n c^{\rm tot} + n^a c_a^{\rm tot} +  \nu z + \nu^a z_a \right),
\end{eqnarray*}
with
\begin{eqnarray*}
  z :=p,\quad {z}_a:=p_a,\quad c^{\rm tot}:=c^{\rm geo}+ c^{\varphi}, \quad c^{\rm tot}_a:=c^{\rm geo}_a+c^{\varphi}_a
\end{eqnarray*}
and
\begin{eqnarray}
\label{Constraints2}
\kappa c^{\rm geo}&=&\frac{1}{\sqrt{q}}\left(q_{ac}q_{bd}-\frac{1}{2}q_{ab}q_{cd}\right)p^{ab}p^{cd}-\sqrt{q}R^{(3)}, \nonumber\\
c^{\varphi}&=&\frac{\delta^{IJ}\pi_I\pi_J}{2\sqrt{q}}+\frac{1}{2}\sqrt{q}\delta_{IJ}q^{ab}\varphi^I_{,a}\varphi^J_{,b}\nonumber, \\
\kappa c^{\rm geo}_a&=&-2q_{ac}D_bp^{bc},\nonumber\\
c^{\varphi}_a&=&\pi_J\varphi^J_{,a},
\end{eqnarray}
here $\kappa = 16 \pi G$ where $G$ is Newton's gravitational constant, $D_a$ is the torsion free metric compatible connection with respect to the ADM metric and $q:=\det(q_{ab})$, $n$ and $n^a$ denote the lapse function and shift vector respectively and $\nu,\nu^a$ are Lagrange multipliers associated with the primary constraints $z$ and $z_a$ respectively. To analyze the time evolution of the primary constraints $z$ and $z_a$ under the primary Hamiltonian we notice that the 
non-vanishing Poisson brackets on the phase space are given by 
\begin{align}
 &\{  q_{cd}(x),p^{ab}(y)\} = \kappa \delta^a_{(c} \delta^b_{d)} \delta^{(3)}\!(x,y), \\
 &\{n(x),p(y)\} = \delta^{(3)}\!(x,y), \\
 &\{ n^b(x), p_a(y)\} = \delta^a_b \delta^{(3)}\!(x,y), \\
 &\{  \varphi^I(x),\pi_J(y)\} = \delta^I_J \delta^{(3)}\!(x,y).
\end{align}
 The analysis of the stability of the primary constraints shows that $c^{\rm tot}$ and $c_a^{\rm tot}$
 are the secondary constraints of the system.
\begin{align}
 & \dot{z} = \{ z,  H_{\rm primary} \} = \{ p,  H_{\rm primary} \} = - c^{\rm tot}, \\
 & \dot{z}_a = \{ z_a,  H_{\rm primary} \} = \{p_a, H_{\rm primary}\} = - c_a^{\rm tot}.
\end{align}
No tertiary constraints arise, since we are in a similar situation as in \cite{Giesel:2007wi}, for a prove see appendix \ref{a1} there.
As expected each of the four reference fields $\varphi^I$ contributes to the Hamiltonian and diffeomorphism constraint with the standard expression of a Klein-Gordon scalar field. The set of constraints $\{{z},{z}_a,c^{\rm \rm tot},c^{\rm \rm tot}_a\}$ is first class. Now we go to the reduced ADM phase space for which ${z}\approx 0$ and ${z}_a\approx 0$ and in this phase space we can treat lapse and shift as Lagrange multipliers. Before we actually discuss the construction of observables in \ref{s22} we will briefly review the general formalism in the next subsection, where we will very closely follow the presentation in \cite{Han:2015jsa}.
\subsection{Brief Review on Observables in the context of the Relational Formalism}
\label{s21}
The relational formalism provides a framework in which the dynamics of general relativity can be formulated in terms of Dirac observables. Their evolution is governed by a so called physical Hamiltonian. In the following we will briefly summarize the main ideas of the formalism and introduce the notation necessary for the work done here. For a more detailed introduction we refer the reader for instance to \cite{Han:2015jsa}.

The starting point is  a system with a set of first class constraints denoted by $\{C_I\}$ labelled by an arbitrary index $I$.  In order to obtain for each $C_I$ a quantity that is at least weakly canonically conjugate to it, we introduce reference fields $T^I$, one per $C_I$ and these need to satisfy $\{T^I,C_J\}=N^{I}_J$ with $N$ being an invertible matrix. This allows to define a set of equivalent constraints that are weakly Abelian and given by
\be
 C'_I:=\sum\limits_{J}(N^{-1})_I^JC_J
\ee
that obviously define the same constraint hypersurface and for which we have $\{T^I,C'_J\}\approx \delta^I_J$. As reviewed in \cite{Han:2015jsa} given the set of Abelian constraints we can define a map that sends each phase space function $f$ to its gauge invariant extension also called Dirac observable as
\be
O_f(\tau):=\left[\sum\limits_{n=0}^\infty\frac{(-1)^n}{n!}X^n_\beta\cdot f\right]_{\beta=\tau-T}~,
\ee
where we introduced the following sum of Hamiltonian vector fields $X_\beta=\sum\limits_I \beta^IX_I$ and $X_I$ denotes the Hamiltonian vector field associated with $C'_I$. Furthermore, $X^n_\beta\cdot f=\{C_\beta,f\}_{(n)}$, with $C_\beta=\sum\limits_I\beta^IC'_I$ and $\{.,.\}_{(n)}$ denoting the iterative Poisson bracket defined
through $\{C_\beta,f\}_{(0)}=f$ and $\{C_\beta,f\}_{(n)}=\{C_\beta,\{C_\beta,f\}_{(n-1)}\}$\footnote{The additional factor $(-1)^n$ in the definition of the map comes from the fact that we use $\{q^A,p_B\}=\delta^A_B$ here, in contrast to \cite{Giesel:2007wi} where $\{q^A,p_B\}=-\delta^A_B$ is used.}. The interpretation of this map is that it returns the values of $f$ at those values where the reference fields $T^I$ take the values $\tau^I$, where we suppressed the index at the $T^I$'s and $\tau^I$'s.

Let us briefly list the main properties of these Dirac observables that have been proven in \cite{Dittrich04,Thiemann2004,Henneaux1992}:
\begin{itemize}
\item[(i)]
$O_f(\tau)$ is a Dirac observable, that is $\{O_f(\tau),C_I\}\approx 0$ for all $I$.
\item[(ii)]
For a given phase space function $f$ where we denote the elementary variables with $(q^A,p_A)$ we have
\be
O_f(\tau)=f(O_{q_A},O_{p^A})(\tau).
\ee
\item[(iii)]
One can show that the Poisson bracket of two observables is weakly equivalent to the observables of the Dirac bracket of the corresponding gauge variant functions, that is 
\ba
\{O_f(\tau),O_{g}(\tau)\}\approx \{O_f(\tau),O_g(\tau)\}^*
\approx O_{\{f,g\}^*}(\tau),
\ea
with the Dirac bracket defined by
\be
\{f,g\}^{\ast}:=\{f,g\}-\sum\limits_{I,J}\{f,\phi_I\}(M^{-1})^{IJ}\{\phi_J,g\},
\ee
where $\phi_I$ denote the constraints of the system and we have the invertible so-called Dirac matrix $M_{IJ}:=\{\phi_I, \phi_J\}$.
In our case the set of constraints is given by the commuting reference fields $T^I$ and the constraints $C_I$. Then the Dirac bracket becomes
\be
\{f,g\}^*:=\{f,g\}-\{f,C_I\}(N^{-1})^I_J\{T^J,g\}+\{g,C_I\}(N^{-1})^I_J\{T^J,f\}.
\ee
\end{itemize}
The second point tells us that it is sufficient to construct observables of the elementary phase space variables and the last point will particularly be important when we consider the quantization of the reduced observable algebra. 
\\
\\
In the special case of deparametrization, that will be also relevant in our work, the situation simplifies. In that case the phase space can be divided into two sets of canonical pairs one for the reference fields $(T^I,P_I)$ and the other for the remaining variables $(q^a,p_a)$ with the property that the set of constraints can be rewritten as
\be
C_I=P_I+h_I(q^a,p_a),
\ee
that is linearly in the reference field momenta and the constraints are independent of the reference fields $T^I$. Note that this might be also obtained only partially that is for a subset of the constraints $C_I$. In the fully deparametrized case we have  
\be
\{T^I,C_J\}=\delta^I_J~.
\ee
In addition due to the fact that constraints are linearly in the clock momenta they form an Abelian algebra and this carries over to the associated Hamiltonian vector fields and they commute and in this case here not only on the constraint surface but on the entire phase space. As consequence all weak equalities mentioned above become strong equalities. For a phase space function that is independent of the reference field degrees of freedom the observable map simplifies to 
\be
\label{OfDep}
O_f(\tau)=\sum\limits_{n=0}^\infty\frac{(-1)^n}{n!}X^n_\tau\cdot f,
\ee
where $X_\tau$ is the Hamiltonian vector field of the function 
\be
H_{\tau}=\sum\limits_{I}(\tau^I-T^I)H_I,
\ee
where $H_I:=O_{h_I}(\tau)$ denotes the observables associated with $h_I$.
Using the property in (ii) we obtain $H_I$ as $H_I=O_{h_I}(\tau)=h_I(Q^a,P_a)(\tau)$. For deparametrization $h_I=H_I$ is already a Dirac observable because we have $\{h_I,C_J\}=0$. Let us denote the observable associated with all non-reference field degrees of freedom  $(q^a,p_a)$ by $Q^a(\tau)$ and $P_a(\tau)$. Then considering the fact that they commute with all $T^I$'s we obtain for their algebra
\be
\{Q^a(\tau),P_b(\tau)\}=\{O_{q^a}(\tau),O_{p_b}(\tau)\}=O_{\{q^a,p_b\}}(\tau)=O_{\delta^a_b}(\tau)=\delta^a_b~,
\ee
Hence, in this case the Poisson algebra of the Dirac observables agrees with the algebra of the gauge variant quantities for these degrees of freedom.
\\
\\
For an observable $O_f(\tau)$ associated with a function that depends only on $q^a$ and $p_a$ the time evolution for $O_f(\tau)$ can be described by the change of $O_f(\tau)$  with respect to $\tau^0$ since this encodes how $O_f(\tau)$ varies with time $\tau^0$ if we choose $T^0$ as the reference field referring to physical time. As shown in \cite{Thiemann2004} this can be written in form of a  Hamilton's equation of motion given by 
\be
\frac{\partial O_f(\tau)}{\partial\tau_0}=\{O_f(\tau),H_0\},
\ee
where $H_0:=\int d^3x O_{h_0}$  is a time independent Hamiltonian in the case of deparametrization. In what follows we call the generator of the dynamics of the observables $H_0$ physical Hamiltonian. Here we restricted our discussion to the case of deparametrization, but as has been shown in \cite{Giesel:2012rb} and will be also important for the models discussed in this paper if the system does not deparametrize but the function $h_0$ depends on the partial derivatives of $T^0$ only, then the final physical Hamiltonian $H_0$ will be still independent of time.
\subsection{Step 1: Construction of Observables}
\label{s22}
Now we will use the formalism introduced in section \ref{s21} and apply it to the four scalar field model in order to
construct observables with respect to the Hamiltonian and spatial diffeomorphism constraint.
For this purpose as a first step we have to rewrite the Hamiltonian as well as the spatial diffeomorphism constraint 
in an equivalent form such that the set of resulting constraints becomes weakly Abelian. To achieve this we will use 
the same strategy as in \cite{Giesel:2007wi}, that is firstly solving the four constraints for the four reference field
momenta $\pi_J$ and then apply the so called Brown-Kucha\v{r} mechanism in order to ensure that the final
physical Hamiltonian is given in deparametrized form. 
\subsubsection{Weakly Abelian Set of Constraints}
We start with the spatial diffeomorphism constraint $c_a^{\rm tot}$ and want to solve it for $\pi_j$.
In order that the scalar fields $\varphi^j$ with $j=1,2,3$ serve as good reference fields we have to
assume that $\varphi: \chi \rightarrow \mathcal{S}$ is a diffeomorphism, where ${\cal S}$ denotes the scalar field manifold consisting of the values the fields $\varphi^j$ can take.
We denote by $\varphi^a_j$ the inverse of $\varphi^j_{,a}$,
such that $\varphi^a_j \varphi^j_{,b} = \delta^a_b$, $\varphi^a_k \varphi^j_{,a} = \delta^j_k$.
Using this we can solve for  $\pi_j$ and get
\begin{equation}
\label{Solvepij}
 c_a^{\rm tot} = 0 \quad \Leftrightarrow \quad  \pi_j = -\varphi^b_j \lr c_b^{\rm geo} + \pi_0 \varphi^0_{,b} \rr=:-\widetilde{h}_j(q_{ab},p^{ab}, \varphi^0,\varphi^j,\pi_0)=:-\widetilde{h}_j.
\end{equation}
 Further, we want to solve $c^{\rm tot}$ for $\pi_0$. 
 Considering the explicit form of $c^{\rm tot}$ in (\ref{Constraints2})
 multiply $c^{\rm tot}$ 
 with $2 \sqrt{q}$ and reinsert the result for the momenta $\pi_j$ from (\ref{Solvepij}) into it,
 where the last step is known as the Brown-Kucha\v{r} mechanism. 
 Note, that we apply the Brown-Kucha\v{r} mechanism not in its standard form here because then we would replace
 $q^{ab}\varphi^0_{,a}\varphi^0_{,b}$ by $\frac{q^{ab}(c^{\rm geo}_a+\pi_j\varphi^j_{,a})(c^{\rm geo}_b+\pi_j\varphi^j_{,b})}{\pi^2_0}$,
 but here we use the spatial diffeomorphism constraint to replace $\pi_j$. 
 The advantage of this is that we get at most a quadratic equation in $\pi_0$ and not a fourth order one as in $\cite{Domagala:2010bm}$ yielding in general to a more complicated form of the final physical Hamiltonian. These steps lead to
\begin{equation*}
  -2 \sqrt{q} c^{\rm \rm geo} = {\pi_0}^2 + \delta^{jk}\varphi^a_j \lr c_a^{\rm geo} + \pi_0 \varphi^0_{,a} \rr \varphi^b_k \lr c_b^{\rm geo} + \pi_0 \varphi^0_{,b} \rr
  + q \delta_{JK} q^{ab} \varphi^J_{,a} \varphi^K_{,b}.
\end{equation*}
This is a quadratic equation for the scalar field momentum $\pi_0$ and can be rewritten as
\begin{eqnarray*} 
 0 &= \lr 1+ \delta^{jk}\varphi^a_j \varphi^b_k  \varphi^0_{,a} \varphi^0_{,b}\rr {\pi_0}^2 
 +\delta^{jk}\varphi^a_j \varphi^b_k \lr c_a^{\rm geo} \varphi^0_{,b} +  c_b^{\rm geo} \varphi^0_{,a} \rr \pi_0 \\
 &+ q \delta_{JK} q^{ab} \varphi^J_{,a} \varphi^K_{,b} +  \delta^{jk}\varphi^a_j \varphi^b_k c_a^{\rm geo} c_b^{\rm geo}  +  2 \sqrt{q} c^{\rm geo}.
\end{eqnarray*}
Let us define the following abbreviations
\begin{eqnarray*}
 a&:=& \lr 1+ \delta^{jk}\varphi^a_j \varphi^b_k  \varphi^0_{,a} \varphi^0_{,b}\rr, \\
 b&:=& \delta^{jk}\varphi^a_j \varphi^b_k \lr c_a^{\rm geo} \varphi^0_{,b} +  c_b^{\rm geo} \varphi^0_{,a} \rr,\\
 c&:= &q \delta_{JK}q^{ab} \varphi^J_{,a} \varphi^K_{,b} +  \delta^{jk}\varphi^a_j \varphi^b_k c_a^{\rm geo} c_b^{\rm geo}  +  2 \sqrt{q} c^{\rm geo},
\end{eqnarray*}
then solving for $\pi_0$ yields
\begin{equation}
\label{SolvePi0}
\pi_0=- \frac{b}{2a} \pm \sqrt{ \lr\frac{b}{2a}\rr^2 - \frac{c}{a}}=: -h(q_{ab}, p^{ab}, \varphi^0, \varphi^j) =:-h.
\end{equation}
Note, that the application of the Brown-Kucha\v{r} mechanism in its standard way does not result in a form of the Hamiltonian constraint that can be written linearly in $\pi_0$ and a function that does not depend on the remaining scalar field momenta $\pi_j$. In order to ensure later on that the physical Hamiltonian density is positive we choose the plus sign in the definition of $h$.
Now we will use the results in (\ref{Solvepij}) and (\ref{SolvePi0}) to write down an equivalent set of constraints that is linearly in the scalar field momenta. We obtain
\begin{eqnarray}
\label{NewConstraints}
c^{\rm tot}&:=&\pi_0+h(q_{ab}, p^{ab}, \varphi^0, \varphi^j), \nonumber\\
c^{\rm tot}_j&:=&\pi_j+h_j(q_{ab}, p^{ab}, \varphi^0, \varphi^j),
\end{eqnarray}
where we used $\pi_0=-h$ to obtain from $\widetilde{h}_j$  a function $h_j$ that no longer depends on the momentum $\pi_0$.
Note, that this result also coincides with \cite{Kuchar:1991}, where a model with eight scalar fields was considered
to implement the harmonic gauge condition. 
This second class model can be reduced to a first class model with four remaining scalar fields of the Klein-Gordon type.
We realize that neither the new Hamiltonian constraint nor the spatial diffeomorphism constraint is in deparametrized form
for the reason that the function $h$ as well as the functions $h_j$ still depend on the scalar fields. 
However, as pointed out in \cite{Giesel:2012rb} in case these functions depend only on spatial derivatives of the
reference fields the final resulting physical Hamiltonian will still be time-independent and this is exactly the
case for the present model as we will show in the next subsection.
In contrast to the old constraints the constraints shown in (\ref{NewConstraints}) are weakly Abelian and can thus
be used to construct observables for the \rm geometric degrees of freedom using the four scalar fields as reference fields.
In the following we will construct the observables in two steps. First we reduce with respect to the spatial diffeomorphism
constraint and afterwards with respect to the Hamiltonian constraint. 
\subsubsection{Explicit Construction of the Observables}
For the construction of the observables we can closely follow \cite{Giesel:2007wi} where four dust reference fields
are used. 
Likewise to the case of the dust reference fields, we will construct the final observable in two steps.
First, we derive spatially diffeomorphism invariant quantities. For this purpose, as in \cite{Giesel:2007wi}, 
we define the smeared constraint 
\be
K_{\beta_1}:=\int\limits_{\chi} d^3x\, \beta^j_1 c_j^{\rm tot}.
\ee
Observables with respect to $K_{\beta_1}$ are given by
\be
O_{f,\{\varphi^j\}}^{(1)}(\sigma)=\sum\limits_{n=0}^{\infty}\frac{(-1)^n}{n!}\Big[\{K_{\beta_1},f\}_{(n)}\Big]_{\beta_1^j=\sigma^j - \varphi^j}.
\ee
For the dust reference fields in \cite{Giesel:2007wi} an explicit form of the inductive Poisson bracket
$\{K_{\beta_1},f\}_{(n)}$ in terms of vector fields $v_j$ acting on a scalar $g$ by $v_j\cdot g(x):=S^a_jg_{,a}$ 
was derived, where $S^j$ denotes the reference dust fields and $S^a_j$ the inverse of $S^j_{,a}$ .
All the steps used \cite{Giesel:2007wi} in order to prove the explicit form of the inductive Poisson bracket go
through also for the scalar field reference fields $\varphi^j$.
We just have to replace $S^a_j$ by $\varphi^a_j$. 
For the benefit of the reader we have reviewed the proof in the appendix in section \ref{a1}. 
Using this result we consequently  obtain for the case that $f$ is a scalar,
e.g. some function $g : \chi \mapsto \mathbb{R}$ on $\chi$
\be
\label{claimdiff}
\{K_{\beta_1}, g(x)\}_{(n)}=\big[\beta^{j_1}_1 ... \beta^{j_n}_1v_{j_1} ... v_{j_n}\cdot g\big](x)
\ee
with $v_j\cdot g(x)=\varphi^a_j g_{,a}(x)$.
Hence the spatially diffeomorphism invariant quantity for $g$ is given by
\be
O_{g,\{\varphi^j\}}^{(1)}(\sigma)=g+\sum\limits_{n=1}^{\infty}\frac{(-1)^n}{n!}\big[\sigma^{j_1} - \varphi^{j_1}\big] ... \big[\sigma^{j_n} - \varphi^{j_n}\big]v_{j_1} ... v_{j_n}\cdot g.
\ee
We have $v_j\cdot \varphi^k=\varphi^a_j\varphi^k_{,a}=\delta^k_j$. In equation (\ref{vkObs}) in \ref{a1} we calculated
the action of the vector field $v_k$ on $O_{g,\{\varphi^j\}}^{(1)}(\sigma)$. The result is given by
\ba
v_k\cdot O_{g,\{\varphi^j\}}^{(1)}(\sigma)&=&
\sum\limits_{n=1}^{\infty}\frac{(-1)^n}{n!}\beta^{j_1}_1 ...\beta^{j_n}_1
\big[v_k\sigma^j\big]v_jv_{j_1} ... v_{j_n}\cdot g.
\ea
As explained in the appendix we are allowed to choose any $\sigma^j$ and a convenient choice is $\sigma^j$ to be constant.
This requires that $\varphi^j$ is invertible for $j=1,2,3$ which is an assumption entering the whole construction and means
that $\varphi^j : \chi\mapsto  {\cal S}$ can be understood as a diffeomorphism, where we denote with $\cal S$ the scalar reference field manifold.
Hence, for a scalar $g$ on $\chi$ we therefore obtain the following explicit integral representation for the spatially diffeomorphism invariant expression
\be
\label{Odiffs}
O_{g,\{\varphi^j\}}^{(1)}(\sigma)=\int\limits_{\chi} d^3x\, \big|\det(\partial\varphi^j/\partial_x)\big|\delta(\varphi^j(x),\sigma^j) g(x).
\ee
Now, as introduced in \cite{Giesel:2007wi} for the quantities that are no scalars on $\chi$ we use the
$(\varphi^j)^{-1} : {\cal S} \mapsto \chi$ to pull back tensors that become scalars on $\chi$ but tensors of 
same rank on ${\cal S}$ where we denote the physical space being the range of $\sigma^j$ within ${\cal S}$.
Explicitly, we construct for all variables that are not reference fields for $c_j^{\rm tot}$ using the abbreviation $J:=|\det(\varphi^j/\partial_x)|$ the following quantities
\be
\varphi^0,\quad \pi_0/J,\quad q_{jk}=q_{ab}\varphi^a_j\varphi^b_k,\quad 
p^{jk}=p^{ab}\varphi^j_{,a}\varphi^k_{,b}/J,
\ee
where  $J$ is used to transform the scalar/tensor densities of weight one $\pi_0$ and $p^{ab}$ into true scalars/tensors.
The integral representations of the corresponding observables are then given by
\ba
\widetilde{\varphi}^0&:=&O_{\varphi^0,\{\varphi^j\}}^{(1)}(\sigma)= 
\int\limits_{\chi} d^3x\, \big|\det(\partial\varphi^j/\partial_x)\big|\delta(\varphi^j(x),\sigma^j)
\varphi^0(x),\nonumber\\
\widetilde{\pi}_0&:=&O_{\pi_0,\{\varphi^j\}}^{(1)}(\sigma)=
\int\limits_{\chi} d^3x\, \delta(\varphi^j(x),\sigma^j)
\pi_0(x),\nonumber\\
\widetilde{q}_{jk}&:=&O_{q_{ab},\{\varphi^j\}}^{(1)}(\sigma)=
\int\limits_{\chi} d^3x\, \big|\det(\partial\varphi^j/\partial_x)\big|\delta(\varphi^j(x),\sigma^j)
\varphi^a_j\varphi^b_kq_{ab}(x),\nonumber\\
\widetilde{p}^{jk}&:=&O_{p^{ab},\{\varphi^j\}}^{(1)}(\sigma)=
\int\limits_{\chi} d^3x\, \delta(\varphi^j(x),\sigma^j)
\varphi_a^j\varphi_b^kp^{ab}(x),
\ea
where we will denote spatially diffeomorphism invariant quantities with a tilde. 
For the degrees of freedom that adopt the role of a reference field for $c_j^{\rm tot}$ we get
\ba
\widetilde{\varphi}^j&=&O_{\varphi^j,\{\varphi^j\}}^{(1)}(\sigma)=\Big[\alpha^{K_{\beta_1}}_{\beta_1}(\varphi^j)\Big]_{\alpha^{K_{\beta_1}}_t(\varphi^j)=\sigma^j}=\sigma^j
=\int\limits_{\chi} d^3x\, \big|\det(\partial\varphi^j/\partial_x)\big|\delta(\varphi^j(x),\sigma^j)
\varphi^j(x),\nonumber\\
\widetilde{\pi}_j&=&O_{\pi_j,\{\varphi^j\}}^{(1)}(\sigma)
=\int\limits_{\chi} d^3x\, \big|\det(\partial\varphi^j/\partial_x)\big|\delta(\varphi^j(x),\sigma^j)
\pi_j(x).\nonumber\\
\ea
For the spatially diffeomorphism invariant version of the constraints $\widetilde{c}^{\rm tot}$ and $\widetilde{c}^{\rm tot}_{a}$
thus we obtain:
\ba
\widetilde{c}^{\rm tot}&=&\widetilde{\pi}_0 +\widetilde{h},\nonumber\\
\widetilde{c}^{\rm tot}_j&=&\widetilde{\pi}_j +\widetilde{h}_j=\widetilde{\pi}_j +\widetilde{c}_j^{\rm geo} - \widetilde{h}\widetilde{\varphi}^0_{,j}
=\widetilde{\pi}_j -2\widetilde{q}_{j\ell}D_{k}\widetilde{p}^{k\ell} - \widetilde{h}\widetilde{\varphi}^0_{,j},
\ea
where we used that 
\be
O_{\varphi^j_{,a}}^{(1)}(\sigma)=\int\limits_{\chi} d^3x\, \big|\det(\partial\varphi^j/\partial_x)\big|\delta(\varphi^j(x),\sigma^j)
\varphi^j_{,a}\varphi^a_k=\delta^j_k,
\ee
 and likewise $O_{\varphi^a_{,j}}^{(1)}(\sigma)=\delta_j^k$ and
with 
\ba
\label{htilde}
\widetilde{h}&=&
\frac{1}{1+\widetilde{\varphi}^0_{,j}\widetilde{\varphi}^0_{,k}\delta^{jk}}
\times\Big(-\widetilde{c}_j^{\rm geo}\widetilde{\varphi}^0_{,k}\delta^{jk}
\nonumber\\
&&+ \sqrt{\Big[\widetilde{c}_j^{\rm geo}\widetilde{\varphi}^0_{,k}\delta^{jk}\big]^2
-\Big(1+\widetilde{\varphi}^0_{,j}\widetilde{\varphi}^0_{,k}\delta^{jk}\Big)\Big[2\sqrt{\det(\widetilde{q})} \widetilde{c}^{\rm geo} 
+ \det(\widetilde{q})\widetilde{q}^{jk}\big(\widetilde{\varphi}^0_{,j}\widetilde{\varphi}^0_{,k}+\delta_{jk}\big) +  \widetilde{c}^{\rm geo}_j\widetilde{c}^{\rm geo}_k\delta^{jk}}\Big]\Big)
,\nonumber\\
\widetilde{c}^{\rm geo}
&=&
\frac{1}{\sqrt{\det(\widetilde{q})}}\Big(\widetilde{q}_{j\ell}\widetilde{q}_{km} - \frac{1}{2}\widetilde{q}_{jk}\widetilde{q}_{\ell m}\Big)\widetilde{p}^{jk}\widetilde{p}^{\ell m} - \sqrt{\det(\widetilde{q})}R(\widetilde{q}).
\ea
Next, we will continue with constructing full observables that are also invariant under $\widetilde{c}^{\rm tot}$.
As before we denote the smeared Hamiltonian constraint as
\be
\widetilde{K}_{\beta_2}:=\int\limits_{\cal S} d^3\sigma\, \beta_2\widetilde{c}^{\rm tot}.
\ee
Then the observables are given by the power series
\ba
\label{Obsfphi}
O_{f,\{\varphi^0,\varphi^j\}}(\sigma,\tau)&=&
O^{(2)}_{\widetilde{f}(\sigma),\widetilde{\varphi}^0}(\sigma)(\tau)\\
&=&\sum\limits_{n=0}^{\infty}\frac{(-1)^n}{n!}\Big[\{\widetilde{K}_{\beta_2},\widetilde{f}\}_{(n)}\Big]_{\beta_2=\tau - \widetilde{\varphi}^0}\nonumber\\
&=&
\widetilde{f}(\sigma) \nonumber\\
&&+\sum\limits_{n=1}^{\infty}\frac{(-1)^n}{n!}\int\limits_{\cal S} d^3\sigma'_1 (\tau - \widetilde{\varphi}^0(\sigma'_1)) ... \int\limits_{\cal S} d^3\sigma'_n (\tau - \widetilde{\varphi}^0(\sigma'_n))
\{\widetilde{c}^{\rm \rm tot}(\sigma'_1), ... \{\widetilde{c}^{\rm \rm tot}(\sigma'_n),\widetilde{f}(\sigma)\} ...\}.\nonumber
\ea
Again we want $\int_{\cal S} d^3\sigma\, (\tau - \widetilde{\varphi}^0(\sigma))\widetilde{h}(\sigma)$  to be spatially diffeomorphism invariant. This requires a constant $\tau$.
We will denote full observables by capital letters, explicitly
\ba
Q_{jk}(\sigma,\tau)&:=&O_{q_{ab},\{\varphi^0,\varphi^j\}}(\sigma,\tau)=O_{\widetilde{q}_{jk}(\sigma),\widetilde{\varphi}^0},\nonumber\\
P^{jk}(\sigma,\tau)&:=&O_{p^{ab},\{\varphi^0,\varphi^j\}}(\sigma,\tau)=O_{\widetilde{p}^{jk}(\sigma),\widetilde{\varphi}^0},\nonumber\\
\Pi_{0}(\sigma,\tau)&:=&O_{\pi_{0},\{\varphi^0,\varphi^j\}}(\sigma,\tau)=O_{\widetilde{\pi}_{0}(\sigma),\widetilde{\varphi}^0},\nonumber\\
\Pi_{j}(\sigma,\tau)&:=&O_{\pi_j,\{\varphi^0,\varphi^j\}}(\sigma,\tau)=O_{\widetilde{\pi}_{j}(\sigma),\widetilde{\varphi}^0}.\nonumber\\
\ea
Note, that $\Pi_0$ and $\Pi_j$ are no independent observables because using the constraints in (\ref{NewConstraints})
these
can be expressed in terms of $Q^{jk}$ and $P_{jk}$. Furthermore, we have
\be
O_{\varphi^0,\{\varphi^0,\varphi^j\}}(\sigma,\tau)=\tau\quad{\rm and}\quad O_{\varphi^j,\{\varphi^0,\varphi^j\}}(\sigma,\tau)=\sigma^j.
\ee
\subsection{Step 2: Dynamics encoded in the physical Hamiltonian}
\label{s23}
Likewise to the dust case in \cite{Giesel:2007wi}  this power series for $O_{f,\{\varphi^0,\varphi^j\}}(\sigma,\tau)$ 
cannot be written down in closed form. However, what is more important is that we know an explicit form of
the physical Hamiltonian ${\bf H}_{\rm phys}$ generating the evolution with respect to the physical time $\tau$. 
Hence, we could derive equations of motion for $O_{f,\{\varphi^0,\varphi^j\}}(\sigma,\tau)$.
Solving these equations yields a possibility to obtain an explicit expression for observables.
When choosing dust fields as reference fields it could be shown that ${\bf H}_{\rm phys}$ is the (physical)
space integral over ${\cal S}$ of the observable associated to the function $h$ in $c^{\rm \rm tot}$,
see  \cite{Giesel:2007wi} for more details. The proof that ${\bf H}_{\rm phys}$ generates $\tau$
-- evolution uses the property that $c^{\rm tot}$ deparametrizes for the dust reference fields. 
Nevertheless, as we will show now also in the scalar field case where deparametrization is not present 
${\bf H}_{\rm phys}$ can be expressed as the integral over the observable associated to $h$. 
Let us consider phase space functions $f$ that are independent of the reference field degrees of freedom used
for $c^{\rm \rm tot}$ that is $f$ is not allowed to depend on $\varphi^0$ and/or $\pi_0$. 
Then by considering the explicit power series for observables in equation (\ref{Obsfphi}) we have 
\ba
\frac{d}{d\tau}O_{f,\{\varphi^0,\varphi^j\}}(\sigma,\tau)&=&
\frac{d}{d\tau}O_{\widetilde{f}(\sigma),\widetilde{\varphi}^0(\sigma)}(\tau)\nonumber\\
&=&\sum\limits_{n=1}^\infty \frac{(-1)^n}{(n-1)!}
\int\limits_{\cal S} d^3\sigma'_1 ... \int\limits_{\cal S} d^3\sigma'_n
\{\widetilde{c}^{\rm \rm tot}(\sigma'_1), ... \{\widetilde{c}^{\rm \rm tot}(\sigma'_n),\widetilde{f}(\sigma)\} ...\}
(\tau - \widetilde{\varphi}^0(\sigma'_2)) ... (\tau - \widetilde{\varphi}^0(\sigma'_n))\nonumber\\
&=&-\sum\limits_{n=0}^{\infty}\frac{(-1)^n}{n!}
\int\limits_{\cal S} d^3\sigma'_1 ... \int\limits_{\cal S} d^3\sigma'_n
\{\widetilde{c}^{\rm \rm tot}(\sigma'_1), ... \{\widetilde{c}^{\rm \rm tot}(\sigma'_n),...\nonumber\\
&&\hspace{4cm}\{\int\limits_{\cal S}d^3\sigma'\widetilde{c}^{\rm \rm tot}(\sigma^{\prime}), \widetilde{f}(\sigma)\} ...\}(\tau - \widetilde{\varphi}^0(\sigma'_1)) ... (\tau - \widetilde{\varphi}^0(\sigma'_n))\nonumber\\
&=&
-O_{\{\int\limits_{\cal S}d^3\sigma'\widetilde{c}^{\rm \rm tot}(\sigma^{\prime}), \widetilde{f}(\sigma)\},\widetilde{\varphi}^0(\sigma)}(\tau)
\nonumber\\
&=&
-O_{\{\int\limits_{\cal S}d^3\sigma'\widetilde{h}(\sigma^{\prime}), \widetilde{f}(\sigma)\},\widetilde{\varphi}^0(\sigma)}(\tau) \nonumber\\
&=&
-O_{\{\int\limits_{\cal S}d^3\sigma'\widetilde{h}(\sigma^{\prime}), \widetilde{f}(\sigma)\}^*,\widetilde{\varphi}^0(\sigma)}(\tau)\nonumber\\
&=&
-\{O_{O^{(1)}_{\int\limits_{\chi}d^3x' h(x'),\varphi^j}(\sigma),\widetilde{\varphi}^0(\sigma)}(\tau),O_{O^{(1)}_{f,\varphi^j}(\sigma),\widetilde{\varphi}^0(\sigma)}(\tau)
\}^*\nonumber\\
&=&
-\{O_{\int\limits_{\chi}d^3x' h(x'),\{\varphi^0,\varphi^j\}}(\tau),O_{f,\{\varphi^0,\varphi^j\}}(\sigma,\tau)\}\nonumber\\
&=&-\{O_{\int\limits_{\cal S}d^3\sigma'\widetilde{h}(\sigma^{\prime}),\{\varphi^0,\varphi^j\}}(\tau),O_{f,\{\varphi^0,\varphi^j\}}(\sigma,\tau)\}
\nonumber\\
&=&-\{\int\limits_{\cal S}d^3\sigma' O_{h,\{\varphi^0,\varphi^j\}}(\sigma'\tau),O_{f,\{\varphi^0,\varphi^j\}}(\sigma,\tau)\}=\{\int\limits_{\cal S}d^3\sigma' H(\sigma',\tau),O_{f,\{\varphi^0,\varphi^j\}}(\sigma,\tau)\}\nonumber\\
&=&-\{ {\bf H}_{\rm phys}(\tau), O_{f,\{\varphi^0,\varphi^j\}}(\sigma,\tau)\}\nonumber \\
&=& \{O_{f,\{\varphi^0,\varphi^j\}}(\sigma,\tau),{\bf H}_{\rm phys}(\tau)\}.
\ea
In the third line we used that $\widetilde{c}^{\rm \rm tot}(\sigma)$ mutually commute and in the fifth line that $f$
is by assumption independent of $\varphi^0$ that allows us to replace $\widetilde{c}^{\rm \rm tot}$ by $\widetilde{h}$.
Furthermore, we could use the Poisson bracket instead of the corresponding Dirac bracket because  $\widetilde{f}$ 
(by assumption) does not depend on the reference field momentum $\pi_0$.
Consequently all terms in the Dirac bracket additional to the Poisson bracket vanish. The Dirac bracket here has the following form for the spatial diffeomorphism invariant quantities
\be
\{\widetilde{f},\widetilde{f}'\}^*:= \{\widetilde{f},\widetilde{f}^{\prime}\} 
- \int\limits_{\cal S} d^3\sigma\,
\Big(\{\widetilde{f},\widetilde{c}^{\rm \rm tot}(\sigma)\}\{\widetilde{f}',\widetilde{\varphi}^{0}(\sigma)\}
- \{\widetilde{f}',\widetilde{c}^{\rm \rm tot}(\sigma)\}\{\widetilde{f},\widetilde{\varphi}^{0}(\sigma)\}\Big)
\ee
and for the unreduced case
\be
\{f,f'\}^*:= \{f,f^{\prime}\} 
- \int\limits_{\chi} d^3x\, \sum\limits_{J=0}^3
\Big(\{f,c_{\mu}^{\rm \rm tot}(x)\}\{f',\varphi^{J}(x)\}
- \{f',c_{\mu}^{\rm \rm tot}(x)\}\{f,\varphi^{J}(x)\}\Big)
\ee
with $c_0^{\rm \rm tot}:=c^{\rm \rm tot}$.
In the last before the last line we used the linearity of the observable map and introduced the abbreviation
$H(\sigma,\tau):=O_h(\sigma,\tau)$.
Thus, the physical Hamiltonian in case of the Klein-Gordon scalar field reference field is given by the following
expression
\be
\label{Hphysphi}
{\bf H}_{\rm phys}:=\int\limits_{\cal S}d^3\sigma\, O_{\widetilde{h}(\sigma),\widetilde{\varphi}^0}(\sigma, \tau)=
\int\limits_{\cal S}d^3\sigma\, H(\sigma,\tau),
\ee
here we denote the (full) observable associated to $h$ according to our notation by $H$ and the latter is explicitly
given by
\be
\label{HamDens1}
H(\sigma)
=\sqrt{-\Big(2\sqrt{\det(Q)}C^{\rm geo} + \det(Q)Q^{jk}\delta_{jk} + \delta^{jk} C^{\rm geo}_jC_k^{\rm geo}}\Big)
\ee
and does not depend on the physical time $\tau$ where
\ba
C^{\rm geo}&:=&\frac{1}{\sqrt{\det(Q)}}\Big(Q_{j\ell}Q_{km} - \frac{1}{2}Q_{jk}Q_{\ell m}\Big)P^{jk}P^{\ell } - \sqrt{\det(Q)}R(Q)+2\sqrt{\det(Q)}\Lambda,\nonumber\\
C_j^{\rm geo}&:=&-2Q_{j\ell}D_k P^{k\ell}.
\ea
The reason why $H$ includes less terms than $\widetilde{h}$ in equation (\ref{htilde}) 
and looks less complicated is that all terms involving spatial derivatives of the reference field $\widetilde{\varphi}^0$
can be dropped because $O_{\widetilde{\varphi}^0_{,j},\widetilde{\varphi}^0}(\sigma,\tau)=d\tau/d\sigma^j=0$. 
A side effect of this is that ${\bf H}_{\rm phys}$ although involving still explicit reference field variable 
dependence $\widetilde{\varphi}^0$, is nevertheless a time independent Hamiltonian since only
derivative terms occur. 
However, the additional explicit dependence on the reference fields $\widetilde{\varphi}^j$
survives because their derivatives give a contribution in terms of Kronecker deltas. From the first impression
it sound astonishing that although we started with a full covariant theory, we end up with a
physical Hamiltonian that looks not covariantly due to the occurring Kronecker deltas. 
However, we should keep in mind that the index $j$ in the equation above refers to the label of the scalar
reference fields and is no spatial index of a space-time index. Thus, the non-covariance of the physical
Hamiltonian refers to the manifold ${\cal S}$ associated to the  spatial reference fields $\varphi^j$ 
and there is no guarantee that ${\bf H}_{\rm phys}$ might be covariant there even 
if we start with a covariant action on $\chi$.
\\
 Furthermore, in contrast to the deparametrized dust case here we cannot conclude from the fact that the
 $c^{\rm tot}$'s mutually commute that also the $h$'s do. For this reason it is more complicated to understand in the scalar field case what precise symmetries ${\bf H}_{\rm phys}$ possesses. This will be discussed more in detail in future work.
\subsection{Step 3: Reduced Phase Space Quantization}
\label{s24}
Finally, we would like to complete the quantization program and find a representation of the observables algebra
whose non-vanishing Poisson brackets are given by
\begin{equation*}
\{Q^{jk}(\sigma,\tau),P_{\ell m}(\widetilde{\sigma},\tau)\}=\delta^{j}_\ell\delta^k_m\delta^{(3)}(\sigma,\widetilde{\sigma}).
\end{equation*}
For the reason that we want to apply the quantization used in loop quantum gravity, we 
formulate the \rm geometry phase space in terms of su(2) connections and canonically conjugate
fields $(A^A_a, E^a_A)$, also known as Ashtekar variables, rather than in terms of the ADM variables
$Q^{jk},P_{jk}$, where $A$ is an su(2) index. This describes the geometrical sector of the phase space as an SU(2)
Yang-Mills theory.
As mentioned above, as a consequence we obtain next to the Hamiltonian and spatial diffeomorphism constraint
the so called SU(2) Gauss constraint on the (extended) phase space.
If we perform a symplectic reduction with respect to the Gauss constraint we get back the usual ADM phase space.
Now in the context of Ashtekar variables the observables constructed in \ref{s22} describe a partially reduced
phase space (only with respect to the Hamiltonian and spatial diffeomorphism constraint) on which we still have to solve
the Gauss constraint given by
\begin{equation*}
G_A:=\partial_j E^j_A+\epsilon^{C\,\,\,\,}_{AB} A^B_j E^j_C. 
\end{equation*}
The introduction of Ashtekar variables allows to rewrite general relativity in terms of the language of
gauge fields and this suggests to formulate the theory in terms of holonomies along one dimensional paths and electric
fluxes through two dimensional surfaces, likewise to the case when one applies Dirac quantization in unreduced
loop quantum gravity. For the unreduced case a uniqueness result \cite{Lewandowski:2005jk,Fleischhack:2005} showing
that cyclic representations of the holonomy – flux algebra which implement a unitary representation of the
spatial diffeomorphism gauge group Diff($\chi$) are unique and are unitarily equivalent to the
Ashtekar – Isham – Lewandowski representation \cite{Ashtekar:1991kc,Ashtekar:1993wf}.
In our case, that considers the (partially) reduced phase space, we do not have the diffeomorphism gauge group
but rather a diffeomorphism symmetry group Diff(${\cal S}$) of the physical Hamiltonian ${\bf H}_{phys}$.
This is physical input enough to also insist on cyclic Diff(${\cal S}$) covariant representations and correspondingly,
like in $\cite{Giesel:2007wn}$ we can copy the uniqueness result. Hence, we choose the background independent
and active diffeomorphism covariant Hilbert space representation of loop quantum gravity that becomes
the representation of the physical Hilbert space here. Thus,  ${\cal H}_{\rm phys}=L_2({\cal A},\mu_{AL})$
can be understood as the space of square integrable function over the set of generalized connections with respect
to Ashtekar-Lewandowski measure, for more details and a pedagogical introduction, see for instance
\cite{Thiemann:2007zz,Ashtekar:2004eh,Rovelli:2010wq,Giesel:2012ws,Giesel:2017wgh,Giesel:2017jzj,Bodendorfer:2016uat} and references therein. 
We solve the remaining Gauss constraint by simply restricting to the gauge invariant sector of that Hilbert space.
This can be achieved by choosing appropriate intertwiners for the vertices of the so called spin network functions
that provide an orthonormal basis in ${\cal H}_{\rm phys}$. For more details see also \cite{Giesel:2007wn}.
\\
As mentioned earlier we are only interested in those representations that also allow to implement the
physical Hamiltonian ${\bf H}_{phys}$ as a well defined operator.
However, looking at the particular form of the physical Hamiltonian density in (\ref{HamDens1}),
we realize that it is exactly this point where the \textbf{reduced phase space quantization cannot be performed}.
Let us explain this in detail: Due to the fact that in the loop quantum gravity representation used for
${\cal H}_{\rm phys}$  the spatial diffeomorphisms are not implemented weakly continuously,
only finite diffeomorphism exists at the quantum level, but the associated infinitesimal generators cannot be defined
as operators on ${\cal H}_{\rm phys}$. In our model this carries directly over to $C^{\rm geo}_{j}$.
As a consequence the expression $\delta^{jk}C^{\rm geo}_jC^{\rm geo}_k$ under the square root cannot be quantized and this 
implies that the physical Hamiltonian ${\bf H}_{phys}$ cannot be implemented as a well defined operator on
${\cal H}_{phys}$. This shows that the four Klein-Gordon scalar fields model is an example for a model where
Dirac quantization and reduced quantization yield very different results.
In case we would use this model and apply Dirac quantization we would meet no technical problem in implementing
the constraint operators on the kinematical Hilbert space that also involve the contribution from the
Klein-Gordon scalar fields.
Therefore, a formulation of the Quantum Einstein Equations in the context of Dirac quantization would be possible,
although the final physical Hilbert space would still need to be derived.
However, in the case of reduced quantization, we are able to construct the physical Hilbert space ${\cal H}_{\rm phys}$,
but then on ${\cal H}_{\rm phys}$ the dynamics encoded in the physical Hamiltonian cannot be formulated as a well defined
operator.
Therefore, the quantization program cannot be completed in the reduced case.
This implies that four Klein-Gordon scalar fields do not provide an appropriate set of reference fields in order to
obtain a reduced phase space quantization of general relativity.
\\
\\
Let us close this section with a few remarks. 
\begin{itemize}
\item[1.] One could ask the question why such issues are not present in any of the other currently
available reference matter models. The reason for this is that in all current available models the
generator $C^{\rm geo}_j$ occurs only in the combination $Q^{jk}C^{\rm geo}_jC^{\rm geo}_k$ 
and it is exactly this combination that can again be quantized in the usual loop quantum gravity representation 
\cite{Giesel:2007wn} used
for ${\cal H}_{\rm phys}$ here. 
\item[2.] In \cite{Laddha:2011mk} a lot of progress was made to formulate an operator that corresponds
to infinitesimal spatial diffeomorphisms at the classical level. However, because this work requires
a particular phase space dependent form of the shift vector, the techniques developed there cannot be
applied here in order to find a suitable quantization of ${\bf H}_{\rm phys}$ on ${\cal H}_{\rm phys}$.
\item[3.] One could take the point of view that this negative result does only occur because we require the theory to be quantizable within the representation used in loop quantum gravity. However, if we drop this requirement and consider for instance Fock quantization, then we could not implement the original constraints and quantities like the volume operator as well defined operators on Fock space. Therefore the situation is even worse in that case.
\end{itemize}
In summary, we conclude that the four Klein-Gordon scalar fields model cannot be used as a natural extension of the APS-model \cite{Ashtekar:2006wn} and the one scalar field model \cite{Domagala:2010bm} to obtain the corresponding reduced quantum theories associated with these models. In the next section we will demonstrate that a slight generalization of the four Klein-Gordon scalar fields model is sufficient enough to get a model for which the dynamics can be implemented and thus the reduced phase space quantization program can be completed. 
\section{Generalized Model with Four Klein-Gordon Scalar Fields}
\label{s3}
In this section we want to extend the former model with four Klein-Gordon scalar fields in order to obtain a model
that is suitable for completing the quantization program in the reduced case. 
The seminal models \cite{Kuchar:1995xn,Kuchar:1991} have a common property,
namely that at first they introduce more than the necessary four scalar fields in addition to general relativity.
It turns out that then these models describe a system with second class constraints.
A symplectic reduction with respect to the second class constraints results in a first class model with
only four additional scalar fields.
For the generalization of the four Klein-Gordon scalar field model we want to follow a similar line. 
We will introduce three additional scalar fields in a particular way such that the final physical Hamiltonian
can be quantized on ${\cal H}_{\rm phys}$. The model we want to consider can be described by the following action
\begin{align}
\label{ActionGen}
S[g,\varphi^0,\varphi^j,M_{ij}]&=\int\limits_M d^4X \sqrt{g}R^{(4)}
-\frac{1}{2}\int\limits_M d^4X \sqrt{g}g^{\mu\nu}\varphi^0_{,\mu}\varphi^0_{,\nu}
-\frac{1}{2}\int\limits_M d^4X \sqrt{g}M_{ij}g^{\mu\nu}\varphi^i_{,\mu}\varphi^j_{,\nu}\\ \nonumber
&=S^{\rm geo}+S^{\varphi^0}+S^{\varphi^j},
\end{align}
here $\mu,\nu$ runs from 0 to 3 whereas $i,j$ runs only from 1 to 3. In principle we have introduced 9 new degrees
of freedom sitting in a not further restricted arbitrary matrix $M_{ij}$ in three dimensions.
However, we will assume further properties of this matrix and this reduces the number of independent degrees
of freedom down to three. Note, that we also could have considered a model with a 4x4 matrix $M_{IJ}$.
However, then the reference field for the Hamiltonian constraint would no longer be a standard Klein-Gordon
field and since we would like to compare our model to the one in \cite{Domagala:2010bm}, 
we will only work with a spatial matrix here.
The first assumption we make is that $M_{ij}$ is a symmetric matrix which reduces the number of degrees of freedom
from 9 to 6. Further, we restrict our model to diagonal matrices for the reason that this is only a minimal
generalization from the former Klein-Gordon scalar field model that can be obtained by choosing $M_{ij}=\delta_{ij}$.
As we will show this extension is already sufficient to get a quantizable model.  
Thus, the form of $M_{ij}$ that we work with is 
\begin{equation*}
\begin{pmatrix}
M_{11}(x) &0 &0 \\
0& M_{22}(x) &0 \\
0&0 & M_{33}(x)
\end{pmatrix}
\end{equation*}
and thus we have three additional degrees of freedom sitting in $M_{jj}(x)$. 
\subsubsection{Equations of Motion for the Generalized Model}
We start with the equations of motion that follow from the Euler-Lagrange equation for the variables $M_{jj}$ 
and obtain for each $j=1,2,3$
\begin{equation}
\label{EOMMjj}
\frac{\delta S}{\delta M_{jj}}=0=-\frac{1}{2}\sqrt{g}g^{\mu\nu}\varphi^j_{,\mu}\varphi^j_{,\nu}.
\end{equation}
If we define for each $j=1,2,3$ a four velocity $U^\mu_{(j)}:=g^{\mu\nu}\varphi^j_{,\nu}$ then the equation
above can be rewritten as
\begin{equation*}
U^\mu_{(j)}\varphi^j_{,\mu}={\cal L}_{U_{(j)}} \varphi^j=0,
\end{equation*}
where ${\cal L}_{U_{(j)}}$ denotes the Lie derivative with respect to $U^\mu_{(j)}$.
Thus, the reference field $\varphi^j$ is constant along the flow of the vector field $U^\mu_{(j)}$.
A similar property can be found in \cite{Kuchar:1995xn}, however there the four velocity is not constructed from
one scalar field $\varphi^j$ only but it is constructed from 7 scalar fields $T,W_j,S^j$ where $j$ runs from 1 to 3.
Next we discuss the equation of motion for $\varphi^0$ which is, as expected, the standard Klein-Gordon equation
as can be seen from 
\begin{equation*}
0=-\frac{\partial}{\partial x^\mu}\frac{\delta S}{\delta \varphi^0_{,\mu}}
=\partial_\mu\left(\sqrt{g}\varphi^0_{,\nu}g^{\mu\nu}\right)=\sqrt{g}\nabla_\mu\left(g^{\mu\nu}\varphi^0_{,\nu}\right)=\sqrt{g}g^{\mu\nu}\nabla_\mu\varphi^0_{,\nu}=\sqrt{g}\Box^{(g)}\varphi^0\quad\Longleftrightarrow\quad \Box^{(g)}\varphi^0=0,
\end{equation*}
here $\nabla_\mu$ defines the torsion free covariant derivative metric compatible with $g$, $\Box^{(g)}$
the d'Alembertian operator and we used how covariant derivatives act on tensor densities.
Finally, we consider the equations of motion for $\varphi^j$. In the former model discussed in section \ref{s2}
the dynamics of $\varphi^j$ was also described by a Klein-Gordon equation.
This will be modified in the generalized model here. We obtain for each $j=1,2,3$ 
\begin{equation*}
0=-\frac{\partial}{\partial x^\mu}\frac{\delta S}{\delta \varphi^j_{,\mu}}
=\partial_\mu\left(\sqrt{g}M_{jj}\varphi^j_{,\nu}g^{\mu\nu}\right)=\sqrt{g}\nabla_\mu\left(g^{\mu\nu}M_{jj}\varphi^j_{,\nu}\right)=\sqrt{g}g^{\mu\nu}\nabla_\mu(M_{jj}\varphi^j_{,\nu})
\end{equation*}
as before no summation over repeated $j$ indices is considered here.
Hence, the equations of motion for each $\varphi^j$ are given by
\begin{equation}
\label{EOMphij}
M_{jj}\sqrt{g}\Box^{(g)}\varphi^j+\sqrt{g} (\nabla_\mu M_{jj})g^{\mu\nu}\varphi^j_{,\nu}=0,
\end{equation}
where again no summation over repeated $j$ indices is assumed. For the reason that the canonical momenta associated with the $M_{jj}$'s vanish and the $M_{jj}$'s themselves enter only linearly into the action, the equations of motion do not determine $M_{jj}$ completely. As we will see in the Hamiltonian framework the equation of motion for $M_{jj}$ still include arbitrary Lagrange multipliers. Depending on the choice of these Lagrange multipliers the fields $\varphi^j$  satisfy the generalized  Klein-Gordon equation shown in (\ref{EOMphij}). Comparing with the Brown-Kucha\v{r} dust model in \cite{Brown:1994py} the role $M_{jj}$ plays in our model is taken by the scalar fields $\rho$ and $W_j$ in the Brown-Kucha\v{r} model. As discussed later, it is exactly this modification for the spatial reference fields that leads to a reduced model whose physical Hamiltonian can be quantized using loop quantum gravity techniques.
In the next section we will show that the model is second class and can be reduced to a first class model with only four instead of seven additional scalar fields.
\subsection{Constraint Stability Analysis} 
\label{s31}
Given the action in (\ref{ActionGen}) we introduce the following canonical momenta 
\begin{equation*}
(q_{ab},p^{ab}),(n,p),(n^a,p_a),(\varphi^J,\pi_J),(M_{jj},\Pi^{jj})
\end{equation*}
Considering the fact that $M_{jj}$  in  diagonal form is invertible we obtain the following
primary constraints 
\begin{eqnarray*}
 \Lambda^{jj} &:=& \Pi^{jj} := \frac{\delta S}{\delta \dot{M}_{jj}} = 0,\\
 z &:=&p := \frac{\delta S}{\delta \dot{n}} = 0, \\ 
 z_a &:= &p_a := \frac{\delta S}{\delta \dot{n}_a} = 0.
\end{eqnarray*}
The action in canonical form reads 
\begin{align*}
&S[q_{ab},p^{ab},n,p,n^a,p_a,\varphi^0,\pi_0,\varphi^j,\pi_j,M_{jj},\Pi^{jj}] \\ 
&=\int\limits_{\mathbb{R}} dt\int\limits_{\chi} d^3x 
\bigg(\frac{1}{\kappa}\dot{q}_{ab}p^{ab}+\dot{\varphi}^0\pi_0+\sum\limits_{j=1}^3\dot{\varphi}^j\pi_j+\dot{n}p+\dot{n}^a p_a+ \sum\limits_{j=1}^3\dot{M}_{jj} \Pi^{jj} \\
&-\left[nc^{\rm tot}+n^ac_a^{\rm tot}+\nu{z}+\nu^a{z}_a+\sum\limits_{j=1}^3 \mu_{jj}\Lambda^{jj}\right]\bigg)
\end{align*}
Note that here we write down the summation over repeated $j$-indices explicitly for later convenience. The associated primary Hamiltonian is given by
\begin{eqnarray*}
 H_{\rm primary} &=& \int\limits_{\chi} \! d^3x \, h_{\rm primary}= \int\limits_{\chi} \! d^3x \,\left(n c^{\rm tot} + n^a c_a^{\rm tot} + \nu z + \nu^a z_a +\sum\limits_{j=1}^3 \mu_{jj}\Lambda^{jj}\right)
\end{eqnarray*}
with 
\begin{eqnarray*}
{z}:=p,\quad {z}_a:=p_a,\quad \Lambda^{jj} := \Pi^{jj},\quad c^{\rm tot}:=c^{\rm geo}+ c^{\varphi}, \quad c^{\rm  tot}_a:=c^{\rm  geo}_a+c^{\varphi}_a
\end{eqnarray*}
and
\begin{eqnarray}
\label{Constraints}
\kappa c^{\rm geo}&=&\frac{1}{\sqrt{q}}\left(q_{ac}q_{bd}-\frac{1}{2}q_{ab}q_{cd}\right)p^{ab}p^{cd}-\sqrt{q}R^{(3)}, \nonumber\\
       c^{\varphi}&=&\frac{\pi_0^2}{2\sqrt{q}}+\frac{1}{2}\sqrt{q}q^{ab}\varphi^0_{,a}\varphi^0_{,b}+\sum\limits_{j=1}^3\left(
	\frac{(M^{-1})^{jj}\pi_j\pi_j}{2\sqrt{q}}+\frac{1}{2}\sqrt{q}M_{jj}q^{ab}\varphi^j_{,a}\varphi^j_{,b}\right),\nonumber \\
\kappa c^{\rm geo}_a&=&-2q_{ac}D_bp^{bc},\nonumber\\
       c^{\varphi}_a&=&\pi_0\varphi^0_{,a}+\sum\limits_{j=1}^3\pi_j\varphi^j_{,a}.
\end{eqnarray}
The non-vanishing Poisson brackets are given by
\begin{eqnarray*}
 \{  q_{cd}(x), p^{ab}(y)\} &=& \kappa \delta^a_{(c} \delta^b_{d)} \delta^{(3)}\!(x,y), \\
 \{ n(x) , p(y) \} &=& \delta^{(3)}\!(x,y), \\
 \{ n^a(x) , p_b(y) \} &=& \delta^a_b \delta^{(3)}\!(x,y), \\
 \{ \varphi^0(x), \pi_0(y) \} &=& \delta^{(3)}\!(x,y),\\
 \{ \varphi^j(x), \pi_k(y) \} &=& \delta^j_k \delta^{(3)}\!(x,y), \\
 \{ M_{jj}(x), \Pi^{kk}(y) \} &=& \delta^k_j \delta^{(3)}\!(x,y).
\end{eqnarray*}
As a first step we need to analyze the stability of the primary constraints under the dynamics of the primary Hamiltonian. For $z$ and $z_a$
this can be easily computed and we obtain
\begin{align}
 & \dot{z} = \{ z,  H_{\rm primary} \} = \{ p,  H_{\rm primary} \} = - c^{\rm tot}, \\
 & \dot{z}_a = \{ z_a,  H_{\rm primary} \} = \{p_a, H_{\rm primary}\} = - c_a^{\rm tot}.
\end{align}
In order to ensure that $z$ and $z_a$ are stable we require $c^{\rm tot}$ and $c^{\rm tot}_a$ to be secondary constraints and these are the Hamiltonian and diffeomorphism constraint. Next we consider the three constraints $\Lambda^{jj}$. Under the primary Hamiltonian $\Lambda^{jj}$ evolves as
\begin{equation*}
\dot{\Lambda}^{jj}=\{\Lambda^{jj},H_{\rm primary}\}=\frac{n}{2}\left[\frac{(M^{-1})^{jk}(M^{-1})^{j\ell}\pi_k\pi_\ell}{\sqrt{q}}-\sqrt{q}q^{ab}\varphi^j_{,a}\varphi^j_{,b}\right],
\end{equation*}
here no summation over repeated $j$-indices is assumed. We realize that we obtain three more secondary constraints that we denote by $c^{jj}$ given by
\begin{equation*}
c^{jj}:=\frac{n}{2}\left[\frac{(M^{-1})^{jk}(M^{-1})^{j\ell}\pi_k\pi_\ell}{\sqrt{q}}-\sqrt{q}q^{ab}\varphi^j_{,a}\varphi^j_{,b}\right].
\end{equation*}
We obtained a set of secondary constraints $\{c^{\rm tot},c_a^{\rm tot},c^{jj}\}$. Now we need to compute whether these constraints are stable or whether tertiary constraints occur. The details of the calculation can be found in 
appendix \ref{b1} , here we summarize only the results. When computing the stability in the case of $c^{\rm tot}_a$ all non-vanishing contributions are proportional to either $c^{\rm tot}$ or $c^{\rm tot}_a$. Thus, we can conclude
\begin{equation*}
\{c^{\rm tot}_a,H_{\rm primary}\}\approx 0.
\end{equation*}
Further, for $c^{\rm tot}$ we have a similar situation. There all non-vanishing contributions are proportional to $c^{\rm tot},c_a^{\rm tot}$ or $c^{jj}$ respectively. Hence, also here we have
\begin{equation*}
\{c^{\rm tot},H_{\rm primary}\}\approx 0.
\end{equation*}
Finally, we consider the stability of $c^{jj}$. Here we consider the individual contributions of the primary Hamiltonian separately. We have
\begin{equation}
\label{StabLambda}
\int\limits_\chi d^3y\{c^{jj}(x),(\mu_{kk}\Lambda^{kk})(y)\}
=-\mu_{jj}\frac{n}{\sqrt{q}}\pi_j^2((M^{-1})^{jj})^3,
\end{equation}
again no summation of $j$ is assumed here. The non-vanishing contributions that are not again proportional
to already existing constraints come from 
\begin{equation*}
\int\limits_\chi d^3y\{c^{jj}(x), (nc^{\rm \rm tot})(y)\}\not=0\quad{\rm and}\quad 
\int\limits_\chi d^3y\{c^{jj}(x), (n^ac^{\rm \rm tot}_a)(y)\}\not=0.
\end{equation*}
However, we do not need to compute these contributions in explicit form because the result in (\ref{StabLambda}) involves the Lagrange multipliers $\mu_{jj}$ in linear form. Therefore, although we have non-vanishing contributions from the Poisson brackets also on the constraint hypersurface we can solve $\{c^{11},H_{\rm primary}\}=0$ for the Lagrange multiplier $\mu_{11}$ and likewise in the cases $j=2,3$ where we can solve the corresponding equations for $\mu_{22}$ and $\mu_{33}$ respectively. 
As a consequence, the stability is also ensured for $c^{jj}$ and thus the model contains no tertiary constraints and the constraint algorithm stops here. The final set of constraints is given by $\{z,z_a,c^{\rm tot}_a,c^{\rm tot},\Lambda^{jj},c^{jj}\}$. Now we need to classify the constraints into first and second class. We define the following linear combination of constraints
\begin{equation*}
\widetilde{c}_a^{\rm tot}:=c_a^{\rm tot}+M_{jj,a}\Pi^{jj}+n_{,a}p+\left({\cal L}_{\vec{n}}p\right)_a=c_a^{\rm tot}+M_{jj,a}\Lambda^{jj}+n_{,a}z+\left({\cal L}_{\vec{n}}z\right)_a.
\end{equation*}
The constraints $\widetilde{c}_a^{\rm tot}$ are the generator of spatial diffeomorphisms on the phase space with
elementary variables $(q_{ab},p^{ab},n,p,n^a,p_a,M_{jj},\Pi^{jj})$ and thus the constraints $\widetilde{c}^{\rm tot}_a$
are first class constraints. For the constraint $c^{\rm tot}$ we consider the following linear combination
\begin{equation*}
\widetilde{c}^{\rm tot}:=c^{\rm tot}+\beta_{jj}\Lambda^{jj}
\end{equation*}
and determine $\beta_{jj}$ such that $\widetilde{c}^{\rm tot}$ and $c^{jj}$ have vanishing Poisson brackets
up to terms proportional to the constraints for all $j=1,2,3$. We have
\begin{equation*}
\{\widetilde{c}^{\rm tot}(x),c^{jj}(y)\}=\{c^{\rm tot}(x),c^{jj}(y)\}+\beta_{jj}\frac{n}{\sqrt{q}}\frac{\pi_j^2}{(M_{jj})^3}\stackrel{!}{=}0.
\end{equation*}
Solving this equation for $\beta_{jj}$ yields
\begin{equation*}
\beta_{jj}(x)=-\int\limits_\chi d^3y\sqrt{q}\frac{(M_{jj})^3}{n\pi^2_j}\{c^{\rm tot}(x),c^{jj}(y)\}.
\end{equation*}
In order to check whether $\beta_{jj}$ is well defined we need to compute
$\{c^{\rm tot}(x),c^{jj}(y)\}$ explicitly.
A rather lengthy but straight forward calculation presented in appendix \ref{c1} shows that
\begin{align*}
\beta_{jj}(x) =&\frac{1}{2}\frac{(M_{jj})^3}{n\pi_j^2}q_{ab}p^{ab}c^{jj}(x)
+\sqrt{q}\varphi^j_{,a}\varphi^j_{,b}p^{ab}\frac{(M_{jj})^3}{\pi_j^2}(x)\\ 
&+\frac{(M_{jj})^2}{n\pi_j} \lr n \sqrt{q} q^{ab} \varphi^j_{,b}\rr_{,a}(x)
+\frac{(M_{jj})}{\pi_j}\lr M_{jj} \sqrt{q} q^{ab} \varphi^j_{,b}\rr_{,a}(x).
\end{align*}
On the constraint surface $c^{jj}=0$ the expression for $\beta_{jj}$ reduces to
\begin{align*}
\beta_{jj}(x)\approx +\sqrt{q}\varphi^j_{,a}\varphi^j_{,b}p^{ab}\frac{(M_{jj})^3}{\pi_j^2}(x)
+\frac{(M_{jj})^2}{n\pi_j} \lr n \sqrt{q} q^{ab} \varphi^j_{,b}\rr_{,a}(x)
+\frac{(M_{jj})}{\pi_j}\lr M_{jj} \sqrt{q} q^{ab} \varphi^j_{,b}\rr_{,a}(x).
\end{align*}
Given this choice of $\beta_{jj}$ also $\widetilde{c}^{\rm tot}$ is a first class constraint. The remaining constraints $\Lambda^{jj}$ and $c^{jj}$ build three second class pairs $(c^{11},\Lambda_{11})$, $(c^{22},\Lambda_{22})$ and $(c^{33},\Lambda_{33})$. Let us shortly summarize. We have extended the four Klein-Gordon scalar fields model by 6 additional degrees of freedom $(M_{jj},\Pi^{jj})$. The constraint analysis showed that our model has four first class constraints $\widetilde{c}^{\rm tot}_a$ and $\widetilde{c}^{\rm tot}$ and six second class constraints $c^{jj},\, \Lambda^{jj}$. Therefore, if we reduce with respect to the second class constraints and consider this partially reduced phase space, we also reduce exactly the six additional degrees of freedom because each second class constraints reduces one degree of freedom in phase space. This partially reduced model consists of gravity plus for scalar fields that we will use as reference fields later in order to derive the reduced phase space with respect to $\widetilde{c}^{\rm tot}$ and $\widetilde{c}^{\rm tot}_a$. To perform the reduction with respect to the second class constraints we need to compute the associated Dirac bracket. For this purpose we define the following set of constraints $c_I$ with $I=1,\cdots, 6$ and $\{c_I\}_{I=1,\cdots, 6}=\{c^{jj},\Lambda^{jj} | j=1,2,3\}$ and introduce the matrix
\begin{equation*}
{\cal N}_{JK}(x,y):=\{c_J(x),c_K(y)\}
=
\begin{pmatrix}
A_{jk}(x,y) & B_{jk}(x,y) \\ C_{jk}(x,y) & 0
\end{pmatrix}
\end{equation*}
where  $A_{jk}(x,y)=\{c^{jj}(x),c^{kk}(y)\}$, $B_{jk}(x,y)=\{c^{jj}(x),\Lambda^{kk}(y)\}$,  $C_{jk}(x,y)=\{\Lambda^{jj}(x),c^{kk}(y)\}$ and we used that $\{\Lambda^{jj}(x),\Lambda^{kk}(y)\}=0$. We have 
\begin{equation*}
\{c^{jj}(x),\Lambda^{kk}(y)\}=-\frac{n}{\sqrt{q}}\frac{\pi_j^2}{(M_{jj})^3}\delta^{kj} \delta^{(3)}\!(x,y).
\end{equation*}
and $\{c^{jj}(x),c^{kk}(y)\}=0$ for $j\not= k$ and as a consequence all $3\times 3$-matrices $A,B,C$ are diagonal matrices. The inverse matrix ${({\cal N}^{-1})}^{IJ}$ is given by
\begin{equation*}
{({\cal N}^{-1})}^{JK}(x,y)=\begin{pmatrix}
 0 & (C^{-1})^{jk}(x,y) \\
 (B^{-1})^{jk}(x,y) & -(B^{-1}AC^{-1})^{jk}(x,y)  \\
\end{pmatrix}
\end{equation*}
The associated inverse matrix satisfies
\begin{equation*}
\int\limits_\chi d^3z\, {\cal N}_{IL}(x,z)({\cal N}^{-1})^{LJ}(z,y)=\delta^{(3)}\!(x,y) \delta_I^J.
\end{equation*}
Given the inverse matrix, we can write down the Dirac bracket that is given by
\begin{eqnarray*}
\{f,g\}^*&=&\{f,g\}-\int\limits_\chi d^3y \int\limits_\chi d^3x  \left(\{f, c_J(x)\}({\cal N}^{-1})^{JK}(x,y)\{c_K(y),g\}\right) \\
&=&
\{f,g\}-\int\limits_\chi d^3y \int\limits_\chi d^3x 
\left(\{f, c^{jj}(x)\}({C}^{-1})^{jk}(x,y)\{\Lambda^{kk}(y),g\}\right. \\
&&
-\int\limits_\chi d^3y \int\limits_\chi d^3x 
\{f, \Lambda^{jj}(x)\}({B}^{-1})^{jk}(x,y)\{c^{kk}(y),g\} \\
&&
+\left. \int\limits_\chi d^3y \int\limits_\chi d^3x 
\{f, \Lambda^{jj}(x)\}({B}^{-1}AC^{-1})^{jk}(x,y)\{\Lambda^{kk}(y),g\}\right)
\end{eqnarray*}
For the reason that the constraints $\Lambda^{jj}=\Pi^{jj}$ are equal to the canonical momenta of $M_{jj}$
we can immediately conclude that the Dirac bracket for the subset of variables
$q_{ab},p^{ab},\varphi^0,\pi_0,\varphi^j,\pi_j$ coincides with the usual Poisson bracket because each of the variables
commutes with $\Lambda^{jj}$. Hence, the Dirac brackets affects the variables $(M_{jj},\Pi^{jj})$ only.
The algebra for this subset has the form
\begin{equation*}
\{M_{jj}(x),M_{kk}(y)\}^*=-({B}^{-1}AC^{-1})^{jk}(x,y),\quad \{\Pi^{jj}(x),\Pi^{kk}(y)\}^*=0,\quad \{M_{jj}(x),\Pi^{kk}(y)\}^*=0.
\end{equation*}
To obtain the partially reduced phase space we can set $\Lambda^{jj}=\Pi^{jj}=0$ and express $M_{jj}$ in terms of the remaining variables using $c^{jj}=0$.
We get
\begin{equation}
\label{Mij}
M_{jj}\Big|_{c^{jj}=0}=\frac{\pi_j}{\sqrt{q}}\frac{1}{\sqrt{q^{ab}\varphi^j_{,a}\varphi^j_{,b}}}\quad{\rm for}\quad j=1,2,3
\end{equation}
and as usual no summation over repeated $j$'s is considered here. On this partially reduced phase space the constraint
$\widetilde{c}_a^{\rm tot}$ has the following form
\begin{equation*}
\widetilde{c}^{\rm tot}_a=c^{\rm tot}_a+{n}_{,a}z+({\cal L}_{\vec{n}}z)_a.
\end{equation*}
In order to rewrite the constraint $\widetilde{c}^{\rm tot}$ on the partially reduced phase space we use $M_{jj}$ in (\ref{Mij}) leading to
\begin{eqnarray*}
\widetilde{c}^{\rm tot}\Big|_{c^{jj}=0, \Lambda^{jj}=0} &=&
c^{\rm geo}+\frac{\pi^2_0}{2\sqrt{q}}+\frac{1}{2}\sqrt{q}q^{ab}\varphi^0_{,a}\varphi^0_{,b}+\sum\limits_{j=1}^3\sqrt{q}M_{jj}q^{ab}\varphi^j_{,a}\varphi^j_{,b} \\
&=& c^{\rm geo}+\frac{\pi^2_0}{2\sqrt{q}}+\frac{1}{2}\sqrt{q}q^{ab}\varphi^0_{,a}\varphi^0_{,b}
+\sum\limits_{j=1}^3\pi_j\sqrt{q^{ab}\varphi^j_{,a}\varphi^j_{,b}} \\
&\approx & c^{\rm geo}+\frac{\pi^2_0}{2\sqrt{q}}+\frac{1}{2}\sqrt{q}q^{ab}\varphi^0_{,a}\varphi^0_{,b}
-\sum\limits_{j=1}^3\varphi^a_j(c^{\rm geo}_{a}+\pi_0\varphi^0_{,a})\sqrt{q^{bc}\varphi^j_{,b}\varphi^j_{,c}}.
\end{eqnarray*}
Now as usual in the context of the ADM formalism we go to the reduced ADM phase space, that is the one where a reduction with respect to the primary constraints $z$ and $z_a$ has been performed. In the reduced ADM phase space we can treat the lapse function $n$ and the shift vector $n^a$ as Lagrangian multipliers. On the reduced ADM phase space we have $\widetilde{c}^{\rm tot}_a=c^{\rm tot}_a$. Summarizing, starting from the model whose action is given in (\ref{ActionGen}), we end up with a reduced ADM phase space with elementary variables $(q_{ab},p^{ab},\varphi^0,\pi_0,\varphi^j,\pi_j)$ which is a model consisting of gravity and four scalar fields and a set of first class constraints given by
\begin{align}
\label{ConsGen}
c^{\rm tot}_a&=c_a^{\rm geo}+\pi_0\varphi^0_{,a}+\pi_j\varphi^j_{,a},\nonumber \\
c^{\rm tot} &= c^{\rm geo}+\frac{\pi^2_0}{2\sqrt{q}}+\frac{1}{2}\sqrt{q}q^{ab}\varphi^0_{,a}\varphi^0_{,b}
-\sum\limits_{j=1}^3\varphi^a_j(c^{\rm geo}_{a}+\pi_0\varphi^0_{,a})\sqrt{q^{bc}\varphi^j_{,b}\varphi^j_{,c}}.
\end{align}
In the next subsection we will discuss the construction of observables for this model.
\subsection{Step 1: Construction of Observables}
\label{s32}
Here we will follow very closely the presentation in section \ref{s22} because most of the steps performed 
for the four Klein-Gordon scalar fields model carry over to the generalized model.
Again we start by rewriting the constraint in Abelianized form. 
\subsubsection{Weakly Abelian Set of Constraints}
For this purpose we start with $c^{\rm tot}$ in (\ref{ConsGen}) and solve it for the reference field momentum $\pi_0$. We get\footnote{We would like to point out that the formula for $c^{tot}$ on page 347 in \cite{Giesel:2017mfc} needs to have a minus sign in front of the fourth term that carries over to the definitions of $b$ and $c$ and finally leads to a plus sign in front of the second term in ${\rm H}_{\rm phys}$ in equation (4.1) on page 348 so that the results in \cite{Giesel:2017mfc} and ours here agree. Moreover in the notation of \cite{Giesel:2017mfc} the functions $h$ and $h_j$ differ by an overall minus sign to notation used here.}
\begin{equation*}
\pi_0^2-\pi_0
\left(2\sqrt{q}\sum\limits_{j=1}^3\varphi^0_{,a}\varphi^a_j\sqrt{q^{cd}\varphi^j_{,c}\varphi^j_{,d}}\right)+q q^{ab}\varphi^0_{,a}\varphi^0_{,b}
-2\sqrt{q}\sum\limits_{j=1}^3\varphi^a_jc_a^{\rm geo}\sqrt{q^{bc}\varphi^{j}_{,b}\varphi^{j}_{,c}}+2\sqrt{q}c^{\rm \rm geo}=0.
\end{equation*}
We define the following abbreviations:
\begin{eqnarray*}
b &:=&-2\sqrt{q}\sum\limits_{j=1}^3 \varphi^0_{,a}\varphi^a_j\sqrt{q^{cd}\varphi^j_{,c}\varphi^j_{,d}}, \\
c &:=& q q^{ab}\varphi^0_{,a}\varphi^0_{,b}-2\sqrt{q}\sum\limits_{j=1}^3\varphi^a_jc_a^{\rm geo}\sqrt{q^{bc}\varphi^{j}_{,b}\varphi^{j}_{,c}}+2\sqrt{q}c^{\rm geo},
\end{eqnarray*}
then solving for the momentum $\pi_0$ yields
\begin{equation}
\label{Solvepi02}
\pi_0=-\frac{b}{2}\pm\sqrt{\left(\frac{b}{2}\right)^2-c}=: -h(q_{ab}, p^{ab}, \varphi^0, \varphi^j) =:-h.
\end{equation}
As before, in order to ensure that the final physical Hamiltonian is positive, we choose the plus sign here in order to define $h$.
The spatial diffeomorphism constraint $c^{\rm tot}_a$ can as in the former model be solved for $\pi_j$ using the inverse $\varphi^a_j$ of $\varphi^j_{,a}$ leading to
\begin{equation}
\label{Solvepij2}
\pi_j=-\varphi^a_j\left(c^{\rm geo}_a+\pi_0\varphi^0_{,a}\right)=: -h_j(q_{ab},p^{ab},\varphi^j,\varphi^0):=-h_j.
\end{equation}
Likewise to the model discussed in section \ref{s2} we can write down the following Abelian set of equivalent
constraints 
\begin{eqnarray}
\label{NewConstraints2}
c^{\rm tot}&:=&\pi_0+h(q_{ab}, p^{ab}, \varphi^0, \varphi^j), \nonumber\\
c^{\rm tot}_j&:=&\pi_j+h_j(q_{ab}, p^{ab}, \varphi^0, \varphi^j),
\end{eqnarray}
where $h$ and $h_j$ are the functions defined in (\ref{Solvepi02}) and (\ref{Solvepij2}). We consider this set of Abelian first class constraints in the section where observables with respect to these constraints are constructed.
\subsubsection{Explicit Construction of the Observables}
We can apply the same procedure as was in detail presented in section \ref{s21}.
Hence, we will first construct observables with respect to the spatial diffeomorphism
constraint $c^{\rm tot}_j$ and afterwards with respect to the Hamiltonian constraint $c^{\rm tot}$.
Since we have explained the individual steps of the construction in section \ref{s22} and these
can be carried over to the generalized model here, we will just present the results here.
As before for all but the reference fields $\varphi^j$ we construct the following quantities:
\be
\varphi^0,\quad \pi_0/J,\quad q_{jk}=q_{ab}\varphi^a_j\varphi^b_k,\quad 
p^{jk}=p^{ab}\varphi^j_{,a}\varphi^k_{,b}/J,
\ee
where $J:=|\det(\varphi^j/\partial_x)|$ is, as before, used to transform scalar/tensor
densities into real scalars/tensors. Then the observables with respect to $c^{\rm tot}_j$
are given by
\ba
\widetilde{\varphi}^0&:=&O_{\varphi^0,\{\varphi^j\}}^{(1)}(\sigma)= 
\int\limits_{\chi} d^3x \big|\det(\partial\varphi^j/\partial_x)\big|\delta(\varphi^j(x),\sigma^j)
\varphi^0(x),\nonumber\\
\widetilde{\pi}_0&:=&O_{\pi_0,\{\varphi^j\}}^{(1)}(\sigma)=
\int\limits_{\chi} d^3x\, \delta(\varphi^j(x),\sigma^j)
\pi_0(x),\nonumber\\
\widetilde{q}_{jk}&:=&O_{q_{ab},\{\varphi^j\}}^{(1)}(\sigma)=
\int\limits_{\chi} d^3x \big|\det(\partial\varphi^j/\partial_x)\big|\delta(\varphi^j(x),\sigma^j)
\varphi^a_j\varphi^b_kq_{ab}(x),\nonumber\\
\widetilde{p}^{jk}&:=&O_{p^{ab},\{\varphi^j\}}^{(1)}(\sigma)=
\int\limits_{\chi} d^3x\, \delta(\varphi^j(x),\sigma^j)
\varphi_a^j\varphi_b^kp^{ab}.
\ea
Here we used the integral representation for the observables introduced in section \ref{s21}. For the reference fields the observable map leads to:
\ba
\widetilde{\varphi}^j&=&O_{\varphi^j,\{\varphi^j\}}^{(1)}(\sigma)=\Big[\alpha^{K_{\beta_1}}_{\beta_1}(\varphi^j)\Big]_{\alpha^{K_{\beta_1}}_t(\varphi^j)=\sigma^j}=\sigma^j,\nonumber\\
\widetilde{\pi}_j&=&O_{\pi_j,\{\varphi^j\}}^{(1)}(\sigma)
=\int\limits_{\chi} d^3x \big|\det(\partial\varphi^j/\partial_x)\big|\delta(\varphi^j(x),\sigma^j)
\pi_j(x).\nonumber\\
\ea
The spatially diffeomorphism invariant observables of the constraints are given by
\ba
\widetilde{c}^{\rm tot}&=&\widetilde{\pi}_0 +\widetilde{h},\nonumber\\
\widetilde{c}^{\rm tot}_j&=&\widetilde{\pi}_j +\widetilde{c}_j^{\rm geo} - \widetilde{h}\widetilde{\varphi}^0_{,j}
=\widetilde{\pi}_j  +\widetilde{c}_j^{\rm geo} - \widetilde{h}\widetilde{\varphi}^0_{,j},
\ea
where we used that 
\be
O_{\varphi^j_{,a}}^{(1)}(\sigma)=\int\limits_{\chi} d^3x \big|\det(\partial\varphi^j/\partial_x)\big|\delta(\varphi^j(x),\sigma^j)
\varphi^j_{,a}\varphi^a_k=\delta^j_k
\ee
 and likewise $O_{\varphi^a_{,j}}^{(1)}(\sigma)=\delta_j^k$.
 The observables with respect to the diffeomorphism constraint associated with $h$ denoted as $\widetilde{h}$ can be easily obtained by
 using the property of the observable map.
 This implies that $\widetilde{h}=h(\widetilde{q}_{jk},\widetilde{p}^{jk},\widetilde{\varphi}^j,\widetilde{\varphi}^j)$. Using this we obtain
\ba
\label{htilde2}
\widetilde{h}&=&-
\sqrt{\widetilde{q}}\widetilde{\varphi}^0_{,j}\sqrt{\widetilde{q}^{jk}\delta_{jk}}+\sqrt{\left(\sqrt{\widetilde{q}}\widetilde{\varphi}^0_{,j}\sqrt{\widetilde{q}^{jk}\delta_{jk}}\right)^2-\widetilde{q}\widetilde{q}^{jk}\widetilde{\varphi}^0_{,j}\widetilde{\varphi}^0_{,k}+2\sqrt{\widetilde{q}}\sum\limits_{j=1}^3\tilde{c}_j^{\rm geo}\sqrt{\widetilde{q}^{jj}}-2\sqrt{\widetilde{q}}\widetilde{c}^{\rm geo}}\nonumber \\
&=&-
\sqrt{\widetilde{q}}\widetilde{\varphi}^0_{,j}\sqrt{\widetilde{q}^{jk}\delta_{jk}}+\sqrt{\left(\sqrt{\widetilde{q}}\widetilde{\varphi}^0_{,j}\sqrt{\widetilde{q}^{jk}\delta_{jk}}\right)^2-\widetilde{q}\widetilde{q}^{jk}\widetilde{\varphi}^0_{,j}\widetilde{\varphi}^0_{,k}+2\sqrt{\widetilde{q}}\sum\limits_{j=1}^3\sqrt{\widetilde{q}^{jj}\tilde{c}_j^{\rm geo}\tilde{c}_j^{\rm geo}}-2\sqrt{\widetilde{q}}\widetilde{c}^{\rm geo}}.\nonumber \\
\ea
Next, we want to derive the observables with respect to $\widetilde{c}^{\rm tot}$ and also here we can exactly follow the construction discussed in section \ref{s22}. For this generalized model the full observables that we as before denote with capital letters are given by
\ba
Q_{jk}(\sigma,\tau)&:=&O_{q_{ab},\{\varphi^0,\varphi^j\}}(\sigma,\tau)=O_{\widetilde{q}_{jk}(\sigma),\widetilde{\varphi}^0},\nonumber\\
P^{jk}(\sigma,\tau)&:=&O_{p^{ab},\{\varphi^0,\varphi^j\}}(\sigma,\tau)=O_{\widetilde{p}^{jk}(\sigma),\widetilde{\varphi}^0},\nonumber\\
\Pi_{0}(\sigma,\tau)&:=&O_{\pi_{0},\{\varphi^0,\varphi^j\}}(\sigma,\tau)=O_{\widetilde{\pi}_{0}(\sigma),\widetilde{\varphi}^0},\nonumber\\
\Pi_{j}(\sigma,\tau)&:=&O_{\pi_j,\{\varphi^0,\varphi^j\}}(\sigma,\tau)=O_{\widetilde{\pi}_{j}(\sigma),\widetilde{\varphi}^0}.\nonumber\\
\ea
Note, that also here $\Pi_0$ and $\Pi_j$ are no independent observables for the reason that these can be
expressed in terms of $Q^{jk}$ and $P_{jk}$ using the constraints in (\ref{NewConstraints2}).
Furthermore, for the four reference fields  we have
\be
O_{\varphi^0,\{\varphi^0,\varphi^j\}}(\sigma,\tau)=\tau\quad{\rm and}\quad O_{\varphi^j,\{\varphi^0,\varphi^j\}}(\sigma,\tau)=\sigma^j.
\ee
Hence, the elementary variables of the reduced phase space are $(Q_{jk},P^{jk})$. This finishes our discussion on the full observables and in the next section we are going to derive the physical Hamiltonian that is generating their dynamics on the reduced phase space. 
\subsection{Step 2: Dynamics encoded in the physical Hamiltonian}
\label{s33}
We have already shown in section \ref{s22} that even if the constraints do not deparametrize the physical Hamiltonian density is given by the full observables associated with the phase space function $h$ that occurs in the rewritten version of the Hamiltonian constraint in (\ref{NewConstraints2}). The same applies to the generalized model considered here. Using that the physical Hamiltonian is as before given by
\be
\label{Hphysphi2}
{\bf H}_{\rm phys}:=\int\limits_{\cal S}d^3\sigma O_{\widetilde{h}(\sigma),\widetilde{\varphi}^0}(\sigma,\tau)
=
\int\limits_{\cal S}d^3\sigma H(\sigma,\tau),
\ee
here we denote the (full) observable associated to $h$ according to our notation by $H$. Now looking into (\ref{htilde2}) and using the property of the observable map we get for the physical Hamiltonian density
\begin{equation}
\label{HamDens2}
H(\sigma)=\sqrt{-2\sqrt{Q}C^{\rm geo}+2\sqrt{Q}\sum\limits_{j=1}^3\sqrt{Q^{jj}C^{\rm geo}_jC^{\rm geo}_j}}(\sigma).
\end{equation}
We realize that the final physical Hamiltonian density is independent of the physical time $\tau$
because the reference field $\widetilde{\varphi}^0$ occurred only via spatial derivatives and as pointed out 
already in \cite{Giesel:2012rb} and also discussed in \ref{s22},
we have $O_{\widetilde{\varphi}^0_{,j},\widetilde{\varphi}^0}(\sigma,\tau)=d\tau/d\sigma^j=0$.
Therefore, all terms that involve $\widetilde{\varphi}^0_{,j}$ in $\widetilde{h}$ in (\ref{htilde2})
will be vanishing at the observable level. Let us compare the form of the physical Hamiltonian density
in the four scalar field model shown in (\ref{HamDens1}). First let us check that the density weight is
correct in both cases.
Each of the terms under the square root has density weight two and hence the physical Hamiltonian density
is of weight one as it should be. The same is true for the physical Hamiltonian density in (\ref{HamDens2})
of our generalized model.
The main difference between the two models is that the term $\delta^{jk}C^{\rm geo}_jC^{\rm geo}_k$ that
occurred in (\ref{HamDens1}) and that prohibited the completion of the reduced quantization program in the case of the four Klein-Gordon scalar fields model, is no longer present in (\ref{HamDens2}). Instead the physical Hamiltonian density for the generalized model contains terms of the form $Q^{jj}C^{\rm geo}_jC^{\rm geo}_j$ for $j=1,2,3$. As we will discuss in the next subsection, it is exactly this feature of the model that allows to complete the reduced quantization program.
\subsection{Step 3: Reduced Quantization}
\label{s34}
Given the fact that we want to quantize the reduced theory using techniques from loop quantum gravity,
we will reformulate the reduced phase space in terms of Ashtekar variables $(A^{A}_j, E^j_A)$. 
Also in the generalized model the observable algebra of the elementary variables $(A^{A}_j, E^j_A)$
is isomorphic to the kinematical subalgebra of $(A^j_a, E^a_j)$ and as discussed in detail in section \ref{s23} 
because of this we can use the usual Ashtekar-Lewandowski representation of loop quantum gravity to obtain
the physical Hilbert space ${\cal H}_{phys}$ of the generalized model.
As before the price to pay when working in the connection formulation instead of the ADM formulation
is an additional SU(2) Gauss constraint. However, this can simply be solved in the quantum theory 
by restricting to only gauge invariant spin networks in ${\cal H}_{phys}$. In the previous attempt with four Klein-Gordon scalar fields the reduced quantization program could not be completed because the physical Hamiltonian ${\bf H}_{\rm phys}$ could not be implemented as an operator on ${\cal H}_{phys}$. Now the situation is different. 
 The individual terms that occur under the square root of the physical Hamiltonian
density in (\ref{HamDens2}) can all be quantized on ${\cal H}_{phys}$ using loop quantum gravity techniques.
Let us consider the first term, that is $-2\sqrt{Q}C^{\rm geo}$. The two individual contributions of $\sqrt{Q}$ and $C^{\rm geo}$ will be quantized as individual operators. The first one, $\sqrt{Q}$  can be quantized by means of the volume operator \cite{Rovelli:1994ge,Ashtekar:1997fb,Thiemann1998}.
The observable associated to the geometric part of the Hamiltonian constraint $C^{\rm geo}$ can be quantized using
the techniques introduced in \cite{Thiemann:1996ay,Thiemann:1996}. For the quantization of the second term $2\sqrt{Q}\sum\limits_{j=1}^3 \sqrt{Q^{jj}C^{\rm geo}_jC^{\rm geo}_j}$,
we will promote the entire term to an operator at the quantum level and this can be done using the usual quantization for holonomies and fluxes in loop quantum gravity.
Note, that the quantization used in \cite{Giesel:2007wn} for the Brown-Kucha\v{r} dust model does not carry over to this model because here the second term does not 
involve a covariant contraction of the spatial indices between the observables associated with the metric $Q^{jk}$ and the geometric part of the spatial diffeomorphism constraint
$C_j^{\rm geo}$. As a consequence, a different regularization procedure needs to be considered.

We start from the classical expression of the physical Hamiltonian given by:
\begin{equation}
\label{HphysClass}
{\bf H}_{\rm phys} = \int\limits_{\cal S}d^3\sigma \sqrt{-2\sqrt{Q}C^{\rm geo}+2\sqrt{Q}\sum\limits_{j=1}^3\sqrt{Q^{jj}C^{\rm geo}_jC^{\rm geo}_j}}(\sigma).
\end{equation}
\\ \\
Likewise to the volume operator or the physical Hamiltonian in \cite{Domagala:2010bm,Giesel:2006uk} the classical expression involves a square root. From the classical point of view, the physical Hamiltonian density is real and this is only true if 
$-2\sqrt{Q}C^{\rm geo}+2\sqrt{Q}\sum\limits_{j=1}^3 \sqrt{Q^{jj}C^{\rm geo}_jC^{\rm geo}_j} \geq 0$. In deriving the form of ${\bf H}_{\rm phys}$ we restrict to the part of the phase space in which $C^{\rm geo}\leq 0$. Moreover, from the classical Hamiltonian constraint equation we get
\begin{equation*}
\pi_0^2-\pi_0
\left(2\sqrt{q} \sum\limits_{j=1}^3\varphi^0_{,a}\varphi^a_j\sqrt{q^{cd}\varphi^j_{,c}\varphi^j_{,d}}\right)+q q^{ab}\varphi^0_{,a}\varphi^0_{,b}-2\sqrt{q}\sum\limits_{j=1}^3\varphi^a_jc_a^{\rm geo}\sqrt{q^{bc}\varphi^{j}_{,b}\varphi^{j}_{,c}}+2\sqrt{q}c^{\rm geo}=0.
\end{equation*}
Applying the observable map to the equation above yields
\begin{equation}
\label{ConstrHphys}
 \Pi_0^2=H^2(\sigma),
\end{equation}
where $\Pi_0$ denotes the observable associated with $\pi_0$ and $H^2(\sigma)$ is the square of the physical Hamiltonian density $H(\sigma)$, that is the expression under the square root in (\ref{HphysClass}). Note, that we applied the observable map with $J:=|\det(\partial\varphi^j/\partial_x)| > 0$. Considering (\ref{ConstrHphys}) we realize that on the physical part of the phase space we have that $H^2(\sigma)$ is non-negative due to the reason that certainly $ \Pi_0^2\geq 0$. However, this does not ensure that the quantized version of $H^2(\sigma)$ is non negative. In principle, we can achieve this by implementing $H^2(\sigma)$ as a self-adjoint operator and project onto the positive part of the spectrum for every $\sigma$. The practical problem that arises here is that we do not know the spectrum of the physical Hamiltonian and hence we cannot follow this way.  Therefore, we choose the same strategy as in \cite{Giesel:2007wn} and consider an absolute value under the square root and quantize
\begin{equation}
{\bf H}_{\rm phys} = \int\limits_{\cal S}d^3\sigma \sqrt{|-2\sqrt{Q}C^{\rm geo}+2\sqrt{Q}\sum\limits_{j=1}^3\sqrt{Q^{jj}C^{\rm geo}_jC^{\rm geo}_j}|}(\sigma).
\end{equation}
In the classical regime the expressions for ${\bf H}_{\rm phys}$ are identical, since we know that 
$|-2\sqrt{Q}C^{\rm geo}+2\sqrt{Q}\sum\limits_{j=1}^3\sqrt{Q^{jj}C^{\rm geo}_jC^{\rm geo}_j}| = -2\sqrt{Q}C^{\rm geo}+2\sqrt{Q}\sum\limits_{j=1}^3\sqrt{Q^{jj}C^{\rm geo}_jC^{\rm geo}_j} \geq 0$
and in the quantum theory we ensure a well defined expression under the square root by taking the absolute value. The general strategy for the quantization within LQG one follows is to introduce a regulator by means of which a regularization of ${\bf H}_{\rm phys}$ can be found. Afterwards one shows that in the limit where the regulator is removed one ends up with a well defined expression for the physical Hamiltonian operator $\hat{\bf H}_{\rm phys}$. As mentioned before in contrast to other physical Hamiltonians that have been quantized so far,  in our case ${\bf H}_{\rm phys}$ is no longer covariant at the observable level because the summation is performed outside the square root in ${\bf H}_{\rm phys}$ and thus we need to introduce a different regularization procedure here. As far as the first term under the square root is considered, we can quantize it by applying a regularization that has already been discussed in the literature for the Hamiltonian constraint in \cite{Thiemann:1996ay} and for the volume operator in \cite{Ashtekar:1997fb}.
To quantize the second term under the square root as a first step we rewrite it in terms of densitized triads. This results in
\begin{align}
 \sqrt{Q}\sum\limits_{j=1}^3\sqrt{Q^{jj}C^{\rm geo}_jC^{\rm geo}_j}
 &= \sum\limits_{j=1}^3\sqrt{Q Q^{jj}C^{\rm geo}_jC^{\rm geo}_j}
 = \sum\limits_{j=1}^3 \sqrt{\frac{Q \delta^{JK} E^j_J E^j_K F_{jk}^L F^{M}_{j\ell} E^k_L E^{\ell}_M}{Q}}\\ \nonumber
 &= \sum\limits_{j=1}^3 \sqrt{ E^j_J E^j_K F_{jk}^L F^{M}_{j\ell} E^k_L E^{\ell}_M \delta^{JK}}
 = \sum\limits_{j=1}^3 \sqrt{F_{jk}^L  E^j_J  E^k_L F^{M}_{j\ell} E^j_K  E^{\ell}_M \delta^{JK}}\\ \nonumber
 &= \sum\limits_{j=1}^3 \sqrt{O^{(j)}_J O^{(j)}_K \delta^{JK}},
\end{align}
where we introduced the quantities $O^{(j)}_J:= F_{jk}^L  E^j_J  E^k_L$ (no summation over $j$) and we used that $Q^{ij} = \frac{\delta^{JK} E^i_J E^j_K}{Q}$, $Q(E) := \det(Q_{ij}(E))$,
$C_j^{\rm geo} = F^L_{jk} E^k_L$ with scalar field manifold indices $i,j,\ldots$ and su(2) Lie algebra indices $I,J,\ldots$ .
At the classical level the order of the curvature $F$ and the densitized triads $E$ is irrelevant, but at the quantum level it is important that $F$ is ordered to the left in order to avoid the creation of infinitely many loops at the vertices of a given graph
when the operator acts on the corresponding cylindrical function. In the next section we will discuss the regularization of the physical Hamiltonian in detail.

\subsubsection{Regularization of ${\bf H}_{\rm phys}$}
For the regularization  of ${\bf H}_{\rm phys}$ we will introduce a point splitting regularization along the lines of \cite{Thiemann:2007zz} where it was applied to quantize the volume operator of LQG. For this purpose we need to introduce a characteristic function associated with some geometrical objects that we denote by $\dummy$.
In principal we can make an arbitrary choice for such geometrical objects, however usually in the existing literature cubes or tetrahedra have been chosen. The only difference between different choices for $\dummy$ will be a constant global factor, called the regularization constant $c_{\dummy}$. This constant is involved in the volume of the considered objects,
i.e. $\text{vol}(\dummy) = c_{\dummy} \epsilon^3$, where $\epsilon > 0$ is the basic length of the object under consideration.
For example, for a cube denoted by $\Box$ we have $c_{\Box} =1$ and for a tetrahedron denoted by $\triangle$ we get $c_{\triangle}= \frac{\sqrt{2}}{12}$. To keep our presentation simple and to be able to compare our results with already existing results we will use tetrahedra in the embedded LQG case and cubes for the AQG framework. The reason for these choices is that then we can carry over already existing quantization techniques for $C^{\rm geo}$ \cite{Giesel:2007wn,Thiemann:2007zz} to the case of our physical Hamiltonian. 
Before we perform the point splitting, we write 
${\bf H}_{\rm phys}$ as
\begin{eqnarray*}
{\bf H}_{\rm phys} &=& \int\limits_{\cal S}d^3p \sqrt{|-2\sqrt{Q}C^{\rm geo}+2\sum\limits_{j=1}^3 \sqrt{O^{(j)}_J O^{(j)}_K \delta^{JK}}|}(p), \\
\end{eqnarray*}
where $p:=\sigma$ denotes the points of the scalar manifold $\cal S$ from now on. For the regularization of $O^{(j)}_J$ we will consider a point splitting regularization for the two densitized triads and the curvature similar to the case of the volume operator where a product of three densitzed triads is involved. Later we will reexpress the curvature in terms of holonomies as usually done in LQG. Let us discuss the individual steps in detail. For simplicity we discuss the case for $j=1$ first, the remaining three cases work similar. Applying the point splitting we regularize $O^{(1)}_J$ as follows
\begin{eqnarray*}
O^{(1)}_J(p)&=&\lim\limits_{\triangle'\triangle\to 0}\frac{1}{{\rm vol}(\triangle'){\rm vol}(\triangle)}
\int\limits_{\cal S}  d^3y\, \chi_{\triangle'}(p,y)F^M_{1k}(y)E^1_J(y)
\int\limits_{\cal S}  d^3x\, \chi_{\triangle}(p,x)E^k_M(x) \\
&=:&\lim\limits_{\triangle'\triangle\to 0}O^{(1)}_J(p,\triangle',\triangle).
\end{eqnarray*}
Here $\chi_{\triangle}(p,x)$ denotes the characteristic function of a tetrahedron $\triangle$ with the limit $\lim\limits_{\triangle\to 0}\frac{\chi_{\triangle}(p,x)}{{\rm vol}(\triangle)}=\delta^{(3)}(p,x)$. 
Due to the Poisson algebra of the Ashtekar connection and the densitized triad which has the form $\{A^I_i(x) , E^j_J(y)\} = \frac{\kappa \beta}{2} \delta^j_i \delta^I_J \delta^{(3)}(x,y)$ 
the operator corresponding to $E^j_J (x)$ can be represented by
\begin{align}
  \hat{E}^j_J(x) = -i \frac{\ell_P^2}{2} \frac{\delta}{\delta A^J_j(x)}
\end{align}
with the Planck length $\ell_P = \sqrt{\hbar \kappa}$ and we set $\beta =1$ for simplicity. Given this, we can define a regularized flux operator by
\begin{align}
 \hat{E}^j_J(p, \triangle) &:= \frac{1}{\text{vol}(\triangle)} \int\limits_{\mathcal{S}} \mathrm{d}^3x \, \chi_{\triangle}(p,x) \hat{E}^j_J(x) \\ \nonumber
			&= -i \frac{\ell_P^2}{2} \frac{1}{\text{vol}(\triangle)} \int\limits_{\mathcal{S}} \mathrm{d}^3x \, \chi_{\triangle}(p,x)  \frac{\delta}{\delta A^J_j(x)}.
\end{align}
Then, we reexpress the regularized operator $O^{(1)}_J(p,\triangle',\triangle)$  as
\begin{equation*}
\hat{O}^{(1)}_J(p,\triangle',\triangle)= \frac{(-i)^2 \ell_P^4}{4} \frac{1}{{\rm vol}(\triangle'){\rm vol}(\triangle)}
\int\limits_{\cal S}  d^3y\, \chi_{\triangle'}(p,y)F^M_{1k}(y)\frac{\delta}{\delta A^J_1(y)}
\int\limits_{\cal S}  d^3x\, \chi_{\triangle}(p,x)\frac{\delta}{\delta A^M_k(x)}.
\end{equation*}
What we still have to analyze is whether the limit in which the regulator is removed leads to a well defined expression for $\hat{\bf H}_{\rm phys}$. For this purpose we will discuss in detail the action  of $\hat{O}^{(1)}_J(p,\triangle',\triangle)$ on cylindrical functions and how the limit can be performed. 

\subsubsection{Action of $\hat{O}^{(j)}_J(p, \triangle', \triangle)$ on cylindrical functions}
For analyzing the action of the regularized operator $\hat{O}^{(j)}_J(p, \triangle', \triangle)$ on a generic cylindrical function $f_{\gamma}$, we first compute the action of the regularized flux operator on $f_{\gamma}$. Afterwards we will discuss how the curvature can be regularized and expressed in terms of holonomy operators. 
We obtain for the action of the regularized flux operator on a generic cylindrical function $f_{\gamma}$
\begin{align}
 \hat{E}^j_J(p, \triangle) f_{\gamma}(h_e[A]) &= -i \frac{\ell_P^2}{2} \frac{1}{\text{vol}(\triangle)} \int\limits_{\cal S} \mathrm{d}^3x \, \chi_{\triangle}(p,x)  \frac{\delta h_e}{\delta A^J_j(x)} \frac{\delta}{\delta h_e} f_{\gamma} \\ \nonumber
					   &= +i \frac{\ell_P^2}{2} \frac{1}{\text{vol}(\triangle)} \sum\limits_{e \in E(\gamma)}
					    \int\limits_0^1 \mathrm{d}t \,\chi_{\triangle}(p,e(t)) \dot{e}^j(t) \frac{1}{2} 
					    \text{Tr}\!\lr \ls h_e(0,t) \tau_J h_e(t,1)\rs \frac{\delta}{\delta h_e^T(0,1)} \rr f_{\gamma},
\end{align}
where we parametrize an edge $e$ by $e:[0,1] \rightarrow \mathcal{S}$, $t \mapsto e(t)$ and $\tau_J=i\sigma_J$ with $\sigma_J$, $J=1,2,3$, being the Pauli matrices. We used the notation $f_{\gamma} =f_{\gamma}(h_e[A])$ to emphasize the dependence of a cylindrical function on the holonomies and the dependence of the latter on the connections. Now we can also apply the second part of the regularized operator leading to an action of $\hat{O}^{(j)}_J(p, \triangle', \triangle)$ on $f_{\gamma}$ given by
\begin{align}
 \hat{O}^{(1)}_J(p, \triangle', \triangle) f_{\gamma} 
 &=  \frac{(+i)^2 \ell_P^4}{4} \frac{1}{\text{vol}(\triangle')}\frac{1}{\text{vol}(\triangle)} \\ \nonumber
 &\bigg\{ \sum\limits_{e,e' \in E(\gamma)} \int\limits_0^1 \mathrm{d}t'  \int\limits_0^1 \mathrm{d}t \,  F^M_{1m}(e'(t')) \, \chi_{\triangle'}(p,e'(t')) \chi_{\triangle}(p,e(t)) 
 \dot{e}'^1(t') \dot{e}^m(t) \\ \nonumber
 &\frac{1}{4} \text{Tr}\!\lr \ls h_{e'}(0,t') \tau_J h_{e'}(t',1)\rs \frac{\delta}{\delta h_{e'}^T(0,1)} \rr
	      \text{Tr}\!\lr \ls h_{e}(0,t) \tau_M h_{e}(t,1)\rs \frac{\delta}{\delta h_{e}^T(0,1)} \rr \\ \nonumber
 &+ \sum\limits_{e \in E(\gamma)} \int\limits_0^1 \mathrm{d}t  \int\limits_0^1 \mathrm{d}t' \, F^M_{1m}(e(t')) \, \chi_{\triangle'}(p,e(t')) \chi_{\triangle}(p,e(t)) 
 \dot{e}^1(t') \dot{e}^m(t)\\ \nonumber
 & \left[ \frac{1}{4} \Theta(t', t) \text{Tr}\!\lr \ls h_{e}(0,t') \tau_J h_{e}(t',t) \tau_M h_{e}(t,1)\rs \frac{\delta}{\delta h_{e}^T(0,1)} \rr \right. \\ \nonumber
 &+ \left. \frac{1}{4} \Theta(t, t') \text{Tr}\!\lr \ls h_{e}(0,t) \tau_M h_{e}(t,t') \tau_J h_{e}(t',1)\rs \frac{\delta}{\delta h_{e}^T(0,1)} \rr  \right] \bigg\} f_{\gamma},
\end{align}
where we again stick to the case $j=1$ here and $E(\gamma)$ denotes the set of all edges of the graph $\gamma$. In the next step we will discuss how the curvature term can be regularized. For this purpose we write it in a more convenient way by introducing for an associated tangent vector of a given edge $e^1(t)$ the following notation:
\begin{align}
  \dot{e}^a_{(1)} := \delta^a_1 \dot{e}^1(t).
\end{align}
This has the advantage that we can express the curvature as 
\begin{align}
 F^M_{1m}(e'(t')) \dot{e}'^1(t') \dot{e}^m(t) = F^M_{nm}(e'(t')) \dot{e}'^n_{(1)}(t') \dot{e}^m(t)
\end{align}
and similarly for the remaining cases $j=2,3$. Considering this we can rewrite $\hat{O}^{(j)}_J(p, \triangle, \triangle') f_{\gamma}$ as
\begin{align}
 \hat{O}^{(j)}_J(p, \triangle', \triangle) f_{\gamma} 
 &=  \frac{(+i)^2 \ell_P^4}{4} \frac{1}{\text{vol}(\triangle')}\frac{1}{\text{vol}(\triangle)} \\ \nonumber
 &\bigg\{ \sum\limits_{e,e' \in E(\gamma)} \int\limits_0^1 \mathrm{d}t'  \int\limits_0^1 \mathrm{d}t \,  F^M_{am}(e'(t')) \, \chi_{\triangle'}(p,e'(t')) \chi_{\triangle}(p,e(t)) 
 \dot{e}'^a_{(j)}(t') \dot{e}^m(t) \\ \nonumber
 &\frac{1}{4} \text{Tr}\!\lr \ls h_{e'}(0,t') \tau_J h_{e'}(t',1)\rs \frac{\delta}{\delta h_{e'}^T(0,1)} \rr
	      \text{Tr}\!\lr \ls h_{e}(0,t) \tau_M h_{e}(t,1)\rs \frac{\delta}{\delta h_{e}^T(0,1)} \rr \\ \nonumber
 &+ \sum\limits_{e \in E(\gamma)} \int\limits_0^1 \mathrm{d}t  \int\limits_0^1 \mathrm{d}t' \, F^M_{am}(e(t')) \, \chi_{\triangle'}(p,e(t')) \chi_{\triangle}(p,e(t)) 
 \dot{e}^a_{(j)}(t') \dot{e}^m(t)\\ \nonumber
 & \left[ \frac{1}{4} \Theta(t', t) \text{Tr}\!\lr \ls h_{e}(0,t') \tau_J h_{e}(t',t) \tau_M h_{e}(t,1)\rs \frac{\delta}{\delta h_{e}^T(0,1)} \rr \right. \\ \nonumber
 &+ \left. \frac{1}{4} \Theta(t, t') \text{Tr}\!\lr \ls h_{e}(0,t) \tau_M h_{e}(t,t') \tau_J h_{e}(t',1)\rs \frac{\delta}{\delta h_{e}^T(0,1)} \rr  \right] \bigg\} f_{\gamma},
\end{align}
where again no summation over $j$ is taken into account. 

\subsubsection{Regularization of $\sqrt{Q}C^{\rm geo}$ and its action on cylindrical functions}
As discussed above for the first term under the outer square root that involves the volume $\sqrt{Q}$ as well as the geometric part of the Hamiltonian constraint $C^{\rm geo}$ we will carry over existing results from the literature where the quantization of both operators has already been presented. In order to be able to perform the limit for both parts of the regularized ${\bf H}_{\rm phys}$ we will use the same strategy that was for instance followed in \cite{Domagala:2010bm}. We introduce the following regularized quantities $\sqrt{Q}(p,\triangle)$ and $C^{\rm geo}(p,\triangle')$ defined through
\begin{eqnarray*}
\sqrt{Q}(p) &:=& \lim\limits_{\triangle\to 0}\sqrt{Q}(p,\triangle)=
\lim\limits_{\triangle\to 0}
\frac{1}{{\rm vol}(\triangle)}\int d^3x\, \sqrt{Q}(x)\chi_{\triangle}(p,x) \\
C^{\rm geo}(p) &:=& \lim\limits_{\triangle'\to 0}C^{\rm geo}(p,\triangle')
	= \lim\limits_{\triangle'\to 0}
	\frac{1}{{\rm vol}(\triangle')} \int d^3y\, C^{\rm geo}(y)\chi_{\triangle'}(p,y)
	 	\end{eqnarray*}
The action of their corresponding regularized operator product on cylindrical functions yields
\begin{eqnarray*}
\widehat{\sqrt{Q}}(p,\triangle) \widehat{C}^{\rm geo}(p,\triangle'))f_\gamma
&=&\frac{1}{{\rm vol}(\triangle'){\rm vol}(\triangle)}\int\limits d^3x\, \int\limits d^3y\, 
\chi_\triangle(p,x)\chi_{\triangle'}(p,y)\widehat{\sqrt{Q}}_x\widehat{C}^{\rm geo}_y f_\gamma, 
\end{eqnarray*}	
where $\widehat{\sqrt{Q}}_x, \widehat{C}^{\rm geo}_y$ denote the usual regularizations of the volume and the geometric Hamiltonian constraint that can for instance be found in \cite{Thiemann:1996ay} and \cite{Ashtekar:1997fb}. 	
In the next section we will discuss in detail how the limit of this regularized operator can be performed and how this can be used to finally define an operator for the physical Hamiltonian ${\bf H}_{\rm phys}$.

\subsubsection{Performing the limit of the regularized physical Hamiltonian}
Let us start with discussing the limit for the regularized operator $\hat{O}^{(j)}(p,\triangle',\triangle)$. Due to the characteristic functions that are involved in the regularized operator, we realize that the first part of the operator involving the sum $\sum_{e,e'}$ will only contribute if $e,e'$ have a point of intersection that we denote by $p$. In case they do not intersect, we can shrink $\triangle',\triangle$ appropriately to some small but finite region and both characteristic functions have support only in a neighborhood of $p$. Hence, if the edges do not intersect the first part vanishes identically. Let us assume that $p$ is the point of intersection of $e,e'$ at parameter values $t_0,t_0'$. For the reason that by assumption the edges are not self-intersecting $t_0,t'_0$ are unique. We parametrize the edges as
\begin{equation*}
	e(t)=p+c(t-t_0), \quad e'(t')=p+c'(t'-t'_0),
\end{equation*}
where $c,c'$ are analytic functions which vanish at $t-t_0=0$, respectively $t'-t'_0=0$. Since $e,e'$ must intersect at $p$ it follows that $p=v=e\cap e'$ must be a common vertex of the edges. By assumption edges can only intersect at their beginning or final points. Without loss of generality we are able to choose an adapted graph $\gamma$ in such a way that it will be possible to classify each edge as an edge of either type up or type down, respectively either type in or type out. If this is not directly given we can subdivide edges appropriately such that we are in this situation.  Further, we divide the edges in such a way that they all have outgoing orientation with respect to a vertex $v$, which is also equal to the intersection point, such that the flux operators can entirely be expressed in terms of right invariant vector fields. In this case the edges intersect in their beginning point and thus the unique value of $t_0,t'_0$ is given by $t_0=t'_0=0$. The general structure of the individual terms in the action of $\hat{O}^{(j)}(p,\triangle',\triangle)$ is of the form $\int dt \int dt'g(t')h(t)f_\gamma$ for appropriately chosen functions $g$ and $h$. Taking the discussion above into account in the limit where the tetrahedra $\triangle$ become smaller and smaller we can expand the individual terms in the action in powers of $\epsilon$ according to:
\begin{equation*}
\int\limits_0^1 dt\int\limits_0^1 dt' g(t')h(t)f_\gamma=\left(g(0)h(0)\frac{\epsilon^2}{4}+o(\epsilon^2)\right)f_\gamma,	
\end{equation*}
where the limit $\triangle\to 0$ corresponds to $\epsilon\to 0$ because ${\rm vol}(\triangle)=c_\triangle\epsilon^3$. Note, that the factor of $\frac{1}{4}$ is due to the fact that $\int_{\mathds{R}_+} dt \delta(0,t)\int_{\mathds{R}_+} dt' \delta(0,t')=\frac{1}{4}$. Additionally, we assumed that the functions $g,h$ have only support in an interval $\epsilon$ which is given due to the characteristic functions involved. If we apply this kind of expansion to the action of $\hat{O}^{(j)}(p,\triangle',\triangle)$ we end up with
\begin{align}
 \hat{O}^{(j)}_J(p, \triangle', \triangle) f_{\gamma} 
 &=  \frac{(+i)^2 \ell_P^4}{4} \frac{1}{c_{\triangle'} c_{\triangle} \epsilon^6} \\ \nonumber
 &\bigg\{ \sum\limits_{e,e' \in E(\gamma)} \frac{\epsilon^2}{4}\Big(  F^M_{am}(e'(0)) \, \chi_{\triangle'}(p,e'(0)) \chi_{\triangle}(p,e(0)) 
 \dot{e}'^a_{(j)}(0) \dot{e}^m(0) \\ \nonumber
 &\frac{1}{4} \text{Tr}\!\lr \ls  \tau_J h_{e'}(0,1)\rs \frac{\delta}{\delta h_{e'}^T(0,1)} \rr
	      \text{Tr}\!\lr \ls \tau_M h_{e}(0,1)\rs \frac{\delta}{\delta h_{e}^T(0,1)}\Big) \rr \\ \nonumber
 &+ \sum\limits_{e \in E(\gamma)} \frac{\epsilon^2}{4}\Big(F^M_{am}(e(0)) \, \chi_{\triangle'}(p,e(0)) \chi_{\triangle}(p,e(0)) 
 \dot{e}^a_{(j)}(0) \dot{e}^m(0)\\ \nonumber
 & \left[ \frac{1}{4} \text{Tr}\!\lr \ls  \tau_J \tau_M h_{e}(0,1)\rs \frac{\delta}{\delta h_{e}^T(0,1)} \rr \right. \\ \nonumber
 &+ \left. \frac{1}{4} \text{Tr}\!\lr \ls  \tau_M \tau_J h_{e}(0,1)\rs \frac{\delta}{\delta h_{e}^T(0,1)}\Big) \rr  \right] +o(\epsilon^2)\bigg\} f_{\gamma}, \\ \nonumber
  \end{align}
where we used that $\Theta(0,0)=1$ as well as $h_e(0,0)=\mathds{1}_{\text{SU(2)}}$. For the approximation of the integrals we did not compute the terms of order $\epsilon^3$ or higher explicitly here because these terms will vanish anyway in the limit where the regulator is removed. 
We can rewrite the second sum over the edges in a more compact form if we introduce the anti-commutator $\{\tau_J,\tau_M\}_+$ and obtain
\begin{align}
\label{ActionO}
 \hat{O}^{(j)}_J(p, \triangle', \triangle) f_{\gamma} 
&=
 \frac{(+i)^2 \ell_P^4}{4} \frac{1}{c_{\triangle'} c_{\triangle} \epsilon^6} \\ \nonumber
 &\bigg\{ \sum\limits_{e,e' \in E(\gamma)} \frac{\epsilon^2}{4}\Big(  F^M_{am}(e'(0)) \, \chi_{\triangle'}(p,e'(0)) \chi_{\triangle}(p,e(0)) 
 \dot{e}'^a_{(j)}(0) \dot{e}^m(0) \\ \nonumber
 &\frac{1}{4} \text{Tr}\!\lr \ls  \tau_J h_{e'}(0,1)\rs \frac{\delta}{\delta h_{e'}^T(0,1)} \rr
	      \text{Tr}\!\lr \ls \tau_M h_{e}(0,1)\rs \frac{\delta}{\delta h_{e}^T(0,1)}\Big) \rr \\ \nonumber
 &+ \sum\limits_{e \in E(\gamma)} \frac{\epsilon^2}{4}\Big(F^M_{am}(e(0)) \, \chi_{\triangle'}(p,e(0)) \chi_{\triangle}(p,e(0)) 
 \dot{e}^a_{(j)}(0) \dot{e}^m(0)\\ \nonumber
 & \left[ \frac{1}{4} \text{Tr}\!\lr \ls  \{\tau_J, \tau_M\}_+ h_{e}(0,1)\rs \frac{\delta}{\delta h_{e}^T(0,1)} \rr\right]
 +o(\epsilon^2)\bigg\} f_{\gamma}. 
\end{align}
Our next steps involve to replace the curvature by appropriate holonomy operators and to use the properties of the Pauli matrices to rewrite the anti-commutator in a convenient way. From our discussion above we know that $e(0)=e'(0)=v$ thus the curvature is evaluated at the vertices $v$ in all terms. Similarly, we can replace $e(0),e'(0)$ by $v$ in all characteristic functions. Using the expansion of a loop $\alpha_{e'(j)e}$ in powers of $\epsilon$ we have
\begin{equation*}
	h_{\alpha_{e'_{(j)}e}}=\mathds{1}_{\text{SU(2)}}+\epsilon^2F^J_{ab}(v)\dot{e}'^a_{(j)}(0)\dot{e}^b(0)\frac{\tau_J}{2}+o(\epsilon^2)
\end{equation*}
and it is simple to show that the following identity holds:
\begin{equation*}
F^M_{ab}(v)\dot{e}'^a_{(j)}(0)\dot{e}^b(0)=-\frac{1}{\epsilon^2}\text{Tr}\!\lr h_{\alpha_{e'_{(j)}e}}\tau^M\rr .
\end{equation*}
The anti-commutator satisfies $\{\tau_J,\tau_M\}_+=-2\delta_{JM}\mathds{1}_{\text{SU(2)}}$. Reinserting both into (\ref{ActionO}) we end up with
\begin{align}
\label{ActionO2}
 \hat{O}^{(j)}_J(p, \triangle', \triangle) f_{\gamma} 
&=
 \frac{(+i)^2 \ell_P^4}{4} \frac{1}{c_{\triangle'} c_{\triangle} \epsilon^6} \\ \nonumber
 &\bigg\{ \sum\limits_{e,e' \in E(\gamma)} (-1)\frac{1}{4}\Big(\text{Tr}\!\lr h_{\alpha_{e'_{(j)}e}}\tau^M\rr  \, \chi_{\triangle'}(p,v) \chi_{\triangle}(p,v) 
  \\ \nonumber
 &\frac{1}{4} \text{Tr}\!\lr \ls  \tau_J h_{e'}(0,1)\rs \frac{\delta}{\delta h_{e'}^T(0,1)} \rr
	      \text{Tr}\!\lr \ls \tau_M h_{e}(0,1)\rs \frac{\delta}{\delta h_{e}^T(0,1)}\Big) \rr \\ \nonumber
 &+ \sum\limits_{e \in E(\gamma)}(-1) \frac{1}{4}\Big(\text{Tr}\!\lr h_{\alpha_{e_{(j)}e}}\tau^M\rr \, \chi_{\triangle'}(p,v) \chi_{\triangle}(p,v) \\ \nonumber
 & \left[(-1) \frac{1}{2} \text{Tr}\!\lr \mathds{1}_{\text{SU(2)}}\delta_{JM}  h_{e}(0,1)
 \frac{\delta}{\delta h_{e}^T(0,1)} \rr\right]
 +o(\epsilon^2)\bigg\} f_{\gamma}.
\end{align}
To further rewrite the action of the operator we introduce the right invariant vector fields $X_0^e$ and $X_L^e$ associated with an edge $e$ by
\begin{align*}
X_0^e &:={\text Tr}\!\lr \tau_0 h_e(0,1)\frac{\delta}{\delta h_e^T(0,1)}\rr,\\
X_L^e &:={\text Tr}\!\lr \tau_L h_e(0,1)\frac{\delta}{\delta h_e^T(0,1)}\rr,  
\end{align*}
where $L$ runs from $1,\ldots 3,$ and we also include the $2\times 2$ unity matrix $\sigma_0 = \mathds{1}_{\text{SU(2)}}$, that is $\tau_0:=i\sigma_0$. Then, $\tau_0$ and $\tau_J$, $J=1,2,3$, are the generators of the group U(2), since every element of U(2) can be written as the exponential of a Hermitian $2 \times 2$ matrix which is equal to $\exp{\lr a\sigma_0 + b^J \sigma_J \rr}$ with $a, b^J \in \mathbb{R}$. In this context we can understand $X^e_0,X^e_L$ as right invariant vector fields associated with U(2). 

For the reason that the edges have to intersect in a common vertex $v$, we can rewrite both sums over edges as a sum over all vertices and a sum over all edges meeting at these vertices. Hence, the action of $\hat{O}^{(j)}_J(p, \triangle', \triangle)$ on $f_{\gamma}$ reads
\begin{align}
\label{ActionO3}
 \hat{O}^{(j)}_J(p, \triangle', \triangle) f_{\gamma} 
&=
- \frac{(+i)^2 \ell_P^4}{4} \frac{1}{c_{\triangle'} c_{\triangle}\epsilon^6} \\ \nonumber
 &\bigg\{\frac{1}{16}\sum\limits_{v\in V(\gamma)} \sum\limits_{e\cap e'=v}  \chi_{\triangle'}(p,v) \chi_{\triangle}(p,v)\, \text{Tr}\!\lr h_{\alpha_{e'_{(j)}e}}\tau^M\rr X_J^{e'}X_M^e\\ \nonumber
 &+\frac{i}{8}\delta_{JM} \sum\limits_{v \in V(\gamma)}\sum\limits_{b(e)=v}
  \chi_{\triangle'}(p,v) \chi_{\triangle}(p,v)\, \text{Tr}\!\lr h_{\alpha_{e_{(j)}e}}\tau^M\rr X^e_0
 +o(\epsilon^2)\bigg\} f_{\gamma}. 
\end{align}
In order to obtain the final operator for $\hat{\bf H}_{\rm phys}$ we need to consider the limit where the regulator is removed explicitly, that is the limit in which the volume of all tetrahedra shrinks to zero or equivalently $\epsilon$ tends to zero. Without loss of generality we can choose $\triangle=\triangle'=\triangle''=\triangle'''=:\triangle$ where $\triangle, \triangle',\triangle'',\triangle'''$ denote the tetrahedra associated to the regularization of the individual operators involved in ${\hat{\bf H}}_{\rm phys}$ and perform all limits simultaneously. Then we can just consider the limit $\epsilon\to 0$. 
Formally, we have
\begin{equation*}
\hat{\bf H}_{\rm phys}f_\gamma:=\lim\limits_{\epsilon\to 0}\hat{\bf H}^\epsilon_{\rm phys}f_\gamma.    
\end{equation*}
With our discussion above, the total regularized physical Hamiltonian is given by
\begin{align}
\label{Hphys}
 \hat{\bf H}_{\rm phys} f_{\gamma} 
& :=\lim\limits_{\epsilon\to 0}
 \int d^3p\Big[2\Big|-\frac{\chi^2_\Delta(p,v)}{c_{\triangle} c_{\triangle}\epsilon^6}\frac{1}{2}
 \widehat{\sqrt{Q}}_v \hat{C}^{\rm geo}_v \\ \nonumber 
&+\sum\limits_{j=1}^3\Big[\sum\limits_{v\in V(\gamma)}
\left(\frac{(+i)^2 \ell_P^4}{4}\right)^2 \frac{\chi^4_{\triangle}(p,v)}{ c_{\triangle}^4\epsilon^{12}}\\ \nonumber
&\Big(\frac{1}{16}\sum\limits_{e\cap e'=v}\text{Tr}\!\lr h_{\alpha_{e'_{(j)}e}}\tau^M\rr X_J^{e'}X_M^e
+\frac{i}{8}\delta_{JM}\sum\limits_{b(e)=v}
 \text{Tr}\!\lr h_{\alpha_{e_{(j)}e}}\tau^M\rr X^e_0 
+o(\epsilon^2)\Big)^\dagger\\ \nonumber
&\Big(\frac{1}{16}\sum\limits_{e''\cap e'''=v}\text{Tr}\!\lr h_{\alpha_{e'''_{(j)}e''}}\tau^M\rr X_K^{e'''}X_N^{e''}
+\frac{i}{8}\delta_{KN}\sum\limits_{b(e'')=v}
 \text{Tr}\!\lr h_{\alpha_{e''_{(j)}e''}}\tau^N\rr X^{e''}_0
\Big)+o(\epsilon^2)\Big)\\ \nonumber
&\delta^{JK}\Big]^{\frac{1}{2}}\Big|\Big]^\frac{1}{2} f_{\gamma}, 
\end{align}
where we chose the operator ordering of $\hat{O}^{(j)}_J(p, \triangle', \triangle)$ and its adjoint in such a way that the square of this operator does not create an infinite number of holonomy loops at the vertices. This is the usual operator ordering that is chosen also for the quantization of the Hamiltonian constraint in \cite{Thiemann:1996ay} and the physical Hamiltonian in \cite{Giesel:2007wn}.

Now in the limit $\epsilon\to 0$ only at most one vertex will contribute because in this limit at most one vertex is contained in the volume of $\triangle$ if these $\triangle$'s are sufficiently small or equivalently if $\epsilon$ is small enough. Given this, we can take the powers of the characteristic functions first out of the inner square root and afterwards out of the remaining square root. In case we further use that these characteristic functions become $\delta$-functions in that limit and also that all $o(\epsilon^2)$-terms vanish we finally obtain
\begin{equation*}
\hat{\bf H}_{\rm phys,\gamma}f_\gamma=\int d^3p\, \hat{\bf h}_{\rm phys,\gamma}(p)f_\gamma
\end{equation*}
with
\begin{equation*}
\hat{\bf h}_{\rm phys,\gamma}(p):=\sum\limits_{v\in V(\gamma)}\delta^{(3)}(p,v)\hat{\bf h}_{\rm phys,\gamma,v},
\end{equation*}
where we added an extra index $\gamma$ as a reminder of the graph dependence and
chose a symmetric ordering of the term involving $\hat{C}^{\rm geo}_{\gamma,v}$ after performing the limit  $\epsilon\to 0$.
So $\hat{\bf h}_{\rm phys,\gamma,v}$ reads
\begin{align}
\hat{\bf h}_{\rm phys,\gamma,v} 
\label{FinalOPHphys} 
&:=\Big[2\Big|-\frac{1}{2}\left(\widehat{\sqrt{Q}}_{\gamma,v}\hat{C}^{\rm geo}_{\gamma,v}+(\hat{C}^{\rm geo}_{\gamma,v})^{\dagger}\widehat{\sqrt{Q}}_{\gamma,v}\right)
+\sum\limits_{j=1}^3\Big[\left(\frac{(+i)^2 \ell_P^4}{4}\right)^2\delta^{JK} \\ \nonumber
&\Big(\frac{1}{16}\sum\limits_{e\cap e'=v}\text{Tr}\!\lr h_{\alpha_{e'(j)e}}\tau^M\rr X_J^{e'}X_M^e
+\frac{i}{8}\delta_{JM}\sum\limits_{b(e)=v}
 \text{Tr}\!\lr h_{\alpha_{e(j)e}}\tau^M\rr X^e_0
\Big)^\dagger\\ \nonumber
&\Big(\frac{1}{16}\sum\limits_{e''\cap e'''=v}\text{Tr}\!\lr h_{\alpha_{e'''(j)e''}}\tau^M\rr X_K^{e'''}X_N^{e''}
+\frac{i}{8}\delta_{KN}\sum\limits_{b(e'')=v}
 \text{Tr}\!\lr h_{\alpha_{e''(j)e''}}\tau^N\rr  X^{e''}_0
\Big)
\Big]^{\frac{1}{2}}\Big|\Big]^\frac{1}{2}.
\end{align}
We will postpone a discussion about the action of the individual parts of this physical Hamiltonian operator to section \ref{s35} where we combine this discussion with a comparison to the physical Hamiltonian operator of the one Klein-Gordon scalar field model in \cite{Domagala:2010bm}. Before doing so we will discuss some aspects of graph-modifying versus graph-preserving quantizations and afterwards show how the operator can be quantized using the framework of Algebraic Quantum Gravity \cite{Giesel:2007wn}. 

\subsubsection{Remarks on the Application of the LQG framework}
There is a conceptual difference when we perform an unreduced or reduced quantization of LQG as far as the spatial diffeomorphism group is considered. In the case of the unreduced quantization spatial diffeomorphism are understood as gauge transformations and one eliminates them via a Dirac quantization procedure. Now in the case of the reduced quantization we look for representations of the observable algebra whose elements are Dirac observables carrying tensor indices of the scalar field manifold. As a consequence, in the reduced case the spatial diffeomorphism group is no longer a gauge group, but should be understood as active diffeomorphisms and hence a symmetry group, for more details see the discussion in \cite{Giesel:2007wn}. Now due to the fact that the observable algebra can also be represented by the standard Ashtekar-Lewandowski representation in the reduced quantization the representation of the physical Hilbert space is chosen to be the standard kinematical representation used in the Dirac quantization approach. As shown in \cite{Ashtekar:2000hv} spatially diffeomorphism invariant operators can only be implemented in a graph-preserving way in this representation. In \cite{Giesel:2007wn} the physical Hamiltonian is also on the dust manifold a spatially diffeomorphism invariant quantity and this led the authors of \cite{Giesel:2007wn} to the conclusion that the resulting physical Hamiltonian in this model must be quantized in a graph-preserving way. However, this constraint is absent in our model because as far as the scalar field manifold is considered ${\bf H}_{\rm phys}$ is not spatially diffeomorphism invariant and therefore needs not necessarily to be quantized graph non-changing. If we would additionally require the operator to be graph-preserving and implement this by introducing similar projectors as has be done in \cite{Giesel:2007wn} then we are in a situation where all contributions of the unusual second term are trivial for the following reason: In cases where the edge $e$ does not point into one of the $j$-directions the way the loop is attached will change the underlying graph $\gamma$ and hence these contributions will be projected out. If $e$ points in one of the $j$-directions then as discussed above the loop operator $h_{\alpha_{e_{(j)}e}}$ become the identity operator and thus the trace involving this loop operator vanishes. Therefore, for a graph-preserving quantization the unusual second term does not contribute to the final action. 
 A similar property can be found for the quantization of ${\bf H}_{\rm phys}$ in the context of Algebraic Quantum Gravity that will briefly discussed in the next section.

\subsubsection{Quantization of the physical Hamiltonian in the AQG framework}
\begin{figure}[h]
    \centering
    \includegraphics[width=0.5\textwidth]{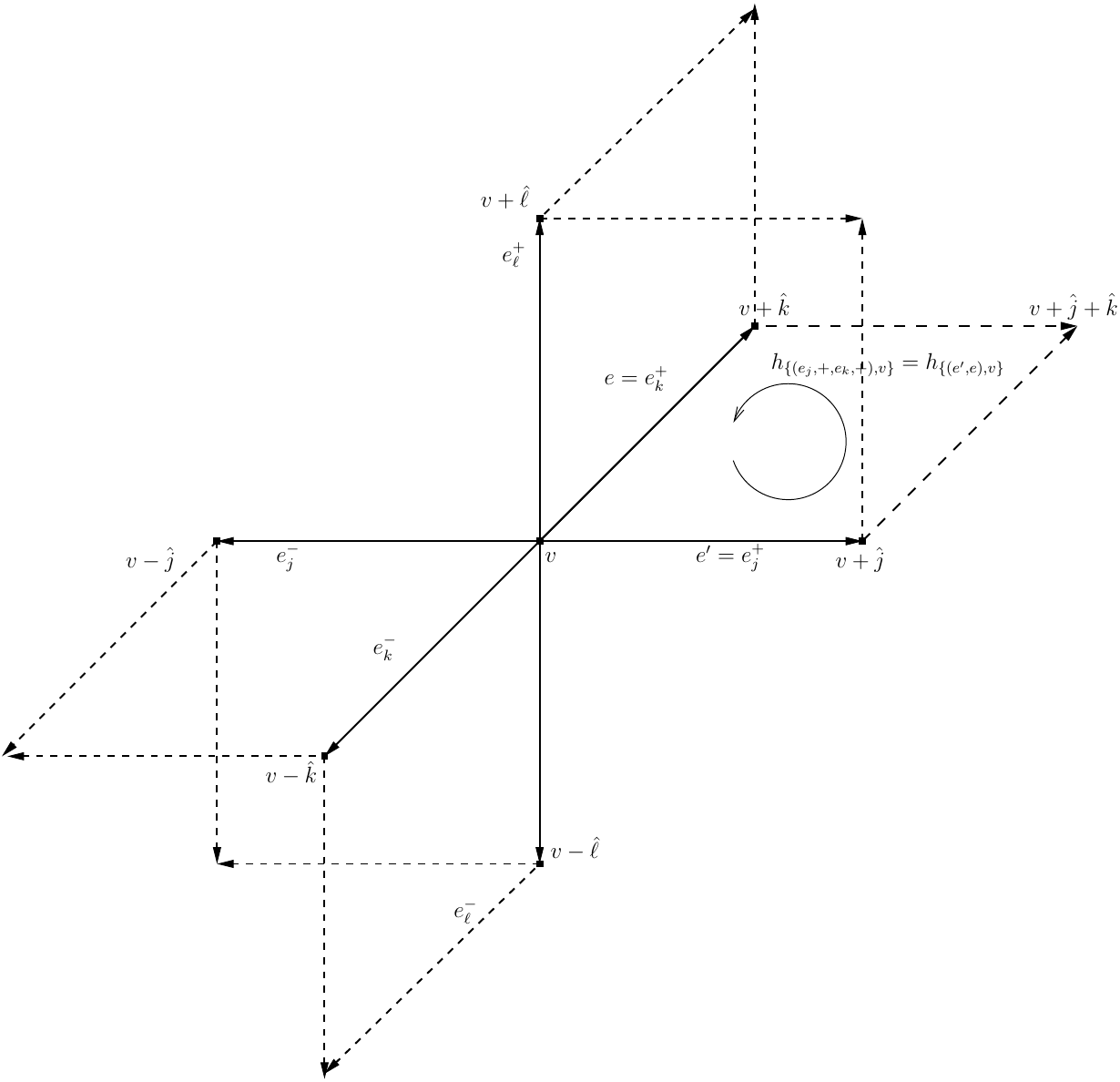}
    \caption{AQG cubic graph}
    \label{fig:AQG_cubic_graph}
\end{figure}

The idea of the Algebraic Quantum Gravity (AQG) framework is to quantize the dynamical operators completely at the algebraic level where no information about the embeddings of the graphs into the spatial manifold is known. This information is encoded in semiclassical states that can only be defined for a given but arbitrary choice of an embedding. Given these semiclassical states the classical limit of the dynamical operators can be computed and their algebraic quantization has to be chosen in such a way that their semiclassical limit has in lowest order in $\hbar$ the correct classical limit of canonical general relativity. Hence, here this will be the guiding principle for choosing an operator at the algebraic level and as far as the semiclassical limit is concerned we use the former results of \cite{Giesel:2007wn} to define the corresponding AQG operator for the four scalar field model analyzed in our work here.
As in \cite{Giesel:2007wn} we consider an AQG model of cubic topology which consists of an infinite algebraic graph with six valent vertices. We choose the orientation of all edges in such a way that all edges have outgoing orientation with respect to a vertex $v$. Using a similar notation to the one that was introduced in \cite{Giesel:2006uk} we label the six edges by $e^{\sigma}_j(v)$, here $\sigma$ stands for the positive $\sigma = +$
or negative direction $\sigma = -$ and $j=\{1,2,3\}$ denotes the edge $e$ whose tangent vector points into the $j$-th direction.
Furthermore, we choose $\{e_1,e_2,e_3\}$ to be right oriented with respect to the orientation of the field manifold $\mathcal{S}$. Note that although we use the same symbol the coordinates $\sigma^j$ of the dust manifold and the $\sigma$ here are completely unrelated. 
In order to implement a quantization of the loop operator at the algebraic level in a graph-preserving way we use the notion of a minimal loop introduced in \cite{Giesel:2007wn}. For this purpose we further define $e^{+}_j := e_j(v)$ and $e^{-}_j:=e_j(v - \hat{j})$ where $v -\hat{j}$ is a translation of the point $v$ along one unit of the $\hat{j}$-axis while the other two directions do not change. Here we  will parametrize the minimal loop $\alpha$ by $i,\sigma,j,\sigma'$ and $v$ denotes the vertex the minimal loop is attached to. Having introduced the definition of the edges above we can then obtain a minimal loop by the composition of the edges in the following way
\begin{align}
\alpha_{\{(i,\sigma, j , \sigma'),v \}}
= e^{\sigma}_i(v) \circ e^{\sigma'}_j(v+\sigma\hat{i}) \circ (e^{\sigma}_i)^{-1}(v+\sigma'\hat{j}) \circ (e^{\sigma'}_j)^{-1}(v).
\end{align}
The holonomy along the minimal loop is then given by
\begin{align}
    h_{\alpha_{\{(i,\sigma, j , \sigma'),v \}}}
    = h_{(i,\sigma),v} \circ h_{(j,\sigma'),v+\sigma\hat{i}} \circ h_{(i,\sigma),v+\sigma'\hat{j}}^{-1} \circ h_{(j,\sigma'),v}^{-1} =:h_{\alpha_{\Box}}.
\end{align}
For a visualization of the notions for an AQG graph, see figure \ref{fig:AQG_cubic_graph}.

Notice that here $h$ still denotes a SU(2) holonomy. 
With this graph-preserving quantization we immediately realize that that the contribution of the second unusual term in the embedded case has a trivial contribution in AQG and therefore does not need to be considered in the final form of $\hat{\bf H}_{\rm phys}$. This also synchronizes well with the fact in the AQG framework the operators are supposed to be embedding independent. In order to illustrate this point more in detail we consider in figure \ref{fig:action_AQG_Term2} as an example the following minimal loop
\begin{align}
\alpha_{\{(j ,+ , j , +),v \}}
= e^{+}_j(v)  \circ e^{+}_{(j)}(v+\hat{j})\circ (e^{+}_j)^{-1}(v+\hat{j})\circ (e^{+}_{(j)})^{-1}(v)= \text{Id},
\end{align}
where \text{Id} stands for the identity map, so that the holonomy along this loop becomes $ h_{\alpha_{\{(j,+,(j), +),v \}}}=\mathds{1}_{\text{SU(2)}}$.
Thus, we realize that in case of a cubic algebraic graph and a graph-preserving quantization the edges $e^\sigma_{(j)}$ and $e^\sigma_j$ can always be identified. Hence, the operator in the AQG framework at each vertex takes the following form:
\begin{figure}[h]
\begin{subfigure}[h]{0.45\textwidth}
     \centering
        \includegraphics[width=\textwidth]{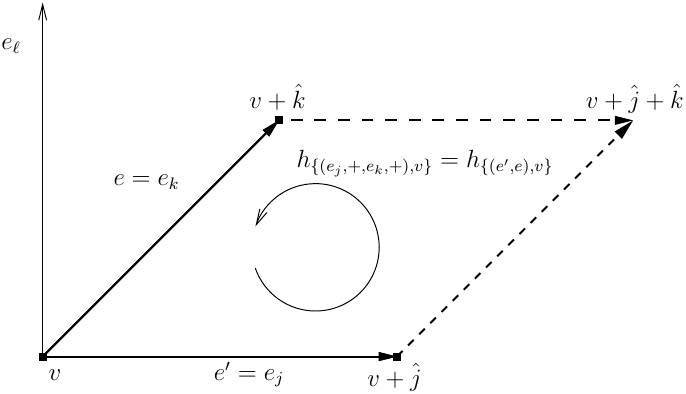}
        \subcaption{First term}
         \label{fig:action_AQG_Term1}
\end{subfigure}  
    \begin{subfigure}[h]{0.55\textwidth}
    \centering
        \includegraphics[width=\textwidth]{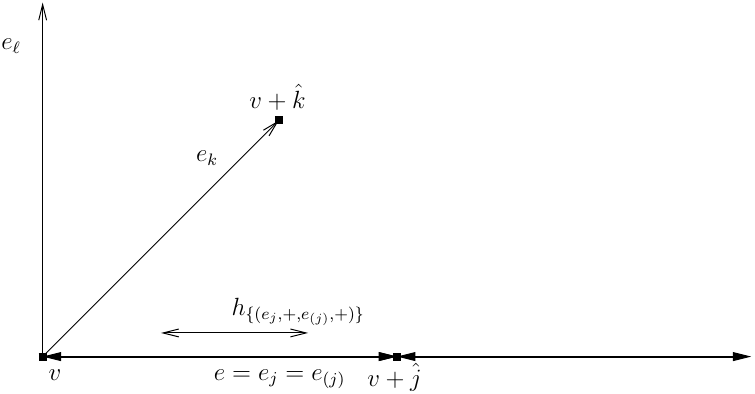}
        \subcaption{Second term}
         \label{fig:action_AQG_Term2}
    \end{subfigure}
\caption{Action of the first and the second term in $\hat{O}^{(j)}_J$ on a cubic AQG graph}
\label{fig:actionAQGterms}
 \end{figure}   
\begin{align}
\hat{\bf h}_{\rm phys,\gamma,v} 
\label{FinalOPHphysAQG} 
&:=\Big[2\Big|-\frac{1}{2}\left(\widehat{\sqrt{Q}}_{\gamma,v}\hat{C}^{\rm geo}_{\gamma,v}+\left(\hat{C}^{\rm geo}_{\gamma,v}\right)^{\dagger}\widehat{\sqrt{Q}}_{\gamma,v}\right)+\sum\limits_{j=1}^3\Big[\left(\frac{(+i)^2 \ell_P^4}{4}\right)^2\delta^{JK}\left(\frac{1}{16}\right)^2 \\ \nonumber
&\Big(\sum\limits_{e\cap e'=v}\Big(\text{Tr}\!\lr h_{\alpha_{\{(j,\sigma',i,\sigma),v\}}}\tau^M\rr X_{J,\{(j,\sigma'),v\}} X_{M,\{(i,\sigma),v\}} \Big)^{\dagger}\\ \nonumber
&\Big(\text{Tr}\!\lr h_{\alpha_{\{(j,\sigma''',k,\sigma''),v\}}}\tau^N\rr X_{K,\{(k,\sigma'''),v\}}
X_{N,\{(\ell,\sigma''),v\}} \Big)\Big]^{\frac{1}{2}}\Big|\Big]^\frac{1}{2}
\end{align}
where the right invariant vector fields are given by $X^e_K = X^{e^{\sigma}_i}_K= X_{K,\{(i,\sigma),v\}}$ and we have
\begin{align}
 \hat{C}^{\rm geo}_{\gamma,v}=\frac{1}{24 \ell_P^2} \sum\limits_{i,j,k}  \sum\limits_{\sigma,\sigma',\sigma'' = \pm} \sigma \sigma' \sigma'' \epsilon^{ijk}\,  \text{Tr}\!\lr h_{\alpha_{\{(i,\sigma,j,\sigma'),v\}}} h_{e_{\{(k,\sigma''),v\}}}\ls h_{e_{\{(k,\sigma''),v\}}}^{-1},  \hat{V}_{\gamma,v}\rs\rr
\end{align}
with $\hat{V}_{\gamma,v}$ the volume operator for a graph $\gamma$ , see also \cite{Giesel:2007wn}, given by
\begin{align}
 \hat{V}_{\gamma,v} = \hat{\sqrt{Q}}_{\gamma,v}= \ell_P^3 \sqrt{\Big| \frac{1}{48} \sum\limits_{i,j,k} \sum\limits_{I,J,K} \sum\limits_{\sigma,\sigma',\sigma'' = \pm} \sigma \sigma' \sigma'' \epsilon^{ijk} \epsilon^{IJK}\, X_{I,\{(i,\sigma),v\}} X_{J,\{(j,\sigma'),v\}} X_{K,\{(k,\sigma''),v\}}\Big|}.
\end{align}
Then the physical Hamiltonian operator becomes
\begin{equation*}
\hat{\bf H}_{\rm phys,\gamma}f_\gamma=\sum\limits_{v\in V(\gamma)}\hat{\bf h}_{\rm phys,\gamma,v}f_\gamma.
\end{equation*}
This finishes our discussion on the quantization of the physical Hamiltonian in the AQG framework. 

\subsection{Comparison of the physical Hamiltonians of the model from \cite{Domagala:2010bm} and the generalized Klein-Gordon scalar field model presented in section \ref{s3}}
\label{s35}
Let us briefly discuss to what kind of contributions the operator in the LQG framework will lead if it acts on a generic spin network function. This also allows us to compare it to the physical Hamiltonian in \cite{Domagala:2010bm} and analyze their differences in detail.

\begin{figure}[H]
\centering
    \begin{subfigure}{0.4\textwidth}
        \includegraphics[width=\textwidth]{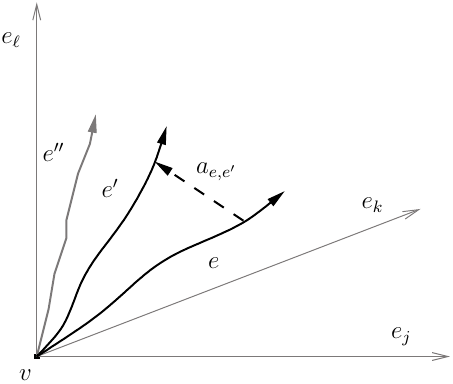}
        \caption{Action of the first term in $\hat{O}^{(j)}_J$.}
        \label{fig:action_LQG_Term1}
    \end{subfigure}
    \begin{subfigure}{0.4\textwidth}
        \includegraphics[width=\textwidth]{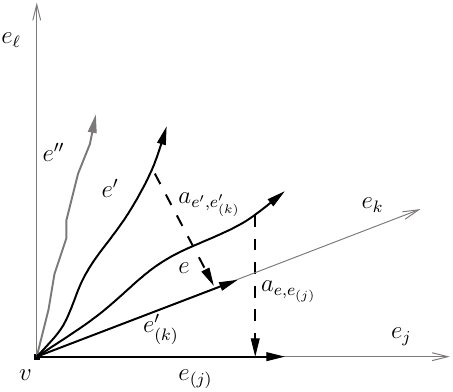}
        \caption{Action of the second term in $\hat{O}^{(j)}_J$.}
        \label{fig:action_LQG_Term2}
    \end{subfigure}
\caption{Action of $\hat{O}^{(j)}_J$ on LQG spin network functions}
\end{figure}

The first term under the square root in eq.\,(\ref{FinalOPHphys}) involving the volume operator as well as the geometric part of the Hamiltonian constraint operator is similar to the contributions that occur in the one Klein-Gordon scalar field model introduced in \cite{Domagala:2010bm}. For that model an additional term which involves $Q^{jk}C_j^{\rm geo}C_k^{\rm geo}$ at the classical level is neglected because in that model the spatial diffeomorphism constraint is solved via Dirac quantization and thus the physical Hamiltonian needs to be implemented on the spatial diffeomorphism invariant Hilbert space ${\cal H}_{diff}$. The operator version of the neglected term is expected to vanish on spatially diffeomorphism invariant states. 
 The final physical Hamiltonian one works with in \cite{Domagala:2010bm} is of the form: 
\begin{equation}
\label{eqHphysOperatorWarsawModel}
\widehat{\bf H}_{phys}=\int d^3x \sqrt{-2\sqrt{\widehat{Q}}\widehat{C}^{\rm geo}}(x).
\end{equation}
Let us now compare our physical Hamiltonian operator shown in eq.\,(\ref{FinalOPHphys}) to the one in \cite{Domagala:2010bm} displayed above in eq.\,(\ref{eqHphysOperatorWarsawModel}). A comparison of both models is possible according to the similarity of the first term under the square root in both models, despite that in our model the situation is different, since after the reduction with respect to the second class constraints we are left with four reference fields for all constraints instead of one Klein-Gordon scalar field as a reference field for the Hamiltonian constraint. At the classical level the term $2\sqrt{Q}\sum_{j=1}^3\sqrt{Q^{jj}C_j^{\rm geo}C_j^{\rm geo}}$ can be understood as a contribution to the physical Hamiltonian density associated with the momentum density of the reference fields $\varphi^j$ which would be absent in case we only consider one instead of four reference fields. Thus, the fingerprint of the spatial reference fields encoded in $2\sqrt{Q}\sum_{j=1}^3\sqrt{Q^{jj}C_j^{\rm geo}C_j^{\rm geo}}$ 
at the classical level, caused by the dynamical coupling of this reference fields, also carries over to the quantum theory and yields to the remaining terms under the square root in eq.\,(\ref{FinalOPHphys}) corresponding the the quantization of the classical term $2\sqrt{Q}\sum_{j=1}^3\sqrt{Q^{jj}C^{\rm geo}_jC^{\rm geo}_j}$. Now the operator $\hat{O}^{(j)}_J$ whose square occurs under the second square root consists of a combination of right invariant vector fields and a loop holonomy operator. For a given vertex $v$ of a graph $\gamma$ associated to a given spin network function, there are two different contributions. The first one considers every pair of edges $e,e'$ at $v$ and acts with two right invariant vector fields $X^{e'}_J X^{e}_M$ onto them and afterwards attaches a loop along the edges $e$ and $e'$ to the graph $\gamma$, see figure \ref{fig:action_LQG_Term1}. The second contribution involves for each vertex $v$ and every edge $e$ attached to it an action of one right invariant vector field $X^e_0$ . In this case the loop that acts afterwards goes along the edges $e$ and the edge that one obtains by projecting the edge $e$ onto the $j$-th tangential direction, as can be seen in figure \ref{fig:action_LQG_Term2}. Note that this second contribution depends crucially on the embedding of the individual edges and is a contribution to the operator that is rather unusual. This can for instance be seen in the specific case where the tangent vector to $e$ has a non-vanishing contribution only in one fixed $j$-direction. In this case the loop $h_{\alpha_{e(j)e}}$ is just the identity and since ${\rm Tr}(\tau^M)=0$ the contribution of the second sum just vanishes identically. However, for a generic embedding of the edges with a tangent vector that has non-zero components in all $j$-directions the contribution from the second sum will in general be non-zero.
Similar expressions occur in the regularization of the volume operator in \cite{Thiemann:2007zz} and would also be involved in a point splitting regularization of the area operator. However, in these cases due to the specific structure of the volume and area operator, in particular both involve only covariant contractions, all terms of these kind vanish in the limit when the regulator is removed. For our physical Hamiltonian this is no longer true and the reason for this seems to be its non-covariant form at the level of observables, that is with respect to the scalar field manifold indices. At first glance this seems unusual, but, as we show in appendix \ref{app:GaugeFixing}, this is caused by the particular choice of gauge fixing associated with this model. As can be seen in the presentation in appendix \ref{app:GaugeFixing} the induced shift vector associated with the choice of clocks in this model has at the observable level the form $N^k=\frac{1}{h(Q,P)}\sum\limits_{j=1}^3\sqrt{Q}\sqrt{Q^{jj}}\delta^k_j$ which naturally explains the second embedding dependent term in the physical Hamiltonian. 

Finally, let us mention that as in the models in \cite{Giesel:2007wn,Giesel:2012rb,Han:2015jsa} the observable $C_j^{\rm geo}$ is a constant of motion with respect to the reference time $\tau$, as can be seen by using the properties of the observable map. We have:
\begin{equation*}
\frac{dC^{\rm geo}_j}{d\tau}=
\{ C_j^{\rm geo}(\sigma,\tau),{\bf H}_{phys}\}
=\{O^{(2)}_{\widetilde{c}_j,\widetilde{\varphi}^0}, O^{(2)}_{\widetilde{h},\widetilde{\varphi}^0}\}
=O^{(2)}_{\{\widetilde{c}_j,\widetilde{h}\}^*}
=O^{(2)}_{\{\widetilde{c}_j,\widetilde{h}\}}=0.
\end{equation*}
Furthermore, in the limit of vanishing momentum density of the reference fields $\varphi^j$  as expected the model in 
\cite{Domagala:2010bm} and our model here posses the same physical Hamiltonian and in this sense the generalized model
introduced in this section can be understood as the corresponding four scalar field model associated with the model
introduced in \cite{Domagala:2010bm}. In the context of cosmology it can also be understood as the natural full loop quantum gravity generalization of the APS-model in \cite{Ashtekar:2006wn}. 
\section{Conclusions}
\label{s4}
In this paper we discussed the reduced phase space quantization in the context of Klein-Gordon scalar fields as reference matter. Such models can be understood as a natural generalization of the the APS-model \cite{Ashtekar:2006wn} in the framework of loop quantum cosmology to full loop quantum gravity. The first model that was derived as a generalization along these lines is the model in \cite{Domagala:2010bm}, for which only one Klein-Gordon scalar field was considered. This allows to deparametrize the Hamiltonian constraint and use this Klein-Gordon scalar field as reference matter for the Hamiltonian constraint, whereas the three spatial diffeomorphism constraints are dealt with using Dirac quantization. If we instead choose to consider the spatial diffeomorphism constraints also in the context of a reduced phase space quantization, we need three more additional reference fields. For this reason we presented in section \ref{s2} a model where we couple gravity to four Klein-Gordon scalar fields. We derived the reduced phase space of this model in terms of the corresponding Dirac observables and also computed the associated physical Hamiltonian which generates their dynamics. We have shown in section \ref{s2} that for such a model the reduced quantization program cannot be completed because we obtain a physical Hamiltonian which cannot be quantized
in the context of loop quantum gravity. The main reason for this is that infinitesimal spatial diffeomorphisms $C^{\rm geo}_j$ cannot be implemented as well defined operators in the standard loop quantum gravity representation. They occur in the combination $\delta^{jk}C^{\rm geo}_j C^{\rm geo}_k$ in the physical Hamiltonian that cannot be promoted to a well defined operator. We have discussed the technical details of this aspect at the end of section \ref{s24}.

If we compare the model in section \ref{s2} to the one in \cite{Domagala:2010bm}, we realize that this is an example for a case where Dirac quantization and reduced quantization lead to very different results. If we chose to quantize the spatial diffeomorphisms via Dirac quantization following \cite{Domagala:2010bm}, the quantization can be completed. In contrast if we just add three more Klein-Gordon scalar fields to the model and aim at performing a reduced phase space quantization, this quantization program cannot be completed because we cannot quantize the dynamics on the physical Hilbert space using the usual loop quantum gravity representation. Hence, the Quantum Einstein Equations of such a model cannot be formulated.

Given this negative result for the four Klein-Gordon scalar field model, we generalized this model in section \ref{s3} with the purpose to obtain a quantizable model where the spatial diffeomorphism constraints are treated via reduced phase space quantization. 
Likewise to the seminal dust model introduced in \cite{Kuchar:1995xn,Brown:1994py} we considered a model that contains next
to the four scalar fields that we want to use as reference fields for the spatial diffeomorphism and the Hamiltonian constraint
three further scalar fields. As discussed in detail in section \ref{s31} this model possesses second class constraints.
When we reduce with respect to the second class constraints, we obtain a model with only first class constraints that involves
gravity and four additional scalar fields. The reference field for the Hamiltonian constraint $\varphi^0$ is a standard
Klein-Gordon scalar field likewise to the model in \cite{Domagala:2010bm}. However, the dynamics of the spatial reference field
$\varphi^j$ describe a generalized dynamics, since they are coupled to three additional scalar fields, whose degrees of freedom are reduced when performing the reduction with respect to the second class constraints. In sections \ref{s32} and \ref{s33} we derive the corresponding reduced phase space and present the explicit construction of Dirac observables as it was also done in sections \ref{s22} and \ref{s23}. It turns out that this generalized model has a physical Hamiltonian that can be quantized using techniques of loop quantum gravity, which is discussed in more detail at the end of section \ref{s32}. For the reason that the resulting physical Hamiltonian has a form which slightly differs from the one in \cite{Domagala:2010bm} and the one in the Brown-Kucha\v{r} dust models from \cite{Giesel:2007wn}, we present in section \ref{s34} in detail the regularization and quantization of the physical Hamiltonian operator. As can be seen from equation (\ref{HamDens2}) at the end of section \ref{s32} the physical Hamiltonian density consists of two main contributions. One involves the gravitational part of the Hamiltonian constraint $C^{\rm geo}$ and for this term we used the already existing quantization in the literature. In the context of a usual loop quantum gravity quantization we used the results in \cite{Thiemann:1996ay} and for the Algebraic Quantum Gravity framework we considered the results from \cite{Giesel:2007wn}. For the second contribution, whose form is determined by the choice of one conventional and three generalized Klein-Gordon scalar fields as reference matter and that involves the geometrical part of the spatial diffeomorphisms $C_j^{\rm geo}$, no quantization was available before. In section \ref{s34} we present a regularization for this second contribution, discuss how the regularized operator acts on spin network function and show that we obtain a well defined operator on the physical Hilbert space when the regulator is removed. It turns out that the final operator depends on the particular embedding of the graph. 
Furthermore, we also illustrate how this second contribution can be quantized in the framework of Algebraic Quantum Gravity (AQG) \cite{Giesel:2007wn} as a corresponding algebraic graph-preserving operator. Interestingly,  the possible problematic unusual term which explicitly depends on the embedding naturally becomes the identity operator in a graph-preserving quantization and therefore the quantization within an AQG model is straightforward.

As a consequence, including the generalized model, we have two models available that can be understood as equally justified generalizations of the APS-model to full loop quantum gravity. Their main difference lies in the fact how the spatial diffeomorphism constraints are handled. The first one from \cite{Domagala:2010bm} is obtained in the Dirac quantization program as far as the spatial diffeomorphism constraints are concerned. The model presented in section \ref{s3} on the other hand uses 
a reduced phase space quantization either in usual loop quantum gravity or in the AQG framework. By comparison of their physical Hamiltonians, as done in section \ref{s35}, we get a first hint towards the question in which sense the final models will differ, if we either use Dirac or reduced phase space quantization to handle the diffeomorphism constraints. A next step is  to work with these models and analyze how the different quantization procedure might influence physical properties of the dynamical models. Due to the technical complexity in the full theory, such an analysis is planed at the level of symmetry reduced models beyond the level of homogeneous and isotropic models for which the second contribution in the physical Hamiltonian, involving the spatial diffeomorphism constraints, just vanishes.  This will be a topic for future research and might also give new insights on the 
role of chosen reference matter (clocks) in the context of a reduced phase space quantization of quantum gravity.
As far as the discussion in \cite{Giesel:2012rb} is concerned the new model introduced in this work extends the possible models of type I and can be used to formulate another dynamical model of the Quantum Einstein Equation in the context of loop quantum gravity.
\section*{ACKNOWLEDGMENTS}
A.V. thanks the Heinrich-B\"oll-Foundation for financial support.
K.G. would like to thank Jerzy Lewandowski and Tomasz Pawlowski for illuminating discussions during the 3rd Conference of the
Polish Society on Relativity in Cracow.
\appendix
\section{Comparison of the reduced model with the corresponding gauge fixed model}
\label{app:GaugeFixing}
In this section we want to compare the reduced generalized four scalar field model with its associated gauge fixed model. In case that we start on the partially reduced phase space with respect to the second class constraints $(c^{jj},\Lambda^{jj})$, then the four gauge fixing conditions associated with the Hamiltonian and spatial diffeomorphism constraints read
\begin{eqnarray}
G^{0}=\tau^0-\varphi^0\quad G^j=\sigma^j-\varphi^j.
\end{eqnarray}
Similar to the Brown-Kucha\v{r} dust model in \cite{Giesel:2007wi} we assume that $\tau^0=\tau^0(t)$ does not depend on the spatial coordinates and we assume $\sigma^j$ to depend on the spatial coordinates only. Considering this and the form of the Hamiltonian and spatially diffeomorphism constraint on the partially reduced phase space the stability requirement for the gauge fixing conditions yields
\begin{eqnarray}
{\rm (i)} \quad \frac{dG^0}{dt}& \stackrel{!}{\approx}& 0=\frac{\partial \tau^0}{\partial t}-\frac{n\pi_0}{\sqrt{q}}-n^a\varphi^0_{,a}, \nonumber \\
{\rm (ii)}\quad\, \frac{dG^j}{dt}&\stackrel{!}{\approx}&0= -n\sqrt{q^{cd}\varphi^j_{,c}\varphi^j_{,d}}-n^a\varphi^j_{,a}.
\end{eqnarray}
The lapse function and shift vector induced by this kind of choice for the gauge fixing are given by
\begin{eqnarray}
n & \approx & \frac{\partial\tau^0}{\partial t}\left(-\frac{h}{\sqrt{q}}-\varphi^0_{,a}\sum\limits_{j=1}^3\varphi^a_j\sqrt{q^{cd}\varphi^j_{,c}\varphi^j_{,d}}\right)^{-1}, \nonumber \\
n^a & \approx & -\frac{\partial\tau^0}{\partial t}\frac{\sum\limits_{j=1}^3\left(\varphi^a_j\sqrt{q}\sqrt{q^{cd}\varphi^j_{,c}\varphi^j_{,d}}\right)}
{\left(-h-\varphi^0_{,a}\sum\limits_{j=1}^3\left(\varphi^a_j\sqrt{q}\sqrt{q^{cd}\varphi^j_{,c}\varphi^j_{,d}}\right)\right)},
\end{eqnarray}
where we used that $\pi_0\approx -h$. At the observable level these weak equalities simplify to 
\begin{eqnarray}
O_{n,\{\varphi^0,\varphi^j\}} &=& -\frac{\sqrt{Q}}{h(Q_{jk},P^{jk})}=:N(Q,P) \nonumber \\
O_{n^a,\{\varphi^0,\varphi^j\}} &=& \frac{1}{h(Q_{jk},P^{jk})}\sum\limits_{j=1}^3\sqrt{Q}\sqrt{Q^{jj}}\delta^k_j=:N^k(Q,P)
\end{eqnarray}
with
\begin{eqnarray}
h(Q_{jk},P^{jk}) &:=& \sqrt{-2\sqrt{Q}C^{\rm geo}+2\sqrt{Q}\sum\limits_{j=1}^3\sqrt{Q^{jj}C^{\rm geo}_jC^{\rm geo}_j}}.
\end{eqnarray}
Let us denote the corresponding quantities in the gauge fixed theory by $n_0(q,p)$, $n^k_0(q,p)$ and $h(q,p)$ respectively whose explicit form is given by
\begin{eqnarray}
n_0(q,p)&=& -\frac{\sqrt{q}}{h(q,p)}\nonumber \\
n^k_0(q,p) &=& \frac{1}{h(q,p)}\sum\limits_{j=1}^3\sqrt{q}\sqrt{q^{jj}}\delta^k_j\nonumber \\
h(q,p) &=& \sqrt{-2\sqrt{q}c^{\rm geo}+2\sqrt{q}\sum\limits_{j=1}^3\sqrt{q^{jj}c_j^{\rm geo}c_j^{\rm geo}}}
\end{eqnarray}
This result is also consistent with the condition following from equation (\ref{EOMMjj}) for the gauge fixing chosen above. 
Given this we obtain for the  dynamics of a function $f$ that does not depend on the clock degrees of freedom in the gauge fixed theory:
\begin{eqnarray}
\frac{df}{d\tau} &=&\left(\frac{\partial\tau^0}{\partial t}\right)^{-1}\frac{df}{dt}\Big|_{G^J=0,c=c_k=0,n=n_0,n^k=n^k_0} \nonumber\\
&=&
\left(\frac{\partial\tau^0}{\partial t}\right)^{-1}\int d^3y \left(n_0(q,p)\{f,c^{\rm tot}(y)\}\Big|_{G^J=0,c=c_k=0}
+n^k_0(q,p)\{f,c^{\rm tot}_k(y)\}\Big|_{G^J=0,c=c_k=0}
\right)
\nonumber\\
&=&
\int d^3y \left(-\frac{\sqrt{q}(y)}{h(q,p)}\{f,c^{\rm tot}(y)\}\Big|_{G^J=0,c=c_k=0}
+\frac{1}{h(q,p)}\sum\limits_{j=1}^3\sqrt{q}\sqrt{q^{jj}}(y)\delta^k_j\{f,c^{\rm tot}_k(y)\}\Big|_{G^J=0,c=c_k=0}
\right)
\nonumber\\
&\approx & 
\int d^3y \frac{1}{2h(q,p)}\left(\{f,-2\sqrt{q}c^{\rm tot}(y)\}\Big|_{G^J=0,c=c_k=0}
+\{f,\sum\limits_{j=1}^32\sqrt{q}\sqrt{q^{jj}}(y)\delta^k_j c^{\rm tot}_k(y)\}\Big|_{G^J=0,c=c_k=0}
\right)
\nonumber\\
&=& 
\int d^3y \frac{1}{2h(q,p)}\left(\{f,-2\sqrt{q}c^{\rm geo}(y)\}
+\{f,\sum\limits_{j=1}^32\sqrt{q}\sqrt{q^{jj}}\delta^k_j c^{\rm geo}_k(y)\}
\right)
\nonumber\\
&=&
\int d^3y \frac{1}{2h(q,p)}\left(\{f,-2\sqrt{q}c^{\rm geo}(y)+\sum\limits_{j=1}^32\sqrt{q}\sqrt{q^{jj}}\delta^k_j c^{\rm geo}_k(y)\}
\right)
\nonumber\\
&=& \int d^3y \frac{1}{2h(q,p)}\left(\{f,-2\sqrt{q}c^{\rm geo}(y)+\sum\limits_{j=1}^32\sqrt{q}\sqrt{q^{jj}c^{\rm geo}_jc^{\rm geo}_j}(y)\}
\right)
\nonumber\\
&=& \int d^3y\left(\{f,\sqrt{-2\sqrt{q}c^{\rm geo}(y)+\sum\limits_{j=1}^32\sqrt{q}\sqrt{q^{jj}c^{\rm geo}_jc^{\rm geo}_j}(y)}\}
\right)
\nonumber\\
&=& \{f,\int d^3y h(q,p)(y)\} \nonumber\\
&=&\{f,H^{\rm GF}\}
\end{eqnarray}
with gauge fixed Hamiltonian 
\begin{equation*}
H^{\rm GF}:=\int d^3y h(q,p)(y).
\end{equation*}
We realize that the dynamics of the observables and the dynamics of the gauge fixed theory are identical. 
\section{Observable Construction Formula}
\label{a1}
If $f$ is a scalar on phase space, e.g. some function $g : \chi \mapsto \mathbb{R}$  we claim
\be
\label{claimdiffd}
\{K_{\beta_1}, g(x)\}_{(n)}=\big[\beta_1^{j_1} ... \beta_1^{j_n}v_{j_1} ... v_{j_n}\cdot g\big](x)
\ee
with $v_j\cdot g(x)=\varphi^a_j g_a(x)$.
In order to proof the claim it is of advantage to use that the vector fields mutually commute, that is $[v_j,v_k]=0$ for all $j,k$. Using that spatial derivatives of $\delta^a_b$ vanish we get $0=\partial_c(\delta^b_a)=\partial_c(\varphi^b_j\varphi_a^j)$ from which we can derive the useful identity
\be
\varphi^b_{k,c}= - \varphi^b_j\varphi^j_{,a}\varphi^a_k
\ee
The commutator of two vector fields yields
\ba
[v_j,v_k]&=&
\varphi^a_j\varphi^b_{\ell}\varphi^{\ell}_{,ac}\varphi^c_{k}\partial_b  
- \varphi^a_k\varphi^b_{j,a}\partial_b.
\nonumber\\
&=&
\varphi^a_j\varphi_{\ell}^b\varphi^{\ell}_{,ac}\varphi^c_k\partial_b
- \varphi^a_k\varphi^b_{j,a}\partial_b
\nonumber\\
&=&\varphi^b_{j,c}\varphi^c_k\partial_b - \varphi^a_k\varphi^b_{j,a}\partial_b\nonumber\\
&=&0.
\ea
We will proof the claim in equation (\ref{claimdiff}) by induction. For this purpose it is of advantage
to express $\{K_{\beta_1},\varphi^a_j(x)\}$  in terms of the vector fields $v_j$.
We have 
\ba
\label{PBtovjd}
\{K_{\beta_1},\varphi^a_j(x)\}&=&
-\varphi^a_k(x)\varphi^b_j(x)\{K_{\beta_1},\varphi^k_{,b}(x)\}\nonumber\\
&=& -\varphi^a_k(x)\varphi^b_j(x)\int\limits_\chi d^3y\beta_1^{\ell}(y)\{c_{\ell}^{\rm tot}(y),\varphi^k_{,b}(x)\}\nonumber\\
&=& -\varphi^a_k(x)\varphi^b_j(x)\int\limits_\chi d^3y\beta_1^{\ell}(y)\{\pi_{\ell}(y) + h_{\ell}(y),\varphi^k_{,b}(x)\}\nonumber\\
&=& -\varphi^a_k(x)\varphi^b_j(x)\int\limits_\chi d^3y\beta_1^{\ell}(y)\{\pi_{\ell}(y) ,\varphi^k_{,b}(x)\}\nonumber\\
&=& - \varphi_k^a(x)\varphi^b_j(x)[\beta_1^k]_{,b}(x)\nonumber\\
&=&-\varphi^a_k[v_j\cdot\beta_1^k].
\ea
Here we used in the fourth line that $h_j$ is independent of the reference field momenta $\pi_j$.
Now we can prove the claim by induction. For $n=1$ we get
\be
\{K_{\beta_1},g(x)\}_{(1)}=[\beta_1^j\varphi_j^ag_{,a}](x)=\beta_1^jv_j\cdot g(x).
\ee
Suppose that the claim in equation (\ref{claimdiffd}) is correct up to order $n$, then 
\ba
\{K_{\beta_1},g(x)\}_{(n+1)}
&=&
\beta_1^{j_1} ... \beta_1^{j_n}\{K_{\beta_1},v_{j_1} ... v_{j_n}\cdot g(x)\}\nonumber\\
&=&
\beta_1^{j_1} ... \beta_1^{j_n}\Big(v_{j_1} ... v_{j_n}\{K_{\beta_1}, g(x)\}
+\sum\limits_{\ell=1}^n v_{j_1}  ... v_{j_{\ell-1}}\{K_{\beta_1},\varphi^a_{j_{\ell}}\}\partial_av_{j_{\ell+1}} ... v_{j_n}\cdot g(x)\Big)\nonumber\\
&=&
\beta_1^{j_1} ... \beta_1^{j_n}\Big(v_{j_1} ... v_{j_n}\beta_1^{j_{n+1}}v_{j_{n+1}}\cdot g(x)
-\sum\limits_{\ell=1}^n v_{j_1}  ... v_{j_{\ell-1}}v_{j_\ell}\beta_1^{j_{n+1}}v_{j_{n+1}}v_{j_{\ell+1}} ... v_{j_n}\cdot g(x)\Big)\nonumber\\
&=&
\beta_1^{j_1} ... \beta_1^{j_n}\Big(v_{j_1} ... v_{j_n}\beta_1^{j_{n+1}}v_{j_{n+1}}\cdot g(x)
-\sum\limits_{\ell=1}^n v_{j_1}  ... v_{j_{\ell-1}}v_{j_\ell}\beta_1^{j_{n+1}}v_{j_{\ell+1}} ... v_{j_n}v_{j_{n+1}}\cdot g(x)\Big)\nonumber\\
&=&
\beta_1^{j_1} ... \beta_1^{j_n}\Big(v_{j_1} ... v_{j_n}\beta_1^{j_{n+1}}v_{j_{n+1}}\cdot g(x)
-\big(v_{j_1} ... v_{j_n}\beta_1^{j_{n+1}} - \beta_1^{j_{n+1}}v_{j_1} ... v_{j_n}\big)v_{j_{n+1}}\cdot g(x)\Big)\nonumber\\
&=&
\beta_1^{j_1} ... \beta_1^{j_{n+1}} v_{j_1} ... v_{j_{n+1}}\cdot g(x).
\ea
In the third line we used equation (\ref{PBtovjd}), in the fourth line that the vector fields mutually commute and in the fifth line the Leibniz rule.
\\
Hence the spatially diffeomorphism invariant quantity for $g$ is given by
\be
O_{g,\{\varphi^j\}}^{(1)}(\sigma)=g+\sum\limits_{n=1}^{\infty}\frac{(-1)^n}{n!}\big[\sigma^{j_1} - \varphi^{j_1}\big] ... \big[\sigma^{j_n} - \varphi^{j_n}\big]v_{j_1} ... v_{j_n}\cdot g
.\ee
We have $v_j\varphi^k=\varphi^a_j\varphi^k_{,a}=\delta^k_j$.
Using the abbreviation $\beta_1^j:=\sigma^j - \varphi^j$ we evaluate the action of $v_k$ on the spatially diffeomorphism invariant quantity $O_g^{(1)}(\sigma)$
\ba
\label{vkObs}
v_k\cdot O_{g,\{\varphi^j\}}^{(1)}(\sigma)&=&
v_k\cdot g +v_k\sum\limits_{n=1}^{\infty}\frac{(-1)^n}{n!}\beta_1^{j_1} ... \beta_1^{j_n} v_{j_1} ... v_{j_n}\cdot g\nonumber\\
&=& v_k\cdot g +\sum\limits_{n=1}^{\infty}\frac{n(-1)^n}{n!}\big[v_k\beta_1^j\big]\beta_1^{j_1} ... \beta_1^{j_{n-1}} v_{j_1} ... v_{j_{n-1}}\cdot g
+\frac{(-1)^n}{n!}\beta_1^{j_1} ... \beta_1^{j_n}v_kv_{j_1} ... v_{j_n}\cdot g
\nonumber\\
&=&
v_k\cdot g +\big[v_k\beta_1^j\big]v_j\cdot g
+\sum\limits_{n=1}^{\infty}\frac{(-1)^n}{n!}\beta_1^{j_1} ...\beta_!^{j_n}\Big(\big[v_k\beta_1^j\big]v_j v_{j_1} ... v_{j_n}\cdot g +v_kv_{j_1} ... v_{j_n}\cdot g\Big)\nonumber\\
&=&
v_k\cdot g +\big[v_k(\sigma^j - \delta^j_k)\big]v_j\cdot g
+\sum\limits_{n=1}^{\infty}\frac{(-1)^n}{n!}\beta_1^{j_1} ...\beta_1^{j_n}\Big(\big[v_k\sigma^j - \delta^j_k\big]v_jv_{j_1} ... v_{j_n}\cdot g + v_k v_{j_1} ... v_{j_n}\cdot g\Big)\nonumber\\
&=&
\big[v_k\sigma^j\big]v_j\cdot g
+\sum\limits_{n=1}^{\infty}\frac{(-1)^n}{n!}\beta_1^{j_1} ...\beta_1^{j_n}
\big[v_k\sigma^j\big]v_jv_{j_1} ... v_{j_n}\cdot g
\nonumber\\
&=&
\sum\limits_{n=1}^{\infty}\frac{(-1)^n}{n!}\beta_1^{j_1} ...\beta_1^{j_n}
\big[v_k\sigma^j\big]v_jv_{j_1} ... v_{j_n}\cdot g.
\ea
We realize that for constant $\sigma^j(x)$ the expression $v_k\cdot O_{g,\{\varphi^j\}}^{(1)}(\sigma)$
vanishes meaning that $O_{g,\{\varphi^j\}}^{(1)}(\sigma)$ does not depend on $x$ at all as expected for
a spatially diffeomorphism invariant quantity.
Consequently we have the freedom to choose any $x$ in the expression for $O{g,\{\varphi^j\}}^{(1)}(\sigma)$.
A convenient choice for which $O_{g,\{\varphi^j\}}^{(1)}(\sigma)$  extremely simplifies is to choose $x_{\sigma}$
such that $\varphi^j(x_{\sigma})=\sigma^j$, since then only the $n=0$ term in the whole summation survives.
This requires that $\varphi^j$ is invertible for $j=1,2,3$ which is true because in order that $\varphi^j$ qualifies as a good reference field we have to assume that  $\varphi^j$ are diffeomorphisms.
For a scalar $g$ on $\chi$ we therefore obtain the following explicit integral representation for the spatially diffeomorphism invariant expression
\be
\label{Odiffd}
O_{g,\{\varphi^j\}}^{(1)}(\sigma)=\int\limits_{\chi} d^3x \big|\det(\partial \varphi^j/\partial_x)\big|\delta(\varphi^j(x),\sigma^j) g(x)
.\ee

\section{Constraint Stability Analysis}
\label{b1}

In the following we need to perform the constraint analysis in order to check whether the primary constraints are stable under time evolution
with respect to $H_{\rm primary}$ or if secondary constraints arise. 
Recall that the non-vanishing Poisson brackets on the phase space are given by 
\begin{align*}
 \{  q_{cd}(x), p^{ab}(y)\} &= \kappa \delta^a_{(c} \delta^b_{d)} \delta^{(3)}\!(x,y), \\
 \{ n(x) , p(y) \} &= \delta^{(3)}\!(x,y), \\
 \{ n^a(x) , p_b(y) \} &= \delta^a_b \delta^{(3)}\!(x,y), \\
 \{ \varphi^0(x), \pi_0(y) \} &= \delta^{(3)}\!(x,y),\\
 \{ \varphi^j(x), \pi_k(y) \} &= \delta^j_k \delta^{(3)}\!(x,y), \\
 \{ M_{jj}(x), \Pi^{kk}(y) \} &= \delta^k_j \delta^{(3)}\!(x,y).
\end{align*}
We calculate the Poisson brackets
\begin{align}
 & \dot{z} = \{ z,  H_{\rm primary} \} = \{ p,  H_{\rm primary} \}, \\
 & \dot{z}_a = \{ z_a,  H_{\rm primary} \} = \{p_a, H_{\rm primary}\},\\ 
 & \dot{\Lambda}^{jj} = \{\Lambda^{jj} ,  H_{\rm primary} \}= \{\Pi^{jj}, H_{\rm primary}\}.
\end{align}

\subsection{Secondary Constraint $\dot{z}$}
We calculate 
\begin{align}
  \dot{z} &= \{ z,  H_{\rm primary} \} = \{ p,  H_{\rm primary} \}\\ \nonumber
	  &= \int_{\chi} \! \mathrm{d^3}x \, \{p, h_{\rm primary}\} 
	    = \int_{\chi} \! \mathrm{d^3}x \, \{p, \nu z + \nu^b z_b + \sum\limits_{j=1}^3 \mu_{jj} \Lambda^{jj} + n c^{\rm tot} + n^b c_b^{\rm tot}\}.
\end{align}
The single terms give rise to the contributions:
\begin{enumerate}
 \item $\int_{\chi} \! \mathrm{d^3}x \, \{ p, \nu z\} = \int_{\chi} \! \mathrm{d^3}x \, \nu \{ p , p \} = 0 $
 \item $\int_{\chi} \! \mathrm{d^3}x \, \{ p, \nu^b z_b\} = \int_{\chi} \! \mathrm{d^3}x \, \nu^b \{ p_b , p \} = 0$
 \item $ \int_{\chi} \! \mathrm{d^3}x \, \{ p, \sum\limits_{j=1}^3 \mu_{jj} \Lambda^{jj}\} 
 = \int_{\chi} \! \mathrm{d^3}x \, \sum\limits_{j=1}^3 \mu_{jj} \{ p, \Pi^{jj} \} = 0$
 \item $ \int_{\chi} \! \mathrm{d^3}x \, \{ p, n c^{\rm tot} \} =  -c^{\rm tot} $
 \item $ \int_{\chi} \! \mathrm{d^3}x \, \{ p, n^b c_b^{\rm tot} \}
 =  \int_{\chi} \! \mathrm{d^3}x \, \{p,  n^b \lr c_b^{\rm geo} + c_b^{\varphi} \rr\}
 =  \int_{\chi} \! \mathrm{d^3}x \, n^b \{ p, \pi_I \varphi^I_{,b}\} = 0$
\end{enumerate}

In summary we obtain the secondary constraint
\begin{align}
 \dot{z} = \{ p, H_{\rm primary}\} = -c^{\rm tot}.
\end{align}

\subsection{Secondary Constraint $\dot{z}_a$}
We calculate 

\begin{align}
  \dot{z}_a &= \{ z_a,  H_{\rm primary} \} = \{ p_a,  H_{\rm primary} \}\\ \nonumber
 &= \int_{\chi} \! \mathrm{d^3}x \, \{p_a, h_{\rm primary}\} 
 = \int_{\chi} \! \mathrm{d^3}x \, \{p_a, \nu z + \nu^b z_b + \sum\limits_{j=1}^3 \mu_{jj} \Lambda^{jj} + n c^{\rm tot} + n^b c_b^{\rm tot}\}.
\end{align}
The single terms give rise to the contributions:
\begin{enumerate}
 \item $\int_{\chi} \! \mathrm{d^3}x \, \{ p_a, \nu z\} = \int_{\chi} \! \mathrm{d^3}x \, \nu \{ p_a , p \} = 0 $
 \item $\int_{\chi} \! \mathrm{d^3}x \, \{ p_a, \nu^b z_b\} = \int_{\chi} \! \mathrm{d^3}x \, \nu^b \{ p_ a, p_b\} = 0$
 \item $ \int_{\chi} \! \mathrm{d^3}x \, \{ p_a , \sum\limits_{j=1}^3 \mu_{jj} \Lambda^{jj}\} 
 = \int_{\chi} \! \mathrm{d^3}x \, \sum\limits_{j=1}^3 \mu_{jj} \{ p_a , \Pi^{jj} \} = 0$
 \item $ \int_{\chi} \! \mathrm{d^3}x \, \{ p_a, n c^{\rm tot} \}
 =  \int_{\chi} \! \mathrm{d^3}x \, \{p_a,  n \lr c^{\rm geo} + c^{\varphi} \rr\} = 0$
  \item $ \int_{\chi} \! \mathrm{d^3}x \, \{ p_a, n c^{\rm tot} \} =  -c^{\rm tot}_a $
\end{enumerate}

In summary we obtain the secondary constraint
\begin{align}
 \dot{z}_a = \{ p_a , H_{\rm primary}\} = -c_a^{\rm tot}.
\end{align}

\subsubsection{Secondary Constraint $\dot{\Lambda}^{jj}$}
We calculate 
\begin{align}
  \dot{\Lambda}^{jj} &= \{\Lambda^{jj},  H_{\rm primary}\} = \{\Pi^{jj}, H_{\rm primary}  \}\\ \nonumber
		     &= \int\limits_{\chi} \! \mathrm{d^3}x \, \{ \Pi^{jj} , h_{\rm primary} \} 
		      = \int\limits_{\chi} \! \mathrm{d^3}x \, \{\Pi^{jj} ,\nu z + \nu^b z_b +\rho^k \Phi_k + \sum\limits_{k=1}^3 \mu_{kk} \Lambda^{kk} + n c^{tot} + n^b c_b^{tot} \}
\end{align}
The single terms give rise to the contributions:
\begin{enumerate}
 \item $\int\limits_{\chi} \! \mathrm{d^3}x \, \{ \Pi^{jj}, \nu z \} 
 = \int\limits_{\chi} \! \mathrm{d^3}x \,\nu \{ \Pi^{jj} , p \} = 0 $
 \item $\int\limits_{\chi} \! \mathrm{d^3}x \, \{ \Pi^{jj}, \nu^b z_b \} 
 = \int\limits_{\chi} \! \mathrm{d^3}x \,\nu^b \{ \Pi^{jj} , p_b \} = 0$
 \item $ \int\limits_{\chi} \! \mathrm{d^3}x \{\Pi^{jj} , \sum\limits_{k=1}^3 \mu_{kk} \Lambda^{kk}  \} 
 = \int\limits_{\chi} \! \mathrm{d^3}x \, \sum\limits_{k=1}^3 \mu_{kk} \{ \Pi^{jj} , \Pi^{kk} \} = 0$
 \item $ \int\limits_{\chi} \! \mathrm{d^3}x \, \{\Pi^{jj}, n c^{\rm tot}  \} 
 = n  \ls \frac{(M^{-1})^{j j} (M^{-1})^{j j} \pi_{j} \pi_{j}}{2 \sqrt{q}} 
 -  \frac{1}{2} \sqrt{q} q^{ab} \varphi^j_{,a} \varphi^j_{,b} \rs$
 \item $ \int\limits_{\chi} \! \mathrm{d^3}x \, \{ \Pi^{jj} ,n^b c_b^{\rm tot} \} 
 = \int\limits_{\chi} \! \mathrm{d^3}x \, n^b \{\Pi^{jj}, c_b^{\rm geo} + \pi_I \varphi^I_{,b}  \} = 0 $
\end{enumerate}

The calculation of the fourth term in detail is given by
\begin{align*}
 &\{\Pi^{jj}, n c^{\rm tot}\} = \{ \Pi^{jj}, n \lr c^{\rm geo} + c^{\varphi} \rr\} 
 = n \underbrace{\{\Pi^{jj} , c^{\rm geo} \}}_{= 0} + n \{ \Pi^{jj} , c^{\varphi} \} \\
 & =  n \{\Pi^{jj},   \frac{\pi_0^2}{2 \sqrt{q}} + \frac{1}{2}  \sqrt{q} q^{ab} \varphi^0_{,a} \varphi^0_{,b} 
 + \sum\limits_{k=1}^3 \lr \frac{(M^{-1})^{kk} \pi_k \pi_k}{2 \sqrt{q}} + \frac{1}{2}  \sqrt{q} M_{kk} \varphi^k_{,a} \varphi^k_{,b}\rr  \} \\
 & =n \sum\limits_{k=1}^3 \frac{\pi_k \pi_k}{2\sqrt{q}} \{ \Pi^{jj} , (M^{-1})^{kk} \}
 + n \sum\limits_{k=1}^3 \frac{1}{2} \sqrt{q} q^{ab} \varphi^k_{,a} \varphi^k_{,b} \underbrace{\{ \Pi^{jj} , M_{kk} \}}_{-\delta^j_k \delta^{(3)}\!(x,y)} \\
 & =-n \sum\limits_{k=1}^3 \frac{\pi_{k} \pi_{k}}{2 \sqrt{q}} (M^{-1})^{k \ell} (M^{-1})^{k \ell}
 \underbrace{\{ \Pi^{jj} , M_{\ell \ell} \}}_{-\delta^j_{\ell} \delta^{(3)}\!(x,y)} 
 - n \frac{1}{2} \sqrt{q} q^{ab} \varphi^j_{,a} \varphi^j_{,b}\delta^{(3)}\!(x,y) \\
 &= n  \sum\limits_{k=1}^3 \frac{(M^{-1})^{k j} (M^{-1})^{k j} \pi_k \pi_k}{2 \sqrt{q}} \delta^{(3)}\!(x,y)
 - n \frac{1}{2} \sqrt{q} q^{ab} \varphi^j_{,a} \varphi^j_{,b} \delta^{(3)}\!(x,y)\\
 &= n  \ls \frac{(M^{-1})^{j j} (M^{-1})^{j j} \pi_{j} \pi_{j}}{2 \sqrt{q}} 
 -  \frac{1}{2} \sqrt{q} q^{ab} \varphi^j_{,a} \varphi^j_{,b} \rs \delta^{(3)}\!(x,y)
\end{align*}

In summary we obtain the secondary constraint
\begin{align}
 \dot{\Lambda}^{jj} = \{ \Pi^{jj} , H_{\rm primary} \} 
 =  n  \ls \frac{(M^{-1})^{j j} (M^{-1})^{j j} \pi_{j} \pi_{j}}{2 \sqrt{q}} 
 -  \frac{1}{2} \sqrt{q} q^{ab} \varphi^j_{,a} \varphi^j_{,b} \rs  =: c^{jj}.
\end{align}

\subsection{Constraint Stability Analysis - Tertiary Constraints}
In the next step we need to check the stability of the secondary constraints $\{ c^{\rm tot}, c_a^{\rm tot}, c^{jj}\}$
with respect to $H_{\rm primary}$. For writing comfort we define $M^{00} := (M^{-1})^{00} := \mathbbm{1}_3$
and $I,J=0,1,2,3$.

\subsubsection{Tertiary Constraint $\dot{c}^{tot}(n)$}
We define the smeared constraint $c^{tot}(n) := \int\limits_{\chi} \! \mathrm{d^3}x \, n(x) c^{tot}(x)$
and calculate
\begin{align*}
 &\{c^{tot}(n) , H_{\rm primary} \}\\
 &= \int\limits_{\chi} \! \mathrm{d^3}x \int\limits_{\chi} \! \mathrm{d^3}y \,
 \lr \{ n(x) c^{\rm tot}(x) , \nu(y) z(y)\} + \{n(x) c^{\rm tot}(x) ,\nu^b(y) z_b(y) \} \rr \\
 &+ \int\limits_{\chi} \! \mathrm{d^3}x \int\limits_{\chi} \! \mathrm{d^3}y \,
 \{  n(x) c^{\rm tot}(x) , \sum\limits_{k=1}^3 \mu_{kk}(y) \Lambda^{kk}(y)\} \\
 &+ \int\limits_{\chi} \! \mathrm{d^3}x \int\limits_{\chi} \! \mathrm{d^3}y \,
 \lr \{  n(x) c^{\rm tot}(x), n'(y) c^{\rm tot}(y)\} + \{ n(x) c^{\rm tot}(x) ,  n'^{b}(y) c_b^{\rm tot}(y)\} \rr.
 \end{align*}

For the single terms we get the expressions:
 \begin{enumerate}
  \item $\int\limits_{\chi} \! \mathrm{d^3}x \int\limits_{\chi} \! \mathrm{d^3}y \, \{ n(x) c^{\rm tot}(x),  \nu(y) z(y) \}
  = \int\limits_{\chi} \! \mathrm{d^3}x \, \nu(x) c^{\rm tot}(x) = c^{\rm tot}(\nu)$
  \item $\int\limits_{\chi} \! \mathrm{d^3}x \int\limits_{\chi} \! \mathrm{d^3}y \, \{ n(x) c^{\rm tot}(x) ,  \nu^b(y) z_b(y)\}= 0$
  \item $\int\limits_{\chi} \! \mathrm{d^3}x \int\limits_{\chi} \! \mathrm{d^3}y \, \{  n(x) c^{tot}(x) , \mu_{kk}(y) \Pi^{kk}(y)\}
        = - \int\limits_{\chi} \! \mathrm{d^3}x \, \sum\limits_{j=1}^3 \mu_{jj}(x) c^{jj}(x) : = -c(\mu)$
  \item $\int\limits_{\chi} \! \mathrm{d^3}x \int\limits_{\chi} \! \mathrm{d^3}y \, \{n(x) c^{\rm tot}(x), n'(y) c^{\rm tot}(y)\}\\
  = \int\limits_{\chi} \! \mathrm{d^3}x \int\limits_{\chi} \! \mathrm{d^3}y \, \lr \{ n(x) c^{\rm geo}(x), n'(y) c^{\rm geo}(y)\} +\{ n(x) c^{\rm geo}(x), n'(y) c^{\varphi}(y)\} \right.\\ 
  \left. +\{ n(x) c^{\varphi}(x), n'(y) c^{\rm geo}(y)\}+\{ n(x) c^{\varphi}(x), n'(y) c^{\varphi}(y)\} \rr$   
  \item $\int\limits_{\chi} \! \mathrm{d^3}x \int\limits_{\chi} \! \mathrm{d^3}y \, \{n(x) c^{\rm tot}(x), n'^b(y) c_b^{\rm tot}(y)\} \\
   = \int\limits_{\chi} \! \mathrm{d^3}x \int\limits_{\chi} \! \mathrm{d^3}y \,
   \lr \{ n(x) c^{\rm geo}(x), n'^b(y) c_b^{\rm geo}(y)\} +\{ n(x) c^{\rm geo}(x), n'^b(y) c_b^{\varphi}(y)\} \right. \\ 
    \left.+\{ n(x) c^{\varphi}(x), n'^b(y) c_b^{\rm geo}(y)\}+\{ n(x) c^{\varphi}(x), n'^b(y) c_b^{\varphi}(y)\}\rr$
 \end{enumerate}
 Since the fourth and the fifth term are rather lengthy, we display them here separately divided again into subterms
\begin{enumerate}
 \item[4.1] $\int\limits_{\chi} \! \mathrm{d^3}x \int\limits_{\chi} \! \mathrm{d^3}y \,\{ n(x) c^{\rm geo}(x), n'(y) c^{\rm geo}(y)\}
	    = \{ c^{\rm geo}(n), c^{\rm geo}(n') \}
	    = \vec{c}^{\rm geo} \lr q^{-1} \ls n \, \mathrm{d}n' - n' \, \mathrm{d}n \rs\rr$
 \item[4.2]$\int\limits_{\chi} \! \mathrm{d^3}x \int\limits_{\chi} \! \mathrm{d^3}y \,\{ n(x) c^{\rm geo}(x), n'(y) c^{\varphi}(y)\}
	    = - \int\limits_{\chi} \! \mathrm{d^3}x \, n n'   \frac{1}{\sqrt{q}} c^{\varphi} q_{ab} p^{ab}  
	      + \int\limits_{\chi} \! \mathrm{d^3}x \, n n'   \frac{4}{\sqrt{q}} \ls \sum\limits_{J=0}^3 \frac{1}{2}  M_{JJ} \sqrt{q} \varphi^J_{,e} \varphi^J_{f}\rs  p^{ef}$
 \item[4.3] $\int\limits_{\chi} \! \mathrm{d^3}x \int\limits_{\chi} \! \mathrm{d^3}y \,\{ n(x) c^{\varphi}(x), n'(y) c^{\rm geo}(y)\}
	    =  \int\limits_{\chi} \! \mathrm{d^3}x \, n n'   \frac{1}{\sqrt{q}} c^{\varphi} q_{ab} p^{ab}  
	      - \int\limits_{\chi} \! \mathrm{d^3}x \, n n'   \frac{4}{\sqrt{q}} \ls \sum\limits_{J=0}^3 \frac{1}{2}  M_{JJ} \sqrt{q} \varphi^J_{,e} \varphi^J_{f}\rs  p^{ef}$
\item[4.4] $\int\limits_{\chi} \! \mathrm{d^3}x \int\limits_{\chi} \! \mathrm{d^3}y \, \{ n(x) c^{\varphi}(x), n'(y) c^{\varphi}(y)\}
	    = \int\limits_{\chi} \! \mathrm{d^3}x \, \lr n \, n'_{,b} - n' \, n_{,b}\rr q^{ab} c_a^{\varphi}
	    = \vec{c}^{\varphi} \lr q^{-1} \ls n \, \mathrm{d}n' - n' \, \mathrm{d}n \rs \rr$ 	  
\end{enumerate}
and
\begin{enumerate}
 \item[5.1]$\int\limits_{\chi} \! \mathrm{d^3}x \int\limits_{\chi} \! \mathrm{d^3}y \,\{ n(x) c^{\rm geo}(x), n'^b(y) c_b^{\rm geo}(y)\}
	= \{ c^{\rm geo}(n), \vec{c}^{\rm geo}(\vec{n}') \} = -c^{\rm geo} \lr \mathcal{L}_{\vec{n}'} n \rr $ 
 \item[5.2] $\int\limits_{\chi} \! \mathrm{d^3}x \int\limits_{\chi} \! \mathrm{d^3}y \, \{ n(x) c^{\rm geo}(x), n'^b(y) c_b^{\varphi}(y)\} = 0$	
 \item[5.3] $\int\limits_{\chi} \! \mathrm{d^3}x \int\limits_{\chi} \! \mathrm{d^3}y \,\{ n(x) c^{\varphi}(x), n'^b(y) c_b^{\rm geo}(y)\} \\
 = \int\limits_{\chi} \! \mathrm{d^3}x \, \ls \frac{n}{2} \sum\limits_{J=0}^3 (M^{-1})^{JJ} \pi_J \pi_J \rs 
 \lr \mathcal{L}_{\vec{n}'} \frac{1}{\sqrt{q}} \rr 
 + \int\limits_{\chi} \! \mathrm{d^3}x \, \ls \frac{n}{2} \sum\limits_{J=0}^3 M_{JJ} \varphi^J_{,a} \varphi^J_{,b} \rs 
  \mathcal{L}_{\vec{n}'}\! \lr \sqrt{q} q^{ab} \rr$
 \item[5.4] $\int\limits_{\chi} \! \mathrm{d^3}x \int\limits_{\chi} \! \mathrm{d^3}y \,\{ n(x) c^{\varphi}(x), n'^b(y) c_b^{\varphi}(y)\} \\
 = \int\limits_{\chi} \! \mathrm{d^3}x \, \frac{n}{\sqrt{q}} \lr \mathcal{L}_{\vec{n}'} \lr \frac{1}{2} \sum\limits_{J=0}^3 (M^{-1})^{JJ} \pi_J \pi_J \rr \rr
 + \int\limits_{\chi} \! \mathrm{d^3}x \, n \sqrt{q} q^{cd} \lr \mathcal{L}_{\vec{n}'}\lr  \frac{1}{2} \sum\limits_{J=0}^3 M_{JJ} \varphi^J_{,c} \varphi^J_{,d} \rr \rr$
\end{enumerate}
The addition of $5.3$ and $5.4$ leads to $-c^{\varphi} \lr \mathcal{L}_{\vec{n}'} n \rr $.

\subsubsection{Tertiary Constraint $\dot{\vec{c}}^{\rm tot}(\vec{n})$}
We define the smeared constraint $\vec{c}^{tot}(\vec{n}) := \int\limits_{\chi} \! \mathrm{d^3}x \, n^a(x) c_a^{\rm tot}(x)$
and calculate
\begin{align*}
 &\{\vec{c}^{\rm tot}(\vec{n}) , H_{\rm primary}\}\\
 &= \int\limits_{\chi} \! \mathrm{d^3}x \int\limits_{\chi} \! \mathrm{d^3}y \,
 \lr \{ n^a(x) c_a^{tot}(x),  \nu(y) z(y) \} + \{ n^a(x) c_a^{\rm tot}(x) ,  \nu^b(y) z_b(y) \} \rr \\
 &+ \int\limits_{\chi} \! \mathrm{d^3}x \int\limits_{\chi} \! \mathrm{d^3}y \,
 \{ n^a(x) c_a^{\rm tot}(x), \sum\limits_{k=1}^3 \mu_{kk}(y) \Lambda^{kk}(y) \} \\
 &+ \int\limits_{\chi} \! \mathrm{d^3}x \int\limits_{\chi} \! \mathrm{d^3}y \,
 \lr \{ n^a(x) c_a^{\rm tot}(x), n'(y) c^{\rm tot}(y)\} + \{n^a(x) c_a^{\rm tot}(x) ,  n'^{b}(y) c_b^{\rm tot}(y) \} \rr.
\end{align*}

\begin{enumerate}
 \item $\int\limits_{\chi} \! \mathrm{d^3}x \int\limits_{\chi} \! \mathrm{d^3}y \, \{ n^a(x) c_a^{\rm tot}(x) , \nu(y) z(y) \} =0$
 \item $\int\limits_{\chi} \! \mathrm{d^3}x \int\limits_{\chi} \! \mathrm{d^3}y \, \{ n^a(x) c_a^{\rm tot}(x) , \nu^b(y) z_b(y)\}
 = \int\limits_{\chi} \! \mathrm{d^3}x \, \nu^a(x) c_a^{\rm tot}(x) = \vec{c}^{\rm tot}(\vec{\nu})$
 \item $\int\limits_{\chi} \! \mathrm{d^3}x \int\limits_{\chi} \! \mathrm{d^3}y \, \{ n^a(x) c_a^{\rm tot}(x) ,  \mu_{kk}(y) \Pi^{kk}(y)\}= 0$
 \item $\int\limits_{\chi} \! \mathrm{d^3}x \int\limits_{\chi} \! \mathrm{d^3}y \, \{ n^a(x) c_a^{\rm tot}(x), n'(y) c^{\rm tot}(y)\} \\
 = \int\limits_{\chi} \! \mathrm{d^3}x \int\limits_{\chi} \! \mathrm{d^3}y \, 
  \lr \{ n^a(x) c_a^{\rm geo}(x), n'(y) c^{\rm geo}(y)\}+\{ n^a(x) c_a^{\rm geo}(x), n'(y) c^{\varphi}(y)\} \right.\\
  \left. +\{ n^a(x) c_a^{\varphi}(x), n'(y) c^{\rm geo}(y)\} +\{ n^a(x) c_a^{\varphi}(x), n'(y) c^{\varphi}(y)\} \rr $
 \item $\int\limits_{\chi} \! \mathrm{d^3}x \int\limits_{\chi} \! \mathrm{d^3}y \, \{n^a(x) c_a^{\rm tot}(x) ,  n'^{b}(y) c_b^{\rm tot}(y) \} \\
 = \int\limits_{\chi} \! \mathrm{d^3}x \int\limits_{\chi} \! \mathrm{d^3}y \, \lr \{n^a(x) c_a^{\rm geo}(x) , n'^{b}(y) c_b^{\rm geo}(y)\} 
 +\{n^a(x) c_a^{\rm geo}(x) , n'^{b}(y) c_b^{\varphi}(y)\}\right. \\
 \left. +\{n^a(x) c_a^{\varphi}(x) , n'^{b}(y) c_b^{\rm geo}(y)\} +\{n^a(x) c_a^{\varphi}(x) , n'^{b}(y) c_b^{\varphi}(y)\}\rr$ 
\end{enumerate}
Since the fourth and the fifth term are rather lengthy, we display them here separately divided again into subterms
\begin{enumerate}
 \item[4.1] $\int\limits_{\chi} \! \mathrm{d^3}x \int\limits_{\chi} \! \mathrm{d^3}y \, \{n^a(x) c_a^{\rm geo}(x) , n'(y) c^{\rm geo}(y)\} 
	= \{ \vec{c}^{\rm geo}(\vec{n}), c^{\rm geo}(n') \} =  c^{\rm geo}(\mathcal{L}_{\vec{n}} n')$
 \item[4.2] $\int\limits_{\chi} \! \mathrm{d^3}x \int\limits_{\chi} \! \mathrm{d^3}y \, \{ n^a(x) c_a^{\rm geo}(x), n'(y) c^{\varphi}(y)\}\\
 = -\int\limits_{\chi} \! \mathrm{d^3}x \, \ls \frac{n'}{2} \sum\limits_{J=0}^3 (M^{-1})^{JJ} \pi_J \pi_J \rs 
 \lr \mathcal{L}_{\vec{n}} \frac{1}{\sqrt{q}} \rr 
 - \int\limits_{\chi} \! \mathrm{d^3}x \, \ls \frac{n'}{2} \sum\limits_{J=0}^3 M_{JJ} \varphi^J_{,a} \varphi^J_{,b} \rs 
  \mathcal{L}_{\vec{n}}\! \lr \sqrt{q} q^{ab} \rr$
 \item[4.3] $\int\limits_{\chi} \! \mathrm{d^3}x \int\limits_{\chi} \! \mathrm{d^3}y \, \{ n^a(x) c_a^{\varphi}(x), n'(y) c^{\rm geo}(y)\} = 0$
 \item[4.4] $\int\limits_{\chi} \! \mathrm{d^3}x \int\limits_{\chi} \! \mathrm{d^3}y \, \{ n^a(x) c_a^{\varphi}(x), n'(y) c^{\varphi}(y)\} \\
 = -\int\limits_{\chi} \! \mathrm{d^3}x \, \frac{n'}{\sqrt{q}} \lr \mathcal{L}_{\vec{n}} \lr \frac{1}{2} \sum\limits_{J=0}^3 (M^{-1})^{JJ} \pi_J \pi_J \rr \rr
   - \int\limits_{\chi} \! \mathrm{d^3}x \, n' \sqrt{q} q^{cd} \lr \mathcal{L}_{\vec{n}}\lr  \frac{1}{2} \sum\limits_{J=0}^3 M_{JJ} \varphi^J_{,c} \varphi^J_{,d} \rr \rr$
\end{enumerate}
and
\begin{enumerate}
 \item[5.1] $\int\limits_{\chi} \! \mathrm{d^3}x \int\limits_{\chi} \! \mathrm{d^3}y \, \{n^a(x) c_a^{\rm geo}(x) , n'^{b}(y) c_b^{\rm geo}(y)\} 
	= \{ \vec{c}^{\rm geo}(\vec{n}), \vec{c}^{\rm geo}(\vec{n'}) \} = \vec{c}^{\rm geo}(\mathcal{L}_{\vec{n}} \vec{n}')$ 
 \item[5.2] $\int\limits_{\chi} \! \mathrm{d^3}x \int\limits_{\chi} \! \mathrm{d^3}y \, \{n^a(x) c_a^{\rm geo}(x) , n'^{b}(y) c_b^{\varphi}(y)\} = 0 $
 \item[5.3] $\int\limits_{\chi} \! \mathrm{d^3}x \int\limits_{\chi} \! \mathrm{d^3}y \, \{n^a(x) c_a^{\varphi}(x) , n'^{b}(y) c_b^{\rm geo}(y)\} = 0$
 \item[5.4] $\int\limits_{\chi} \! \mathrm{d^3}x \int\limits_{\chi} \! \mathrm{d^3}y \, \{n^a(x) c_a^{\varphi}(x) , n'^{b}(y) c_b^{\varphi}(y)\} \\
	=\int\limits_{\chi} \! \mathrm{d^3}x \, \sum\limits_{J=0}^3 
	  \lr  n^a   n'^{b}_{,a} \varphi^J_{,b} \pi_J  - n^a_{,b}  n'^{b} \varphi^J_{,a}  \pi_J  \rr
	= \int\limits_{\chi} \! \mathrm{d^3}x \, \lr  n^b   n'^{a}_{,b}   -  n'^{b} n^a_{,b} \rr c^{\varphi}_a
	 = \vec{c}^{\varphi}\lr \mathcal{L}_{\vec{n}} \vec{n}' \rr $
\end{enumerate}
The addition of $4.3$ and $4.4$ leads to $c^{\varphi} \lr \mathcal{L}_{\vec{n}} n' \rr $.

\subsubsection{Tertiary Constraint $\dot{c}(r)$}
We define the smeared constraint $c(r) := \int\limits_{\chi} \! \mathrm{d^3}x \, \sum\limits_{j=1}^3  r_{jj}(x) c^{jj}(x)$
and calculate
\begin{align*}
 &\{ c(r) , H_{\rm primary}\}\\
 &= \int\limits_{\chi} \! \mathrm{d^3}x \int\limits_{\chi} \! \mathrm{d^3}y \,
 \lr \{\sum\limits_{j=1}^3 r_{jj}(x) c^{jj}(x) ,  \nu(y) z(y)\} + \{\sum\limits_{j=1}^3  r_{jj}(x) c^{jj}(x) ,  \nu^b(y) z_b(y)\} \rr \\
 &+ \int\limits_{\chi} \! \mathrm{d^3}x \int\limits_{\chi} \! \mathrm{d^3}y \,
 \{ \sum\limits_{j=1}^3 r_{jj}(x) c^{jj}(x) , \sum\limits_{k=1}^3 \mu_{kk}(y) \Lambda^{kk}(y) \} \\
 &+ \int\limits_{\chi} \! \mathrm{d^3}x \int\limits_{\chi} \! \mathrm{d^3}y \,
 \lr \{ \sum\limits_{j=1}^3  r_{jj}(x) c^{jj}(x) ,  n'(y) c^{\rm tot}(y) \} 
 + \{\sum\limits_{j=1}^3  r_{jj}(x) c^{jj}(x) ,  n'^{b}(y) c_b^{\rm tot}(y)\} \rr.
\end{align*}
step by step.

\begin{enumerate}
 \item $\int\limits_{\chi} \! \mathrm{d^3}x \int \! \mathrm{d^3}y \,
      \{\sum\limits_{j=1}^3  r_{jj}(x) c^{jj}(x) , \nu(y) z(y) \} \\
      = \int\limits_{\chi} \nu(x)\sum\limits_{j=1}^3  r_{jj}(x) \ls  \frac{1}{2} \ls \frac{ (M^{-1})^{j j} (M^{-1})^{j j} \pi_{j} \pi_{j}}{\sqrt{q}} 
      -\sqrt{q} q^{ab} \varphi^j_{,a} \varphi^j_{,b} \rs \rs(x)\\
      = \int\limits_{\chi} \! \mathrm{d^3}x \, \frac{\nu}{n}(x) \sum\limits_{j=1}^3  r_{jj}(x) c^{jj}(x) := c(\frac{\nu}{n}r)$
 \item  $\int\limits_{\chi} \! \mathrm{d^3}x \int\limits_{\chi} \! \mathrm{d^3}y \,
      \{ \sum\limits_{j=1}^3  r_{jj}(x) c^{jj}(x) , \nu^b(y) z_b(y) \} = 0$    
 \item $\int\limits_{\chi} \! \mathrm{d^3}x \int\limits_{\chi} \! \mathrm{d^3}y \,
 \{ \sum\limits_{j=1}^3 r_{jj}(x) c^{jj}(x) , \sum\limits_{k=1}^3 \mu_{kk}(y) \Lambda^{kk}(y) \} \\
  = -\sum\limits_{j=1}^3   r_{jj}(x) \mu_{jj}(x)  \ls n \frac{ \lr (M^{-1})^{j j} \rr^3  \pi_{j} \pi_{j}}{\sqrt{q}} \rs(x) $
 \item $\int\limits_{\chi} \! \mathrm{d^3}x \int\limits_{\chi} \! \mathrm{d^3}y \, \sum\limits_{j=1}^3  r_{jj}(x) c^{jj}(x) ,  n'(y) c^{\rm tot}(y) \} \neq 0$
 \item $\int\limits_{\chi} \! \mathrm{d^3}x \int\limits_{\chi} \! \mathrm{d^3}y \,  \{\sum\limits_{j=1}^3  r_{jj}(x) c^{jj}(x) ,  n'^{b}(y) c_b^{\rm tot}(y)\} \neq 0$
\end{enumerate}
The terms $4.$ and $5.$ need not to be calculated explicitly, since we can get rid of them with the help of the
Lagrange multiplier $\mu_{jj}$.

In summary all contributions are proportional to $c^{\rm tot}, \vec{c}^{\rm tot}(\vec{n}), c(r)$ or can be
eliminated with the help of the Lagrange multiplier $\mu_{jj}$. Therefore, no tertiary constraints arise.

\section{Calculation of $\beta_{jj}$}
\label{c1}
We calculate $\beta_{jj}$ in more detail, starting from
\begin{align*}
\beta_{jj}(x) &=-\int\limits_\chi d^3y\sqrt{q}\frac{(M_{jj})^3}{n\pi^2_j} \{c^{\rm tot}(x),c^{jj}(y)\} \\
	     &=-\int\limits_\chi d^3y\sqrt{q}\frac{(M_{jj})^3}{n\pi^2_j}\left( \{c^{\rm geo}(x),c^{jj}(y)\} + \{c^{\phi}(x),c^{jj}(y)\} \right)\\
	     &= -\sqrt{q}\frac{(M_{jj})^3}{n\pi^2_j}
	     \left(- \frac{1}{2\sqrt{q}} c^{jj} \left(p^{ab} q_{ab}\right)- n \varphi^j_{,a}\varphi^j_{,b} \; p^{ab}\right.\\
	     &\left.-\left[ \frac{(M^{-1})^{jj} \pi_j}{\sqrt{q}}\right]
	      \left[n \sqrt{q} q^{ab}\varphi^j_{,b}\right]_{,a} -\left[ \sqrt{q}M_{jj} q^{ab}\varphi^j_{,b}\right]_{,a}
	      \left[n\frac{(M^{-1})^{jj}(M^{-1})^{jj} \pi_j} {\sqrt{q}} \right]\right)\\
	      &= \frac{(M_{jj})^3}{2n\pi^2_j} c^{jj} \left(p^{ab} q_{ab}\right)
	      + \sqrt{q}\frac{(M_{jj})^3}{\pi^2_j}\varphi^j_{,a}\varphi^j_{,b} \; p^{ab}
	      +\frac{(M_{jj})^2}{n\pi_j} \left[n \sqrt{q} q^{ab}\varphi^j_{,b}\right]_{,a} 
	      +\frac{M_{jj}}{\pi_j} \left[ \sqrt{q}M_{jj} q^{ab}\varphi^j_{,b}\right]_{,a},
\end{align*}
where we used the results of the calculations of the Poisson brackets, displayed in the following.

First we calculate

\begin{align*}
 \int\limits_\chi d^3y \{\kappa c^{\rm geo}(x),c^{jj}(y)\} 
	      =& \int\limits_\chi d^3y 
	      \Big\{\left[\frac{1}{\sqrt{q}} \left( q_{ac} q{bd} - \frac{1}{2} q_{ab} q_{cd}\right) p^{ab} p^{cd} \right.
	      \left. - \sqrt{q} R^{(3)} + 2 \sqrt{q} \Lambda \right](x), \\
	      &  \left[\frac{n}{2}\left[\sum_{k=1}^3\frac{(M^{-1})^{jk}(M^{-1})^{jk}\pi_k\pi_k} {\sqrt{q}}
	      -\sqrt{q}q^{ef}\varphi^j_{,e}\varphi^j_{,f}\right] \right] (y) \Big\} \\
	      \stackrel{M^{jk} \neq 0 \text{ for } j = k }{=} & \int\limits_\chi d^3y 
	      \left[\frac{1}{\sqrt{q}} \left( q_{ac} q{bd} - \frac{1}{2} q_{ab} q_{cd}\right)  \right](x)
	      \left[\frac{n}{2} (M^{-1})^{jj}(M^{-1})^{jj}\pi_j\pi_j\right] (y)\\
	      &\{p^{ab}(x) p^{cd} (x), \frac{1}{\sqrt{q}} (y)\}\\
	      &- \int\limits_\chi d^3y 
	      \left[\frac{1}{\sqrt{q}} \left( q_{ac} q{bd} - \frac{1}{2} q_{ab} q_{cd}\right)  \right](x)
	      \left[\frac{n}{2}\varphi^j_{,e}\varphi^j_{,f}\right] (y)\\
	      &\{p^{ab}(x) p^{cd} (x), \sqrt{q}(y) q^{ef}(y)\}\\
	      &= \int\limits_\chi d^3y 
	      \left[\frac{1}{\sqrt{q}} \left( q_{ac} q{bd} - \frac{1}{2} q_{ab} q_{cd}\right)  \right](x)
	      \left[\frac{n}{2} (M^{-1})^{jj}(M^{-1})^{jj}\pi_j\pi_j\right] (y)\\
	      &\left(p^{ab}(x) \{p^{cd} (x), \frac{1}{\sqrt{q}} (y)\}+ p^{cd} (x) \{p^{ab}(x), \frac{1}{\sqrt{q}} (y)\}\right)\\
	      &- \int\limits_\chi d^3y 
	      \left[\frac{1}{\sqrt{q}} \left( q_{ac} q{bd} - \frac{1}{2} q_{ab} q_{cd}\right)  \right](x)
	      \left[\frac{n}{2}\varphi^j_{,e}\varphi^j_{,f}\right] (y)\\
	      &\left(p^{ab}(x) q^{ef}(y) \{p^{cd} (x), \sqrt{q}(y) \} +p^{ab}(x) \sqrt{q}(y) \{p^{cd} (x), q^{ef} (y)\}\right.\\
	      &\left. +  p^{cd}(x) q^{ef}(y)\{p^{ab}(x), \sqrt{q}(y) \}+  p^{cd}(x) \sqrt{q}(y) \{p^{ab}(x),  q^{ef}(y)\}\right)\\
	      &= \int\limits_\chi d^3y 
	      \left[\frac{1}{\sqrt{q}} \left( q_{ac} q{bd} - \frac{1}{2} q_{ab} q_{cd}\right)  \right](x)
	      \left[\frac{n}{2} (M^{-1})^{jj}(M^{-1})^{jj}\pi_j\pi_j\right] (y)\\
	      &\left(p^{ab}(x) (-\frac{1}{2\sqrt{q}} q^{ef})(y) (- \kappa \delta^c_{(e}\delta^d_{f)}) \delta^{(3)}(x,y) 
	      + p^{cd} (x) (-\frac{1}{2\sqrt{q}} q^{ef})(y) (- \kappa \delta^a_{(e}\delta^b_{f)}) \delta^{(3)}(x,y)\right)\\
	      &- \int\limits_\chi d^3y 
	      \left[\frac{1}{\sqrt{q}} \left( q_{ac} q{bd} - \frac{1}{2} q_{ab} q_{cd}\right)  \right](x)
	      \left[\frac{n}{2}\varphi^j_{,e}\varphi^j_{,f}\right] (y)\\
	      &\left(p^{ab}(x) (\frac{1}{2}\sqrt{q} q^{ef} q^{gh})(y) (- \kappa \delta^c_{(g}\delta^d_{h)}) \delta^{(3)}(x,y) 
	      +p^{cd}(x) (\frac{1}{2}\sqrt{q} q^{ef} q^{gh})(y) (- \kappa \delta^a_{(g}\delta^b_{h)}) \delta^{(3)}(x,y) \right.\\
	      &\left. + p^{ab}(x) (-\sqrt{q} q^{ge} q^{hf})(y) (- \kappa \delta^c_{(g}\delta^d_{h)}) \delta^{(3)}(x,y)
	      +  p^{cd}(x) (-\sqrt{q} q^{ge} q^{hf})(y) (- \kappa \delta^a_{(g}\delta^b_{h)}) \delta^{(3)}(x,y) \right)\\
	      &\stackrel{q^{ab}= q^{ba}}{=} \kappa \frac{1}{2\sqrt{q}}\left[\frac{1}{\sqrt{q}} \left( q_{ac} q_{bd} - \frac{1}{2} q_{ab} q_{cd}\right)  \right]
	      \left[\frac{n}{2} (M^{-1})^{jj}(M^{-1})^{jj}\pi_j\pi_j\right]
	      \left(p^{ab} q^{cd} + p^{cd} q^{ab} \right)\\
	      &+\kappa \frac{1}{2}\sqrt{q} \left[\frac{1}{\sqrt{q}} \left( q_{ac} q_{bd} - \frac{1}{2} q_{ab} q_{cd}\right)  \right]
	      \left[\frac{n}{2}\varphi^j_{,e}\varphi^j_{,f}\right] \left(p^{ab}  q^{ef} q^{cd}   +p^{cd} q^{ef} q^{ab} \right)\\
	      &-\kappa \sqrt{q} \left[\frac{1}{\sqrt{q}} \left( q_{ac} q_{bd} - \frac{1}{2} q_{ab} q_{cd}\right)  \right]
	      \left[\frac{n}{2}\varphi^j_{,e}\varphi^j_{,f}\right] \left(p^{ab}  q^{ce} q^{df}   +p^{cd} q^{ae} q^{bf} \right)\\
	      &= \kappa \frac{1}{2\sqrt{q}}\frac{1}{\sqrt{q}} \left[\frac{n}{2} (M^{-1})^{jj}(M^{-1})^{jj}\pi_j\pi_j\right]
	      \left(-p^{ab} q_{ab}\right)\\
	      &+\kappa\sqrt{q}  \frac{1}{\sqrt{q}} \left[\frac{n}{2}\varphi^j_{,e}\varphi^j_{,f}\right] 
	      \left(- \frac{1}{2} p^{ab}  q_{ab} q^{ef}\right)
	      +\kappa \sqrt{q} \frac{1}{\sqrt{q}} 
	      \left[\frac{n}{2}\varphi^j_{,e}\varphi^j_{,f}\right] \left( -2 p^{ef}+p^{ab} q_{ab} q^{ef}  \right)\\
	      &= -\kappa \frac{1}{2\sqrt{q}} \frac{n}{2} \left[ \frac{(M^{-1})^{jj}(M^{-1})^{jj}\pi_j\pi_j}{\sqrt{q}} 
	      +\sqrt{q} q^{ef} \varphi^j_{,e}\varphi^j_{,f}\right]
	      \left(p^{ab} q_{ab}\right)-\kappa n \varphi^j_{,a}\varphi^j_{,b} \; p^{ab} \\
	      &= -\kappa \frac{1}{2\sqrt{q}} c^{jj} \left(p^{ab} q_{ab}\right)-\kappa n \varphi^j_{,a}\varphi^j_{,b} \; p^{ab}. \\
\end{align*}

Next we calculate

\begin{align*}
 \int\limits_\chi d^3y \{c^{\varphi}(x),c^{jj}(y)\} 
	      =& \int\limits_\chi d^3y \Big\{\left[\frac{\pi_0^2}{2\sqrt{q}}+\frac{1}{2}\sqrt{q}q^{ab}\varphi^0_{,a}\varphi^0_{,b}
	      +\sum\limits_{\ell=1}^3\left(\frac{(M^{-1})^{\ell \ell}\pi_{\ell}\pi_{\ell}}{2\sqrt{q}}
	      +\frac{1}{2}\sqrt{q}M_{\ell \ell}q^{ab}\varphi^{\ell}_{,a}\varphi^{\ell}_{,b}\right)\right](x), \\
	      &  \left[\frac{n}{2}\left[\sum_{k=1}^3\frac{(M^{-1})^{jk}(M^{-1})^{jk}\pi_k\pi_k} {\sqrt{q}}
	      -\sqrt{q}q^{cd}\varphi^j_{,c}\varphi^j_{,d}\right] \right] (y) \Big\} \\
	      = &- \int\limits_\chi d^3y  \Big\{ \left[ \sum\limits_{\ell=1}^3 \frac{(M^{-1})^{\ell \ell}\pi_{\ell}\pi_{\ell}}{2\sqrt{q}}\right](x),
	      \left[\frac{n}{2} \sqrt{q} q^{cd}\varphi^j_{,c}\varphi^j_{,d} \right] (y)\Big\}	\\
	      &+ \int\limits_\chi d^3y \Big\{\left[\sum\limits_{\ell=1}^3 \frac{1}{2}\sqrt{q}M_{\ell \ell}q^{ab}\varphi^{\ell}_{,a}\varphi^{\ell}_{,b}\right](x),
	      \left[\frac{n}{2}\sum_{k=1}^3\frac{(M^{-1})^{jk}(M^{-1})^{jk}\pi_k\pi_k} {\sqrt{q}} \right](y)\Big\}\\
	      = &- \int\limits_\chi d^3y  \left[ \sum\limits_{\ell=1}^3 \frac{(M^{-1})^{\ell \ell}}{2\sqrt{q}}\right](x),
	      \left[\frac{n}{2} \sqrt{q} q^{cd}\right] (y) \{\pi_{\ell}(x) \pi_{\ell}(x), \varphi^j_{,c}(y) \varphi^j_{,d}(y)\}\\
	      &+ \int\limits_\chi d^3y \left[\sum\limits_{\ell=1}^3 \frac{1}{2}\sqrt{q}M_{\ell \ell}q^{ab}\right](x),
	      \left[\frac{n}{2}\sum_{k=1}^3\frac{(M^{-1})^{jk}(M^{-1})^{jk}} {\sqrt{q}} \right](y)
	      \{\varphi^{\ell}_{,a}(x) \varphi^{\ell}_{,b}(x), \pi_k(y) \pi_k (y) \}\\
	      \stackrel{q^{ab}=q^{ba}}{=} &- \int\limits_\chi d^3y  \left[ \sum\limits_{\ell=1}^3 \frac{(M^{-1})^{\ell \ell}}{2\sqrt{q}}\right](x),
	      \left[\frac{n}{2} \sqrt{q} q^{cd}\right] (y) \,
	      4 \pi_{\ell}(x) \varphi^j_{,d} (y) \{\pi_{\ell}(x), \varphi^j_{,c} (y)\}\\
	      &+ \int\limits_\chi d^3y \left[\sum\limits_{\ell=1}^3 \frac{1}{2}\sqrt{q}M_{\ell \ell}q^{ab}\right](x),
	      \left[\frac{n}{2}\sum_{k=1}^3\frac{(M^{-1})^{jk}(M^{-1})^{jk}} {\sqrt{q}} \right](y)\,
	      4 \varphi^{\ell}_{,b}(x) \pi_k (y) \{\varphi^{\ell}_{,a}(x), \pi_k (y) \}\\
	      = &- \int\limits_\chi d^3y  \left[ \sum\limits_{\ell=1}^3 \frac{(M^{-1})^{\ell \ell}}{\sqrt{q}}\right](x)
	      \left[n \sqrt{q} q^{cd}\right] (y) \,
	      \pi_{\ell}(x) \varphi^j_{,d} (y) \left( - \delta^j_{\ell} \frac{\partial}{\partial y^c} \delta^{(3)}(x,y) \right)\\
	      &+ \int\limits_\chi d^3y \left[\sum\limits_{\ell=1}^3 \sqrt{q}M_{\ell \ell}q^{ab}\right](x)
	      \left[n \sum_{k=1}^3\frac{(M^{-1})^{jk}(M^{-1})^{jk}} {\sqrt{q}} \right](y)\,
	      \varphi^{\ell}_{,b}(x) \pi_k (y) \left(\delta^{\ell}_k \frac{\partial}{\partial x^c} \delta^{(3)}(x,y) \right)\\
	      = & \int\limits_\chi d^3y  \left[ \frac{(M^{-1})^{jj} \pi_j}{\sqrt{q}}\right](x)
	      \left[n \sqrt{q} q^{cd}\varphi^j_{,d}\right] (y) \,
	        \left(\frac{\partial}{\partial y^c} \delta^{(3)}(x,y) \right)\\
	      &+ \int\limits_\chi d^3y \left[\sum\limits_{k=1}^3 \sqrt{q}M_{kk} q^{ab}\varphi^k_{,b}\right](x)
	      \left[n\frac{(M^{-1})^{jk}(M^{-1})^{jk} \pi_k} {\sqrt{q}} \right](y)\,
	        \left( \frac{\partial}{\partial x^c} \delta^{(3)}(x,y) \right)\\
	      = &-\left[ \frac{(M^{-1})^{jj} \pi_j}{\sqrt{q}}\right]
	      \left[n \sqrt{q} q^{ab}\varphi^j_{,b}\right]_{,a} -\left[\sum\limits_{k=1}^3 \sqrt{q}M_{kk} q^{ab}\varphi^k_{,b}\right]_{,a}
	      \left[n\frac{(M^{-1})^{jk}(M^{-1})^{jk} \pi_k} {\sqrt{q}} \right]\\
	      \stackrel{M^{jk} \neq 0 \text{ for } j = k }{=} &-\left[ \frac{(M^{-1})^{jj} \pi_j}{\sqrt{q}}\right]
	      \left[n \sqrt{q} q^{ab}\varphi^j_{,b}\right]_{,a} -\left[ \sqrt{q}M_{jj} q^{ab}\varphi^j_{,b}\right]_{,a}
	      \left[n\frac{(M^{-1})^{jj}(M^{-1})^{jj} \pi_j} {\sqrt{q}} \right].
\end{align*}


\begin{thebibliography}{99}

\parskip -5pt

\bibitem{Giesel:2007wn}
  K.~Giesel and T.~Thiemann,
  ``Algebraic quantum gravity (AQG). IV. Reduced phase space quantisation of loop quantum gravity,''
  Class.\ Quant.\ Grav.\  {\bf 27} (2010) 175009
  doi:10.1088/0264-9381/27/17/175009
  [arXiv:0711.0119 [gr-qc]].  

  
\bibitem{Domagala:2010bm}
  M.~Domagala, K.~Giesel, W.~Kaminski and J.~Lewandowski,
  ``Gravity quantized: Loop Quantum Gravity with a Scalar Field,''
  Phys.\ Rev.\ D {\bf 82} (2010) 104038
  doi:10.1103/PhysRevD.82.104038
  [arXiv:1009.2445 [gr-qc]].  
  
  
\bibitem{Husain:2011tk}
  V.~Husain and T.~Pawlowski,
  ``Time and a physical Hamiltonian for quantum gravity,''
  Phys.\ Rev.\ Lett.\  {\bf 108} (2012) 141301
  doi:10.1103/PhysRevLett.108.141301
  [arXiv:1108.1145 [gr-qc]].
  
  
\bibitem{Domagala:2012tq}
  M.~Domagala, M.~Dziendzikowski and J.~Lewandowski,
  ``The polymer quantization in LQG: massless scalar field,''
  arXiv:1210.0849 [gr-qc].  
         
\bibitem{Rovelli:all} C. Rovelli.  What is observable in classical and quantum
gravity? {\it Class. Quantum Grav.} {\bf 8} (1991), 297-316.\\
C. Rovelli.  Quantum reference systems. {\it Class. Quantum Grav.} {\bf 8}
(1991), 317-332.\\
C. Rovelli.  Time in quantum gravity: physics beyond
the Schrodinger regime. {\it Phys. Rev.} {\bf D43} (1991), 442-456.\\
C. Rovelli. Quantum mechanics without time: a model. {\it Phys. Rev.}
{\bf D42} (1990), 2638-2646.

\bibitem{Dittrich04} B. Dittrich. Partial and complete observables for Hamiltonian constrained systems,
Gen. Rel. Grav. {\bf 39}, (2009) 1891-1927.\\
B. Dittrich. Partial and complete observables for canonical general relativity,
Class. Quant. Grav. {\bf 23} (2006) 6155-6184.

\bibitem{Thiemann2004}  T. Thiemann,
Reduced phase space quantization and Dirac observables,
Class. Quant. Grav.  {\bf 23} (2006) 1163-1180.  

\bibitem{Giesel:2012rb}
  K.~Giesel and T.~Thiemann,
  ``Scalar Material Reference Systems and Loop Quantum Gravity,''
  Class.\ Quant.\ Grav.\  {\bf 32} (2015) 135015
  doi:10.1088/0264-9381/32/13/135015
  [arXiv:1206.3807 [gr-qc]].
  
 
\bibitem{Kuchar:1995xn}
  K.~V.~Kuchar and J.~D.~Romano,
  ``Gravitational constraints which generate a lie algebra,''
  Phys.\ Rev.\ D {\bf 51} (1995) 5579
  doi:10.1103/PhysRevD.51.5579
  [gr-qc/9501005].    
  
\bibitem{Bicak:1997bx}
  J.~Bicak and K.~V.~Kuchar,
  ``Null dust in canonical gravity,''
  Phys.\ Rev.\ D {\bf 56} (1997) 4878
  doi:10.1103/PhysRevD.56.4878
  [gr-qc/9704053]. 
  
  
\bibitem{Brown:1994py}
  J.~D.~Brown and K.~V.~Kuchar,
  ``Dust as a standard of space and time in canonical quantum gravity,''
  Phys.\ Rev.\ D {\bf 51} (1995) 5600
  doi:10.1103/PhysRevD.51.5600
  [gr-qc/9409001].  
 
  
\bibitem{Ashtekar:2006wn} 
  A.~Ashtekar, T.~Pawlowski and P.~Singh,
  "Quantum Nature of the Big Bang: Improved dynamics,''
  Phys.\ Rev.\ D {\bf 74}, 084003 (2006)
  doi:10.1103/PhysRevD.74.084003
  [gr-qc/0607039].  
  
 
\bibitem{Giesel:2006uj}
  K.~Giesel and T.~Thiemann,
  "Algebraic Quantum Gravity (AQG). I. Conceptual Setup,''
  Class.\ Quant.\ Grav.\  {\bf 24} (2007) 2465
  doi:10.1088/0264-9381/24/10/003
  [gr-qc/0607099].

\bibitem{Giesel:2007wi}
  K.~Giesel, S.~Hofmann, T.~Thiemann and O.~Winkler,
  {\it Manifestly Gauge-Invariant General Relativistic Perturbation Theory. I. Foundations},
  Class.\ Quant.\ Grav.\  {\bf 27} (2010) 055005
  [arXiv:0711.0115 [gr-qc]].

\bibitem{Dapor:2017rwv}
  A.~Dapor and K.~Liegener,
  "Cosmological Effective Hamiltonian from full Loop Quantum Gravity Dynamics,"
  Phys.\ Lett.\ B {\bf 785} (2018) 506
  doi:10.1016/j.physletb.2018.09.005
  [arXiv:1706.09833 [gr-qc]].  
  
\bibitem{Han:2015jsa}
  Y.~Han, K.~Giesel and Y.~Ma,
  ``Manifestly gauge invariant perturbations of scalar–tensor theories of gravity,''
  Class.\ Quant.\ Grav.\  {\bf 32} (2015) 135006
  doi:10.1088/0264-9381/32/13/135006
  [arXiv:1501.04947 [gr-qc]].

\bibitem{Henneaux1992}  M. Henneaux, C. Teitelboim,
\textit{Quantization of gauge systems},
Princeton University Press (1992).

\bibitem{Lewandowski:2005jk}
  J.~Lewandowski, A.~Okolow, H.~Sahlmann and T.~Thiemann,
  ``Uniqueness of diffeomorphism invariant states on holonomy-flux algebras,''
  Commun.\ Math.\ Phys.\  {\bf 267} (2006) 703
  doi:10.1007/s00220-006-0100-7
  [gr-qc/0504147].
  
\bibitem{Fleischhack:2005}
  C.~Fleischhack
  ``Representations of the Weyl Algebra in Quantum Geometry,''
  Commun.\ Math.\ Phys.\  {\bf 285} (2009) 67
  doi: 10.1007/s00220-008-0593-3
   [gr-qc/0504147].

  
\bibitem{Ashtekar:1991kc}
  A.~Ashtekar and C.~J.~Isham,
  ``Representations of the holonomy algebras of gravity and nonAbelian gauge theories,''
  Class.\ Quant.\ Grav.\  {\bf 9} (1992) 1433
  doi:10.1088/0264-9381/9/6/004
  [hep-th/9202053].

\bibitem{Ashtekar:1993wf} 
  A.~Ashtekar and J.~Lewandowski,
  ``Representation theory of analytic holonomy C* algebras,''
  gr-qc/9311010.  

\bibitem{Thiemann:2007zz}
  T.~Thiemann,
  ``Modern canonical quantum general relativity,''
  gr-qc/0110034. 

\bibitem{Ashtekar:2004eh}
  A.~Ashtekar and J.~Lewandowski,
  ``Background independent quantum gravity: A Status report,''
  Class.\ Quant.\ Grav.\  {\bf 21} (2004) R53
  doi:10.1088/0264-9381/21/15/R01
  [gr-qc/0404018].
  
 \bibitem{Rovelli:2010wq}
  C.~Rovelli,
  ``A new look at loop quantum gravity,''
  Class.\ Quant.\ Grav.\  {\bf 28} (2011) 114005
  doi:10.1088/0264-9381/28/11/114005
  [arXiv:1004.1780 [gr-qc]].
 
\bibitem{Giesel:2012ws}
  K.~Giesel and H.~Sahlmann,
  ``From Classical To Quantum Gravity: Introduction to Loop Quantum Gravity,''
  PoS QGQGS {\bf 2011} (2011) 002
  [arXiv:1203.2733 [gr-qc]].     
  
  \bibitem{Giesel:2017wgh}
  K.~Giesel,
  ``The kinematical Setup of Quantum Geometry: A Brief Review,''
  arXiv:1707.03059 [gr-qc].
  
  \bibitem{Giesel:2017jzj} 
  K.~Giesel,
  'Quantum Geometry,'
  doi:10.1142/9789813220003 0002.
  
  \bibitem{Bodendorfer:2016uat}
  N.~Bodendorfer,
  ``An elementary introduction to loop quantum gravity,''
  arXiv:1607.05129 [gr-qc].
  
\bibitem{Laddha:2011mk}
  A.~Laddha and M.~Varadarajan,
  Class.\ Quant.\ Grav.\  {\bf 28} (2011) 195010
  doi:10.1088/0264-9381/28/19/195010
  [arXiv:1105.0636 [gr-qc]].
  

 \bibitem{Kuchar:1991}
  K.~V.~Kuchar and C.~G.~Torre,
"The Harmonic gauge in canonical gravity" 
Phys.\ Rev.\ D {\bf 44} (1991) 3116
doi: 10.1103/PhysRevD.44.3116 
 
  \bibitem{Giesel:2017mfc}
  K.~Giesel and A.~Oelmann,
  Acta Phys.\ Polon.\ Supp.\  {\bf 10} (2017) 339.
  doi:10.5506/APhysPolBSupp.10.339
  

          
\bibitem{Rovelli:1994ge}
  C.~Rovelli and L.~Smolin,
  ``Discreteness of area and volume in quantum gravity,''
  Nucl.\ Phys.\ B {\bf 442} (1995) 593
   Erratum: [Nucl.\ Phys.\ B {\bf 456} (1995) 753]
  doi:10.1016/0550-3213(95)00150-Q, 10.1016/0550-3213(95)00550-5
  [gr-qc/9411005].
  
\bibitem{Ashtekar:1997fb}
  A.~Ashtekar and J.~Lewandowski,
  ``Quantum theory of geometry. 2. Volume operators,''
  Adv.\ Theor.\ Math.\ Phys.\  {\bf 1} (1998) 388
  [gr-qc/9711031]. 

\bibitem{Thiemann1998}
  T.~Thiemann,
  ''Closed formula for the matrix elements of a volume operator in canonica quantum gravity.''
  J.\ Math.\ Phys.{\bf 39} (1998) 3347-71.
  [gr-qc/9606091]
  
\bibitem{Thiemann:1996ay}
  T.~Thiemann,
  "Anomaly - free formulation of nonperturbative, four-dimensional Lorentzian quantum gravity,''
  Phys.\ Lett.\ B {\bf 380} (1996) 257
  doi:10.1016/0370-2693(96)00532-1
  [gr-qc/9606088].  
  
\bibitem{Thiemann:1996}   
 T.~Thiemann, 
 "Quantum Spin Dynamics (QSD)"
 Class.\ Quant. \ Grav.\ {\bf 15} (1998) 839-873
 doi: 10.1088/0264-9381/15/4/011
 [arXiv:9606089 [gr-qc]].   
 
\bibitem{Giesel:2006uk}
  Giesel~K. and Thiemann~T.,
   "Algebraic Quantum Gravity (AQG). II. Semiclassical Analysis"
    Class.\ Quant.\ Grav.\ {\bf 24} (2007) 2499-2564
    doi: 10.1088/0264-9381/24/10/004
    [gr-qc/0607100].

\bibitem{Ashtekar:2000hv}
  A.~Ashtekar, D.~Marolf, J.~Mourao and T.~Thiemann,
  "Constructing Hamiltonian quantum theories from path integrals in a diffeomorphism-invariant context,''
  Class.\ Quant.\ Grav.\  {\bf 17} (2000) 4919
  doi:10.1088/0264-9381/17/23/310
  [quant-ph/9904094].
  
 

\end{thebibliography}
\end{document}